\documentclass[a4paper,12pt]{article}
\usepackage[english]{babel}
\usepackage[utf8x]{inputenc}
\usepackage[T1]{fontenc}

\usepackage[top=1.25in, bottom = 1.25in, right=1.25in,left=1.25in]{geometry} %ecta

\usepackage{amsmath}
\usepackage{graphicx}
\usepackage[colorlinks=true, allcolors=black]{hyperref}
\usepackage{setspace}
\usepackage{apacite}
\usepackage{dsfont}
\usepackage{booktabs}
\usepackage{threeparttablex} % for Notes with longtable
\usepackage{threeparttable}
\usepackage{chngcntr}		% For A.1 numbering of graphs in Appendix
\usepackage{microtype}     % to avoid that apacite does not respect margins
\usepackage{caption}        % caption same width as figure
\usepackage{rotating}
\usepackage{pdflscape}    % Or lscape
\usepackage{afterpage}
\usepackage{geometry}
\usepackage[titletoc]{appendix}
\usepackage{amssymb}
\usepackage{subcaption}
\usepackage{ragged2e}
\usepackage{tabularx}
\usepackage{bbm}
\usepackage{cancel}
\usepackage{threeparttable}
\usepackage{enumitem}

\usepackage{tikz}
\usetikzlibrary{shapes,decorations,arrows,calc,arrows.meta,fit,positioning}
\tikzset{
	-Latex,auto,node distance =1 cm and 1 cm,semithick,
	state/.style ={ellipse, draw, minimum width = 0.7 cm},
	point/.style = {circle, draw, inner sep=0.04cm,fill,node contents={}},
	bidirected/.style={Latex-Latex,dashed},
	el/.style = {inner sep=2pt, align=left, sloped}
}
\tikzset{
	vertex/.style = {
		circle,
		fill            = black,
		outer sep = 2pt,
		inner sep = 1pt,
	}
}
\tikzstyle{line} = [draw, -latex']
\usetikzlibrary{arrows,decorations.markings,patterns,calc}

\usepackage{changes}

\usepackage[hang,flushmargin]{footmisc}

\newcommand{\bigCI}{\mathrel{\text{\scalebox{1.07}{$\perp\mkern-10mu\perp$}}}} %% CIA symbol

\newcommand\blfootnote[1]{%
  \begingroup
  \renewcommand\thefootnote{}\footnote{#1}%
  \endgroup
}

\newcommand\IF{
\mathbbm{IF}
}

\newcommand{\la}{\mathcal{L}^2_a}
\newcommand{\ip}[1]{\langle #1\rangle}
\newcommand{\xg}{\mathcal{X}_g}
\newcommand{\ta}{\mathcal{T}_a}
\newcommand{\tap}{\mathcal{T}_{a'}}

\newcommand{\xgp}{\mathcal{X}_{g'}}
\newcommand{\1}{\mathbbm{1}}

\newtheorem{assump}{Assumption}
\newtheorem{thm}{Theorem}[section]

\newtheorem{corr}{Corollary}[section]
\usepackage{enumitem}

\usepackage{titlesec}

\titleformat{\section}
  {\normalfont\large\bfseries} % font styling
  {\thesection}{1em}{}         % numbering, spacing, title

\titleformat{\subsection}
  {\normalfont\normalsize\bfseries}
  {\thesubsection}{1em}{}

\title{
\vspace{-30pt}
Heterogeneity Analysis with Heterogeneous Treatments
\blfootnote{The paper was circulated and presented previously under different titles. We would like to thank Yingying Dong, Clint Harris, Han Hong, Martin Huber, Guido Imbens, Anders Kock, Michael Lechner, Yingying Lee, Elena Manresa, Jana Mareckova, Pedro Sant'Anna, Julian Sch\"ussler, Tymon S{\l}oczy{\'{n}}ski, Jeffrey Smith, Jann Spiess, Anthony Strittmatter, Stefan Wager, and the participants of various seminars and workshops for valuable comments and discussions. ChatGPT 4o mini-high, 5.1 and 5.2 were used for proofreading and parts of Section 8 and Appendices C and G. All remaining errors are ours.}
} 

\author{Phillip Heiler\thanks{Aarhus University. Aarhus Center for Econometrics (ACE), Department of Economics and Business Economics, TrygFonden's Centre for Child Research, Universitetsbyen 51, DK-8000 Aarhus C, Denmark, \href{mailto:pheiler@econ.au.dk}{pheiler@econ.au.dk}. Part of this research was conducted during main affiliation with Harvard University, Department of Economics. 02138 Cambridge MA, United States. The research is supported by the Danish National Research Foundation grant DNRF186.} \and Michael C. Knaus\thanks{University of T\"ubingen, Mohlstra{\ss}e 36, 72074 T\"ubingen, Germany. Michael C. Knaus is also affiliated with IZA@LISER, \href{mailto:michael.knaus@unisg.ch}{michael.knaus@uni-tuebingen.de}. }}

\date{\today}
\begin{document}
\maketitle

% \onehalfspacing
\doublespacing

\vspace{-13pt}
\begin{abstract}
\singlespacing
Analysis of effect heterogeneity at the group level is standard practice in empirical treatment evaluation. However, treatments analyzed are often aggregates of multiple underlying treatments which are themselves heterogeneous, e.g.~different modules of training programs. In these settings, conventional approaches such as comparing (adjusted) differences-in-means across groups can produce misleading conclusions when underlying treatment propensities differ systematically between groups. This paper develops a novel decomposition framework that disentangles contributions of effect heterogeneity and distinct components of treatment heterogeneity to observed group-level differences. We propose debiased machine learning estimators that adapt to many discrete and/or continuous treatments and limited overlap. We revisit a widely documented gender gap in training returns of an active labor market policy. The decomposition reveals that it is almost entirely driven by women being treated differently rather than different returns from identical treatments. In particular, women are disproportionately targeted towards vocational training tracks with lower unconditional returns.
\\[4ex]

\end{abstract}

\thispagestyle{empty}

\clearpage
\setcounter{page}{1}
\setstretch{1.45}
% \section{Wording}

% The general setup and in particular the wording:
% \begin{itemize}
%     \item Effective treatment $T \in \mathcal{T}$
%     \item Potential outcomes $Y(t)$
%     \item Treatment aggregation / aggregated treatments / analyzed treatment $\mathcal{T}_a \subset \mathcal{T}$ for $a \in \mathcal{A}$.%\{1,...,A\}$
%     \item Confounding and potential heterogeneity variables variables $X \in \mathcal{X}$ ensuring $Y(t) \bigCI T | X$ Think more about role of X    
%     \item Groups / Heterogeneity group / covariate aggregates $\mathcal{X}_g \subset \mathcal{X}$ for $g \in \mathcal{G}$. %\{1,...,G\}$
% \end{itemize}

% Elese
% \begin{enumerate}
% \item Comment Section 5.1 - aggregate UC
% \item Layout equations x...(name form)[right]
% \item Woman and female
%     \item findings immediately apply to model based approaches to $E[Y|\ta,\xg]$ (even when correctly specified). Discuss.   
% \end{enumerate}

\section{Introduction}

Heterogeneous causal effects are ubiquitous in applied economic research. They are used to evaluate and design policies or interventions for units with different background characteristics. A large literature develops and applies identification and estimation strategies for causal parameters that explicitly take such heterogeneity into account, see e.g.~\citeA{Athey2017} or \citeA{Imbens2024CausalSciences} for recent overviews. 

Common heterogeneity analysis compares treatment effects for different groups. 
Identification, estimation and interpretation of such group heterogeneity at different granularity are well-understood for homogeneous treatments, i.e.~when each treatment level corresponds to a single potential outcome. In practice, however, the \textit{analyzed treatment} is often heterogeneous in the sense that it represents aggregations of a more complex \textit{effective treatment}. For example, an aggregated treatment could be given by an ex-ante design, e.g.~access to a training program that then offers different modules. It can also stem from manual ex-post aggregation, e.g.~by merging categories or binning different doses of a continuous treatment. Both are prevalent, though often implicit, in applied research.  \citeA{Caetano2025CausalTreatment} discuss 19 studies in labor, urban, environmental, and health economics as well as peer effects and time use research falling into these categories.\footnote{Additionally, ex-ante designs are often observed in experimental studies \cite<e.g.>{Schochet2008DoesStudy,Avdeenko2025CostPakistan}. Ex-post aggregation is common in observational studies \cite<e.g.>{Carneiro2011EstimatingEducation,Dobbie2013GettingCity,Aizer2015JuvenileJudges,Carneiro2017AverageIndonesia, Cornelissen2018WhoAttendance,Kurtz2022DoesDepoliticization,Arteaga2023PARENTALATTAINMENT}. In either case, heterogeneity analyses usually focus on heterogeneity between groups defined by characteristics such as age, gender, or household income.}
% In both cases, the dimension of the effective treatment can often be large. 
The consequences of heterogeneous treatments for interpretation of estimands aimed at average, i.e.~unconditional, effects are well-studied \cite<e.g.>{VanderWeele2013CausalTreatment,Caetano2025CausalTreatment}.
In which sense heterogeneity analyses are informative about policy mechanisms and effect heterogeneity once treatments are also heterogeneous, however, is an open question.

This paper studies the causal content of canonical heterogeneity estimands in the presence of treatment heterogeneity. 
We focus on comparisons of differences-in-means (DiM) between groups used frequently in randomized controlled trials (RCTs) and the respective covariate-adjusted DiM used in observational studies \cite{Imbens2015CausalSciences,Chernozhukov2024AppliedAIb}. 
These statistical estimands are scrutinized in a setting in which the analyzed treatment is defined by aggregating effective treatments and heterogeneity groups are defined by aggregated confounders of the effective treatment. This \textit{double aggregation} setting spans a variety of empirically relevant cases. In particular, it nests heterogeneity analyses at different granularity under homogeneous and heterogeneous treatments as well as a variety of other estimands from the literature.

We provide a detailed decomposition showing how canonical heterogeneity estimands in the double aggregation setting capture that groups react differently to the same treatment (effect heterogeneity), groups are treated differently (treatment heterogeneity), and their non-trivial interactions. In particular, the decomposition highlights that group differences may solely be driven by treatment heterogeneity and therefore not necessarily reflect effect heterogeneity. Similarly, effect heterogeneity could be masked by treatment heterogeneity. Consequently, interpretation and policy conclusions derived from group differences obtained from, e.g.,~interacted regressions, matching or causal machine learning methods can be misleading.

The decomposition consists of several estimands capturing qualitatively distinct sources of group heterogeneity: 1) A single parameter that can vary if and only if there is effect heterogeneity across groups in the effective treatment and 2) multiple parameters stemming from the presence of heterogeneous treatments. In particular, the latter are informative about different forms of \textit{targeting}, i.e.~over- or under-representation in treatments based on potential outcomes. More specifically, there are three distinct targeting parameters: (i) group targeting with respect to average outcomes, (ii) group targeting with respect to group outcomes, and (iii) individualized targeting with respect to individualized outcomes beyond group membership. This suggests that researchers who analyze group
heterogeneity in the presence of heterogeneous treatments should square
their interpretations with the potential presence of all of these channels even if the effective treatment and its confounders are not observed. The decomposition provides the key elements for a principled discussion of the interpretation of group differences that
goes beyond loosely appealing to selection into the effective treatment.

All decomposition components are identified if the effective treatment is observed and a standard unconfoundedness assumption applies. They can then be used to quantify the contributions of effect and treatment heterogeneity as well as to test necessary conditions for ``real'' effect homogeneity across groups, i.e.~effect homogeneity on the level of the effective treatment. The targeting parameters further enable ex-post evaluation of group specific assignment mechanisms and their difference as measured by their contribution to the (adjusted) DiM.

\begin{figure}[!h]
    \centering
    \caption{Decomposition of Interaction Coefficient in Job Corps Application} \label{fig:decomp-intro}
    \begin{subfigure}{1\textwidth}
        \centering
        \includegraphics[width=0.99\textwidth]{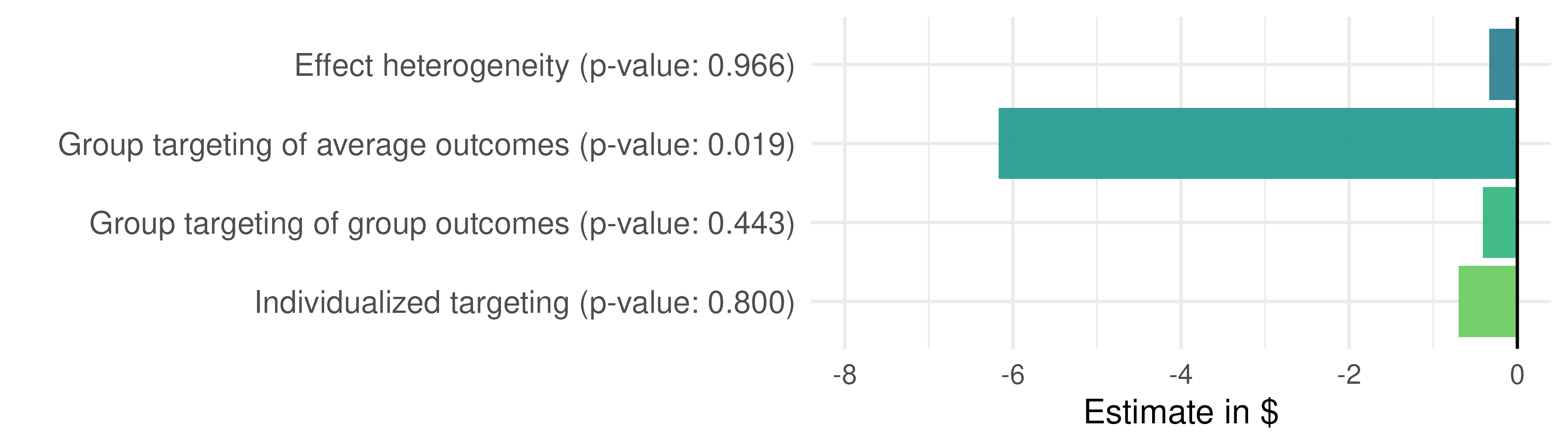}
    \end{subfigure}
    \subcaption*{\textit{Notes:} The graph shows the estimated decomposition introduced in Section \ref{sec_decompositionALL1} for interaction coefficient $\beta_3$ in regression \eqref{eq_REG1_INTRO}. Point estimates and $p$-values are obtained using the estimation and inference procedures discussed in Section \ref{sec_estimation1}. Consider Section \ref{sec_jobcorps1} and the \href{https://mcknaus.github.io/assets/code/RNB_HK2_decomposition.nb.html}{Supplementary R Notebook} for the details of the empirical study.}
\end{figure}

We apply the decomposition method to an experimental evaluation of the Job Corps active labor market policy where the data contains detailed information about concrete program experience such as the content of (vocational) training. We study its well-documented gender gap in training returns. Consider the following interacted regression model for weekly earnings (in US\$) with a binary treatment $A$ (access to training) and a female indicator \begin{align}
    earnings = \beta_0 + \beta_1 A + \beta_2 female + \beta_3 (A \times female) + \varepsilon. \label{eq_REG1_INTRO}
\end{align}
Estimating the model via OLS yields an interaction coefficient $\hat{\beta}_3 = 7.6$. Figure \ref{fig:decomp-intro} presents a decomposition of this estimate.

 \vspace{-12pt}

We can see that \$6.2 out of \$7.6 (80\%) of the total gap can be explained by treatment heterogeneity stemming from different vocational training tracks. Moreover, this difference is purely driven by (i) targeting of women into trainings that have, on average, lower returns. It is neither explained by (ii) group targeting based on higher/lower group returns nor (iii) targeting based on individual characteristics that are correlated with being female/male and would predict higher returns from specific vocational trainings. In particular, the gender differences are not driven by effect heterogeneity in the form of heterogeneous returns to different vocational tracks. A nuanced policy conclusion is therefore that training allocation of women within Job Corps could potentially be improved. This is quite different to the conclusion that Job Corps, regardless of the assignment mechanism, yields the highest returns when targeted to men, which could be a conclusion if detected gender differences are attributed to effect heterogeneity. 

For estimation, we provide orthogonal/debiased moments for all decomposition estimands. These can be used in conjunction with flexible estimation methods such as machine learning for the nuisance quantities. We study the large sample properties of the resulting decomposition estimators in an asymptotic framework where the dimensionality of the effective treatment is allowed to grow with the sample size, albeit at a slower rate.\footnote{In practice, the number of measured effective treatments is always finite. However, this asymptotic framework provides a more suitable approximation to the finite sample properties of the involved estimators when there is a moderately large number of treatments for a given sample size compared to standard asymptotics that treat their dimension as fixed \cite{Chernozhukov2018}}  We provide high-level sufficient conditions for asymptotic normality that explicitly take the dimension of the effective treatment and the therefore increasing nuisance parameter space into account. The results can be used for conventional statistical hypothesis tests regarding any (combination of) decomposition estimands.

In the context of testing for effect homogeneity, we contrast the local power properties of our decomposition-based procedure to commonly applied norm-based tests within a multi-valued treatment effect (MVTE) setup. We find that conventional MVTE based tests can have trivial power when the dimension of the effective treatment is large, while the power of our decomposition-based testing approach is often invariant to the dimensionality. 

The proposed decompositions as well as the estimation and inference method apply to very general aggregates that are defined over potentially complicated effective treatment spaces. We provide explicit results for treatments that can contain (multiple) continuous and (many) discrete components. Our estimator can then directly be interpreted as a weighted local partitioning approach over the generalized treatment space. Under suitable growth and complexity restrictions on the effective treatment and potential outcomes, the estimator is still asymptotically normal around the generalized decomposition estimands without any asymptotic bias from partitioning.

The paper is structured as follows: Section \ref{sec_literature1} discusses methodological literature. Section \ref{sec:toy-exmpl} provides a motivating example. Section \ref{sec_decompositionALL1} introduces the double aggregation framework, the first decomposition, and discusses policy relevance in general and in the empirical application. Section \ref{sec:extensions} discusses extensions and how several results from the literature are nested. Section \ref{sec_indirect1} discusses hypothesis testing in comparison to the MVTE approach. Section \ref{sec_estimation1} introduces the debiased machine learning estimators, technical assumptions, and large sample properties. Section \ref{sec:generalized1} generalizes decompositions and estimation to more general effective treatment spaces. Section \ref{sec_conclusion1} contains additional concluding remarks. The \href{https://mcknaus.github.io/assets/code/RNB_HK2_decomposition.nb.html}{Supplementary R Notebook} and \href{https://hub.docker.com/repository/docker/mcknaus/hk2_decomposition/general}{Docker Hub} provide additional results and replication code. The method is implemented as part of the \href{https://github.com/MCKnaus/causalDML}{\texttt{causalDML}} R package.

\section{Literature} \label{sec_literature1}
To our knowledge, this is the first paper that formally studies group heterogeneity estimands when treatments are not homogeneous.\footnote{This paper supersedes \citeA{Heiler2021EffectTreatments} who consider conditional average treatment effects and their projections under heterogeneous treatments in a more restrictive two-fold decomposition. The decomposition provided here is more detailed, additionally covers group average treatment effects, and more general effective treatment spaces.}
It connects two topics that have been studied extensively in the literature: first, heterogeneity analyses with homogeneous treatments and therefore simple interpretation; and second, complex interpretation of simple average effect estimands with heterogeneous treatments.

Flexible estimation of heterogeneous treatment effects at different granularity is an active research area \cite{Athey2016,Athey2017a,Kunzel2017,Zimmert2019NonparametricConfounding,Knaus2021,Nie2021,Semenova2021DebiasedFunctions,Fan2022EstimationData,Kennedy2023TowardsEffects,Chernozhukov2024AppliedAIb,Lechner2024ComprehensiveLearning}. 
Interpretation of the respective estimands is straightforward if analyzed treatment values correspond to a single potential outcome, i.e.~treatments are assumed to be homogeneous.
This paper provides a detailed decomposition showing how these estimands capture effect heterogeneity, treatment heterogeneity and their interaction when treatments are instead heterogeneous.

Treatment heterogeneity in the sense of continuous or multi-valued effective treatments are discussed in \citeA{Imbens2000}, \citeA{Lechner2001}, or \citeA{Hirano2004TheTreatments}. They consider identification and estimation of dose-response functions or average effects when researchers analyze each effective treatment separately. 
This paper instead contributes to the literature on interpretation and external validity as a consequence of analyzing aggregates of heterogeneous treatments \cite{Hotz2005PredictingLocations,Hotz2006EvaluatingProgram,Stitelman2010TheVariables,Hernan2011CompoundInference,Caetano2025CausalTreatment}. 
Our double aggregation framework nests many results from the literature as special cases \cite{VanderWeele2013CausalTreatment,vanderLaan2023NonparametricIntervention,Lee2024BridgingExposures}. In particular, the latter focus on standard unconditional target estimands like average treatment effects or average potential outcomes. We instead focus on heterogeneity estimands aiming to uncover heterogeneous treatment effects. We discuss how the decomposition relates to existing results in detail in Section \ref{sec:lit-rel}. 

Consequences of treatment aggregation for average estimands are also discussed for instrumental variables \cite{Angrist1995Two-stageIntensity,Marshall2016CoarseningEstimates,Andresen2021Instrument-basedRestriction,Harris2022InterpretingEducation,Rose2024OnIndicators}, regression discontinuity designs \cite{Cattaneo2016InterpretingCutoffs}, difference-in-differences \cite{Callaway2021Difference-in-DifferencesTreatment}, and models with spillovers and interactions \cite{Manski2013IdentificationInteractions,Vazquez-Bare2023IdentificationExperiments}. We focus on heterogeneity estimands in experimental and unconfounded settings and provide an extension to instrumental variable designs. We expect that many results of this paper will have counterparts in alternative research designs.

% The paper is also related to, but distinct from, the literature on the causal content of common estimands in regression, panel data, and instrumental variable models in the presence of effect heterogeneity
% \cite<e.g.>{deChaisemartin2020Two-WayEffects,Goodman-Bacon2021Difference-in-differencesTiming,Sun2021EstimatingEffects,Soczynski2022InterpretingWeights,Goldsmith-Pinkham2024ContaminationRegressions,Poirier2024QuantifyingEstimands}. There, complications in the interpretation arise if parametric models are not flexible enough to suitably capture effect heterogeneity.
% In our setup, heterogeneity estimands are the primary goal of analysis and complications in the interpretation of common target estimands arises due to the mere presence of heterogeneous treatments and not as a consequence of misspecification.

The conventional literature on estimation of MVTE and group means usually assumes propensity scores that are bounded away from zero \cite{Cattaneo2010EfficientIgnorability,Farrell2015,Heiler2021ShrinkageRegressors}. In contrast, the theory in this paper allows all or many propensity scores to converge to zero, which is inevitable when considering a growing multi-valued or a continuous effective treatment as the propensities have to integrate up to one. Thus, on the level of the effective treatment, there is an overlap problem by construction. 
This connects our paper to the literature on inference under many extreme propensity scores, limited overlap, and irregular identification \cite{Khan2010IrregularEstimation,Rothe2017RobustOverlap,Hong2020InferenceOverlap,Heiler2021ValidScores,Ma2024TestingOverlap}. Our assumption on the dimensionality is closest to the limited overlap assumption for binary treatments made in \citeA{Hong2020InferenceOverlap}. In particular, we limit the growth of the effective treatment dimension such that consistent estimation of its average potential outcomes is technically still possible, albeit at very slow rates as in \citeA{Hong2020InferenceOverlap}. Our fixed-dimensional decomposition estimands, however, are regularly identified and thus can be relatively precisely estimated even when the effective treatment is growing with the sample size or continuous. 

The dimension-invariance of the decomposition also translates immediately to more dimension-robust power properties compared to using MVTE estimates in $\ell_p$-norm based tests for homogeneity. In particular, our decomposition yields testable \textit{necessary} conditions for group homogeneity.
Testing group effect homogeneity in the growing effective treatment setup induces a problem similar to but different from testing a mean vector or a difference between vectors in high dimensions. In iid settings, these problems have been studied extensively \cite{Zhong2011TestsDesigns,Srivastava2013AData,Steinberger2016TheRegression}. Our setup is most reminiscent of the slowly growing high-dimensional problem also studied in \citeA{Kock2019PowerProblems}. However, it is further complicated by its two-sample nature and dependence of potential outcomes across treatment versions. We provide approximate local power results for asymptotically pivotal tests based on $\ell_2$ and $\ell_{\infty}$ distance as well as our decomposition. They demonstrate the relative dimension robustness of the decomposition-based test over the MVTE approach.\footnote{
Inference on treatment effects with two versions has also been analyzed by \citeA{Hasegawa2020CausalTreatment} from a randomization perspective. They show that, for sharp nulls of constant unit-level effects, more powerful inference can be obtained by combining two tests. We instead study frequentist inference with many versions under group average nulls.}

Our estimation method builds on the semiparametric double/debiased machine learning framework \cite{Robins1994,Chernozhukov2018,Chernozhukov2022LocallyEstimation}. 
We provide all debiased influence functions for the novel decomposition estimands. A key technical difference to \citeA{Chernozhukov2018} is that our theory allows for an increasing dimension of the effective treatment which encompasses growing nuisance parameter spaces and limited overlap. This complicates inference and requires somewhat stronger, but still feasible rate conditions to control the first stage nuisance/machine learning bias. These requirements, however, are ameliorated by the local superefficiency properties of some of the involved frequency-based estimators.
Our theory nests the results in \citeA{Chernozhukov2018} when applied to finite dimensional multi-valued treatments.

\section{Motivating Example} \label{sec:toy-exmpl}

Consider a stylized data generating process (DGP) with randomized access to treatment but potentially non-random selection into two treatment versions. We are interested in understanding heterogeneity with respect to female group indicator $G \in \{0,1\}$. 
We formalize the setting using effective treatment $T \in \{0,1,2\}$ with 0 the control condition, and 1 (2) representing treatment version 1 (2). Assume a true noiseless outcome model
\begin{align}\label{eq:toy-y}
    Y = \tau_1 T_1 + \tau_2(X) T_2,
\end{align}
where $T_1 = \mathbbm{1}(T=1)$, $T_2 = \mathbbm{1}(T=2)$, and $X$ are some pre-treatment variables, possibly including $G$.
%\footnote{The outcome model can be motivated by potential outcomes $Y(0) = 0$, $Y(1) = \tau_1$, and $Y(2) = \tau_2(X)$ being plugged into observational rule $Y = \sum_t \mathbbm{1}(T=t) Y(t)$.}
Analysts interested in understanding the gender difference in treatment effects would usually run the following regression with randomization  indicator $A = \mathbbm{1}(T>0)$, the female indicator, and their interaction
\begin{align} \label{eq:reg-interact}
    Y = \beta_0 + \beta_1 A + \beta_2 G + \beta_3 (A \times G) + \varepsilon.
\end{align}
A non-zero $\beta_3$ would commonly be interpreted as ``effect heterogeneity''. However, the interpretation becomes more nuanced with multiple treatment versions. To see this note that $\beta_3$ in \eqref{eq:reg-interact} under outcome model \eqref{eq:toy-y} is
\begin{align} \label{eq:toy-beta3}
    \beta_3 
    & = E[Y | A=1, G=1 ] - E[Y | A=0, G=1 ] \\&\quad - (E[Y | A=1, G=0 ] - E[Y | A=0, G=0 ]) \nonumber \\   
    % & = E[\tau_1 T_1 + \tau_2(X) T_2 | A=1, F=1 ] - E[\tau_1 T_1 + \tau_2(X) T_2 | A=0, F=1 ] \\ 
    % & \quad - (E[\tau_1 T_1 + \tau_2(X) T_2 | A=1, F=0 ] - E[\tau_1 T_1 + \tau_2(X) T_2 | A=0, F=0 ]) \\ 
    % & = E[\tau_1 T_1 + \tau_2(X) T_2 | A=1, F=1 ] \\ 
    % & \quad - (E[\tau_1 T_1 + \tau_2(X) T_2 | A=1, F=0 ]) \\ 
    % & = E[\tau_1 T_1 | A=1, F=1] + E[\tau_2(X) T_2 | A=1, F=1] \\ 
    % & \quad - (E[\tau_1 T_1 | A=1, F=0] + E[\tau_2(X) T_2 | A=1, F=0]) \\ 
    % & = \tau_1 E[T_1 | A=1, F=1] + E[\tau_2(X) | A=1, F=1] E[T_2 | A=1, F=1] + Cov(\tau_2(X), T_2 | A=1, F=1) \\ 
    % & \quad - (\tau_1 E[T_1 | A=1, F=0] + E[\tau_2(X) | A=1, F=0] E[T_2 | A=1, F=0] + Cov(\tau_2(X), T_2 | A=1, F=0)) \\ 
    & = \tau_1 (E[T_1 | A=1, G=1] - E[T_1 | A=1, G=0]) \nonumber \\ 
    & \quad + E[\tau_2(X) | A=1, G=1] E[T_2 | A=1, G=1] \nonumber \\
    &\quad - E[\tau_2(X) | A=1, G=0] E[T_2 | A=1, G=0] \nonumber \\  
    & \quad + Cov(\tau_2(X), T_2 | A=1, G=1) - Cov(\tau_2(X), T_2 | A=1, G=0). 
\end{align}
%where the first equality follows from the standard representation of interaction coefficients as difference of conditional differences-in-means and the second equality by plugging in \eqref{eq:toy-y} and rearranging. 
We use this expression to discuss two phenomena that motivate the decompositions provided in the following sections.

\subsection{Finding Heterogeneity Without Heterogeneous Effects} \label{sec:toy-exmpl1}
Consider now the special case where $\tau_2(X) = \tau_2$ in \eqref{eq:toy-y} (effect homogeneity). Still,
\begin{align}
    \beta_3 
    & = (\tau_1 - \tau_2) (E[T_1 | A=1, G=1] - E[T_1 | A=1, G=0])
\end{align}
and therefore $\beta_3$ can be nonzero due to treatment heterogeneity. In particular, $\beta_3 \neq 0$ if the treatment versions have different constant effects ($\tau_1 \neq \tau_2$) and the probabilities to receive the versions depend on $G$, i.e.~$E[T_1 | A=1, G=1] \neq E[T_1 | A=1, G=0]$.

To make this more concrete, let $\tau_1 = 1$, $\tau_2 = 2$, $E[T_1 | A=1, G=1] = 2/3$, and $E[T_1 | A=1, G=0] = 1/3$. Both treatment versions are beneficial but version 2 yields larger positive effects and women are less likely to receive the best treatment version 2. Under this parametrization $\beta_3 = -1/3$, indicating that the effect is smaller for women. A policy maker could use this evidence to prioritize treatment away from women. However, this would only reinforce that women are already less likely to receive the best treatment. It would not be justified by real effect heterogeneity as there is simply no real effect heterogeneity to be exploited in this example. This motivates the question of how we can isolate real effect heterogeneity and treatment heterogeneity in such and related settings.

% R code
% tau1 = 1
% tau2 = 2
% gamma = 0

% n = 100000
% x = runif(n,-1,1)
% f = rbinom(n,1,0.5)
% p = 1/2
% e2 = 2/3 - 1/3 * f
% probs = cbind(rep(p,n),
%               p * (1-e2),
%               p * e2)
% t = apply(probs, 1, function(p) sample(0:2, 1, prob = p))
% T1 = (t == 1)
% T2 = (t == 2)
% Y0 = 0
% Y1 = tau1
% Y2 = tau2 + gamma * x
% mean(Y2 - Y0)
% mean(Y2[f==0])
% mean(Y2[f==1])
% Y = (t == 0) * Y0 + (t == 1) * Y1 + (t == 2) * Y2
% A = (t > 0) * 1
% mean(Y[A==1&f == 0]) - mean(Y[A==0&f == 0])

% summary(lm(Y ~ A*f))

\subsection{Finding Heterogeneity When Group Effects Are Homogeneous} \label{sec:toy-exmpl2}

Now consider the case where $\tau_1 = E[\tau_2(X) | A=1, G=1] = E[\tau_2(X) | A=1, G=0] = 0$, i.e.~all group specific effects for the treated, and, by randomization of $A$, also the group average treatment effects $E[\tau_2(X) | G=g]$, are zero and therefore homogeneous. Still, we may detect ``effect heterogeneity'' if the covariance between receiving the second treatment version and its effect differs between the analyzed heterogeneity groups:
\begin{align}
    \beta_3 
    & = Cov(\tau_2(X), T_2 | A=1, G=1) - Cov(\tau_2(X), T_2 | A=1, G=0) \notag \\
    & =  Cov(\tau_2(X), E[T_2|A=1,G=1,X] | A=1, G=1) \notag \\ &\quad -  Cov(\tau_2(X), E[T_2|A=1,G=0,X] | A=1, G=0).
\end{align}
For example, let $X = (X_1,G)$, where $X_1$ is an indicator whether an individual has children. Further, let $\tau_2(X) = X_1-E[X_1]$, $E[T_2 | A=1, G=0, X] = 1/2$, and $E[T_2 | A=1, G=1, X] = 2/3 - 1/3 X_1$ such that only individuals with children benefit from treatment 2, but women have a lower probability to receive the program if they have children. Then, $
    \beta_3 
    % = Cov(X-E[X], 2/3 - 1/3 X| A=1, G=1) \\
    % = Cov(X, - 1/3 X| A=1, G=1) \\
    % = - 1/3 Cov(X,  X| A=1, G=1) \\
    = - 1/3 \times Var(X_1 | A=1, G=1) $. We find a negative interaction unless all women in the population have either children or no children despite homogeneous group effects. This phenomenon appears as $\beta_3$, implicitly, is a comparison of outcomes under differently targeted treatments based on the numbers of children. 
In general, we can detect group heterogeneity even if group effects are homogeneous if treatments are heterogeneous at the same time. By the same token, true effect heterogeneity could be masked or attenuated. The following provides a formal way to understand the underlying mechanisms.

\section{Framework and Decomposition} \label{sec_decompositionALL1}

This section introduces the double aggregation framework for the analysis of heterogeneity estimands in the presence of heterogeneous treatments. We first focus on the group differences-in-means corresponding to interaction coefficients like \eqref{eq:toy-beta3} in the motivating example of Section \ref{sec:toy-exmpl}. This allows us to develop a decomposition to isolate different sources of group heterogeneity in a relatively simple setting. 
Later sections generalize the decomposition to more complex settings and show that the framework covers multiple results in the literature as special cases. All derivations are in Appendix \ref{app:deriv-both}.

\subsection{Double Aggregation Framework} \label{sec_setting1}
Let $T \in \mathcal{T}$ denote the effective treatment.\footnote{\citeA{Manski2013IdentificationInteractions} uses the term ``effective treatment'' in the context of interference. Like in our setting, it describes the treatments creating variation in potential outcomes. In the following the term ``treatment'' refers to effective treatment if not stated differently.}  $\mathcal{T}$ denotes the underlying set of possible treatment values; the associated probability measure may be discrete, continuous, or mixed. For exposition, we first focus on the case of discrete effective treatments. The extension to continuous and mixed effective treatments is provided in Section \ref{sec:generalized1}. We assume existence of potential outcomes
$\{Y(t): t \in \mathcal{T}\}$. The realized outcome is then obtained as $Y = Y(T)$. For simplicity, we treat larger outcomes as preferable throughout the discussion. Let $X \in \mathcal{X}$ be variables that ensure conditional independence:
\begin{assump}[Effective Treatment Unconfoundedness] For all $t \in \mathcal{T}$, the potential outcome is conditionally independent 
   % \begin{align*} 
    $Y(t) \bigCI T | X$. 
   % \end{align*}
%\deleted{and has common support
%$P(T=t|X=x) > 0 \quad\forall~ x \in \mathcal{X}$}.
\label{ass_CIA}
\end{assump}

\paragraph{Remark 1.} For deriving the decomposition results below $X$ and $T$ may be unobserved. $X$ could include all confounders or determinants of $T$. Thus, Assumption \ref{ass_CIA} is essentially non-restrictive, see \citeA{Goldsmith-Pinkham2024ContaminationRegressions} for a similar discussion around their Equation (11). For identification and estimation, however, $X$ and $T$ need to be observed, see Section \ref{sec:identification}. \hfill $\square$

We now define the analyzed aggregate treatments and heterogeneity groups. The mechanics of our approach are agnostic to whether aggregate treatments are obtained via ex-ante or ex-post aggregation, see \citeA{VanderWeele2013CausalTreatment} for differences in interpretation.
Let $a:\mathcal{T}\rightarrow\{0,1\}$ and $g:\mathcal{X}\rightarrow \{0,1\}$ be known measurable treatment and group aggregation functions that indicate whether an observation belongs to a treatment or covariate group. Let $\mathcal{T}_a$ and $\xg$ denote the level sets (pre-images) of 1 under $a$ and $g$, respectively: 
%\begin{align}
    $\ta = \{t \in \mathcal{T}: a(t) = 1\}$, 
    $\xg = \{x \in \mathcal{X}: g(x) = 1\}$,
%\end{align}
and equivalently for other aggregation functions $a'$, $g'$.\footnote{For simplicity, any aggregations are defined as mutually exclusive in what follows, meaning that $\mathcal{T}_a \cap \mathcal{T}_{a'} = \emptyset$ for all $a \neq a'$ and $\mathcal{X}_g \cap \mathcal{X}_{g'} = \emptyset$ for all $g \neq g'$. This corresponds to empirical practice, but can in principle be relaxed at the expense of more involved notation.} 

The group differences-in-means corresponding to the interaction term in \eqref{eq:toy-beta3} that is decomposed in this section reads in this general notation
\begin{align}
    & E[Y| T \in \mathcal{T}_a, X \in \mathcal{X}_g] - E[Y| T \in \mathcal{T}_{a'}, X \in \mathcal{X}_g]  \notag \\
     - \Big( &E[Y| T \in \mathcal{T}_a, X \in \mathcal{X}_{g'}] - E[Y| T \in \mathcal{T}_{a'}, X \in \mathcal{X}_{g'}] \Big) \label{eq_DiM}
\end{align}
for some known $a,a'$ and $g,g'$. For \eqref{eq_DiM} to be well defined we assume throughout that $P(T \in \ta, X \in \xg) > 0$ (and analogously for the other aggregation events).

\textit{Example continued:} In Section \ref{sec:toy-exmpl}, $\mathcal{T} = \{0,1,2\}$, $a(T) = \mathbbm{1}(T > 0) = A$, $a'(T) = \mathbbm{1}(T = 0) = 1-A$,
 $\mathcal{T}_{a} = \{1,2\}$, $\mathcal{T}_{a'} = \{0\}$,
$\mathcal{X} = \{0,1\} \times \{0,1\}$. $g(X) = G$, $g'(X) = 1-G$,
$\mathcal{X}_{g} = \{0,1\} \times \{1\}$, 
$\mathcal{X}_{g'} = \{0,1\} \times \{0\}$.
\hfill $\square$

% $T \in \mathcal{T}$ indicates the effective treatment creating potential outcomes $Y(t)$ for each $t \in \mathcal{T}$. We assume consistency such that the observed outcome becomes $Y = \sum_t \mathbbm{1}(T =t) Y(t)$. We assume the existence of a set of confounding variables $X \in \mathcal{X}$ ensuring conditional independence $Y(t) \bigCI T | X$. $T$ and $X$ may be unobserved and we will be explicit when they being observed becomes relevant. 

% We consider settings where the analyzed treatments are aggregations of effective treatment $\mathcal{T}_a \subset \mathcal{T}$ for $a \in \mathcal{A}$, and the analyzed heterogeneity groups aggregate confounders $\mathcal{X}_g \subset \mathcal{X}$ for $g \in \mathcal{G}$.\footnote{Both aggregations are mutually exclusive meaning that $\mathcal{T}_a \cap \mathcal{T}_{a'} = \emptyset$ for all $a \neq a'$ and $\mathcal{X}_g \cap \mathcal{X}_{g'} = \emptyset$ for all $g \neq g'$.} 
% Data is generated by independent distributions $(Y,A,G)$ where $A = \sum_d d \mathbbm{1}(T \in \mathcal{T}_a)$ and $G = \sum_g g \mathbbm{1}(X \in \mathcal{X}_g)$.

Further, for any $x \in \mathcal{X}$, denote
\begin{alignat*}{3}
    \mu_t(x) &= E[Y(t)|X =  x], 
    &\mu_t &= E[\mu_t(X)], & \\
    e_t(x) &= P(T =t|X=x), 
    &e_t &= E[e_t(X)], & \\
 %   e_{ta}(x) &= P(T=t|X=x,T\in\ta),
%    &\quad e_{ta} &= E[e_{ta}(X)|T\in\ta], \\
    \tau_{t,t'}(x) &= \mu_{t}(x) - \mu_{t'}(x),
    &\quad \tau_{t,t'} &= \mu_{t} - \mu_{t'}.
\end{alignat*}
For any generic function (possibly depending on $t$ and $a$), we write $g(\mathcal{X}_g) = E[g(X)|X\in \mathcal{X}_g]$, e.g.~$\mu_t(\mathcal{X}_g) = E[\mu_t(X)|X\in \mathcal{X}_g]$. 
We also define conditional (group) propensities given treatment aggregate $a$ as
 
\begin{align*}
e_{ta}(x)
&= P(T=t | X=x, T\in\mathcal T_a) 
= \begin{cases}
\frac{\mathbbm{1}\{t\in\mathcal T_a\}e_t(x)}{P(T\in\mathcal T_a | X=x)} & \text{ if } P(T\in\mathcal T_a | X=x) > 0 \\
0 & \text{ else } 
\end{cases}  \\
%\end{align*}
%\begin{align*}
e_{ta}(\mathcal X_g)
&= P(T=t | X\in\mathcal X_g, T\in\mathcal T_a)
= \begin{cases}
\frac{\mathbbm{1}\{t\in\mathcal T_a\}e_t(\mathcal X_g)}{P(T\in \ta | X\in\mathcal X_g)}
      & \text{ if } P(T\in\mathcal T_a | X\in\mathcal X_g) > 0 \\
0 & \text{ else } 
\end{cases}  
\end{align*}

% Define the following quantities: 
% \begin{align*}
%     \mu_t(x) &:= E[Y(t)|X=x] \\
%     \mu_t(\xg) &:= E[Y(t)|X\in\xg] \\
%     \mu_t &:= E[Y(t)] \\
%     e_t(x) &:= P(T=t|X=x) \\
%     e_t(\xg) &:= P(T=t|X\in\xg) \\
%     e_t &:= P(T=t) \\
%     e_{ta}(x) &:= P(T=t|X=x,T\in\mathcal{T}_a) = \frac{\mathbbm{1}(t\in\mathcal{T}_a)P(T=t|X=x)}{P(T\in\mathcal{T}_a|X=x)} \\
%     e_{ta}(\xg) &:= P(T=t|X\in\xg,T\in\mathcal{T}_a) = E[e_{ta}(X)|X\in\xg,T\in\mathcal{T}_a] \\
%     e_{ta} &:= P(T=t|T\in\mathcal{T}_a) = \frac{\mathbbm{1}(T\in\mathcal{T}_a)P(T=t)}{P(T\in\mathcal{T}_a)}    
% \end{align*}

\subsection{Application: Training Specializations in Job Corps} \label{sec_jobcorps1}

We use data of an experimental evaluation of the Job Corps (JC) program in 1994-1996 \cite{Schochet2019ReplicationStudy} to illustrate the practical relevance of the decomposition terms and to support their intuition by linking them to a well-known application.
JC operates since 1964 and is the largest training program for disadvantaged youth aged 16-24 in the US \cite<see>[for a detailed description]{Schochet2001NationalOutcomes,Schochet2008DoesStudy}. 
The roughly 50,000 participants per year receive an intensive treatment as a combination of different components like academic education, vocational training, and job placement assistance. 
% Participants plan their curriculum together with counselors. 
This means that although the variable ``access to JC'' is a binary indicator, different tracks of JC participation are conceivable.

Several studies document that women benefit less than men from JC access \cite<e.g.>{Schochet2001NationalOutcomes,Schochet2008DoesStudy,Flores2012EstimatingCorps,Eren2014WhoProgram,Strittmatter2019HeterogeneousApproach}. 
A potential explanation is differential assignment to vocational tracks: men are more often trained for higher-paying craft occupations, while women concentrate in service-sector tracks \cite{Quadagno1995TheCorps,Inanc2017GenderGap}.

The decomposition developed in the following allows to quantify the mechanisms behind this finding. Specifically, we decompose the gender gap in the intention-to-treat effect of the binary variable indicating random access to JC on weekly earnings four years after random assignment estimated as $\beta_3$ in \eqref{eq_REG1_INTRO} and consider 11 versions of the treatment: (i) \textit{No JC} if eligible individuals did not participate (non-compliers), (ii) \textit{JC without vocational training} if eligible individuals entered JC but did not receive vocational training, (iii-ix) training in clerical, health, auto mechanics, welding, electrical/electronics, construction, or food sectors, (x) other vocational training, (xi) training in multiple sectors.

We control for 57 pre-treatment variables (see Section \ref{sec:identification} for a detailed discussion of identification) and analyze 9,708 observations. Estimation follows Section \ref{sec_estimation1}, using 5-fold cross-fitted random forests for nuisance parameters.\footnote{We consider a method that is able to flexibly model heterogeneity in the assignment between men and women like random forest as crucial in this application. This is evident by the propensity scores' bimodal patterns based on gender, see Supplementary R Notebook S.2.2 for more details.} 
The \href{https://mcknaus.github.io/assets/code/RNB_HK2_decomposition.nb.html}{Supplementary R Notebook} provides additional descriptives, robustness checks using sampling weights, and further results.\footnote{Code to replicate the analysis is stored on \href{https://hub.docker.com/repository/docker/mcknaus/hk2_decomposition/general}{Docker Hub}. The data is available as public use file via \href{https://doi.org/10.3886/E113269V1}{https://doi.org/10.3886/E113269V1}.}

\subsection{Decomposing a Single Group Mean} \label{sec:decomp-single-mean}

The main goal in the following is to understand the causal content of group differences-in-means (DiM) as defined in \eqref{eq_DiM} in the double aggregation framework. For compactness, we first decompose one group mean $E[Y|T\in\mathcal{T}_a,X\in \xg]$ before combining them in Section \ref{sec:decomp-diffs} and Section \ref{sec:decomp-diff-in-diffs}.

We first isolate the part that only depends on group heterogeneity from parts also depending on additional covariates (see Section \ref{sec:toy-exmpl2} how the latter may drive heterogeneity). Under Assumption \ref{ass_CIA}, which, again, is essentially non-restrictive (see Remark 1), the group mean can be decomposed as
\begin{align} \label{eq:decomp-cm1}
    E[Y|T\in\mathcal{T}_a,X\in \xg] %& = \sum_{t\in\ta} E[\mathbbm{1}(T=t)Y(t)|T\in\mathcal{T}_a,X\in \xg]  \nonumber \\
    & = \sum_{t\in\ta} 
    \underbrace{e_{ta}(\xg) \mu_t(\xg)}_{\text{Stratified RCT}}
  + \underbrace{\frac{Cov(e_t(X),\mu_t(X) | X\in \xg)}{P(T\in\mathcal{T}_a | X\in \xg)}}_{\text{Individualized targeting}} \nonumber \\
  & =: d(a,g) + d_{4}(a,g).
\end{align}

The first component $d(a,g)$ corresponds to a group mean that would be obtained in a hypothetical DGP with treatments assigned as in a group stratified RCT while leaving the potential outcome distributions untouched. The treatment probabilities in this synthetic RCT estimand correspond to the observed group probabilities $e_{ta}(\xg)$ and therefore vary only at the group-level. The second term $d_{4}(a,g)$ represents contributions stemming from variables beyond group membership. The numerator is the covariance of the actual treatment probabilities and the respective potential outcomes. It captures individualized targeting going beyond heterogeneity group specific targeting. It is positive (negative) if subgroups with higher potential outcomes for a particular treatment tend to have a higher (lower) probability of receiving it. The denominator normalizes by the probability to be observed in the particular treatment aggregation $a$. %and is constant over all $t\in\ta$. 

The decomposition in \eqref{eq:decomp-cm1} contains the estimand of an RCT stratified using only group information. However, even in such a setup, treatment heterogeneity may be mistaken for effect heterogeneity as illustrated in Section \ref{sec:toy-exmpl1}. To further separate the different sources of heterogeneity, we define three synthetic estimands:
\begin{align}
    s_{0}(a) &:= \sum_{t\in\ta} e_{ta} \mu_t, \quad   
    s_{1}(a,g) := \sum_{t\in\ta} e_{ta} \mu_t(\xg), \quad  s_{2}(a,g) := \sum_{t\in\ta} e_{ta}(\xg) \mu_t. \label{eq_synth_s}
\end{align}
The parameters in \eqref{eq_synth_s} are the expected results of a group mean under specific counterfactual DGPs: $s_{0}(a)$ if both group treatment probabilities and group average potential outcomes were constant, $s_{1}(a,g)$ if group probabilities were constant but group outcomes varied, and $s_{2}(a,g)$ if group probabilities varied but group outcomes were constant.
Actual DGPs corresponding to these synthetic estimands, in particular $s_{0}(a)$ and $s_{2}(a,g)$, are hypothetical and restrictive. However, they allow us to further decompose the stratified RCT:\footnote{The mechanics of this decomposition is similar to the 3-fold Blinder-Oaxaca decomposition \cite{Blinder1973WageEstimates}, which is obtained by adding and subtracting synthetic estimands corresponding to expected group outcomes under hypothetical covariate distributions. The centering could also be around other reasonable synthetic propensities and averages in relation to other thought experiments, see Appendix \ref{sec_centering1} for an example and additional discussion.}
\begin{align}
    &\sum_{t\in\ta}e_{ta}(\xg) \mu_t(\xg) \pm 2 s_{0}(a) \pm s_{1}(a,g) \pm s_{2}(a,g) \nonumber \\ % \multicolumn{2}{l}{$\displaystyle e_{ta}(\xg) \mu_t(\xg) \pm 2 s_{0}(a) \pm s_{1}(a,g) \pm s_{2}(a,g)$}  \nonumber \\
    &= \sum_{t\in\ta} e_{ta} \mu_t  \tag{baseline}  \\
    & + e_{ta} [\mu_t(\xg) - \mu_t]  \tag{outcome heterogeneity}   \\
    & + [e_{ta}(\xg) - e_{ta}] \mu_t  \tag{group targeting of average outcomes}   \\
    & + [e_{ta}(\xg) - e_{ta}] [\mu_t(\xg) - \mu_t]  \tag{group targeting of group outcomes}  \\
    &=: d_{0}(a) + d_{1}(a,g) + d_{2}(a,g) + d_{3}(a,g). \label{eq:decomp-srct}
\end{align}
$d_{0}(a) = s_{0}(a)$ is the constant baseline and does not depend on $g$. $d_{1}(a,g)$ varies only due to outcome heterogeneity and is positive (negative) if group $g$ tends to have higher (lower) than average outcomes. $d_{2}(a,g)$ varies only due to treatment heterogeneity and is positive (negative) if group $g$ is more likely to receive treatments with higher (lower) average potential outcomes. It captures if group $g$ is targeted towards on average better treatments. $d_{3}(a,g)$ captures group targeting based on heterogeneous group outcomes and therefore the interaction between treatment and outcome heterogeneity. It is positive (negative) for group $g$ if its over-represented treatments ($e_{ta}(\xg) - e_{ta} > 0$) are aligned with relatively higher (lower) group outcomes.\footnote{For example, take $\mathcal{T}_a = \{1,2\}$ with $e_{1a}(\xg) - e_{1a} = -1/4$ and $e_{2a}(\xg) - e_{2a} = 1/4$. Then $d_{3}(a,g) = 1$ for groups with $\mu_1(\xg) - \mu_1 = 4$ and $\mu_2(\xg) - \mu_2 = 8$, but also if $\mu_1(\xg) - \mu_1 = -8$ and $\mu_2(\xg) - \mu_2 = -4$ showcasing that also groups with consistently lower than average outcomes can benefit from group targeting on group outcomes if distances to the outcome averages differ within group.} The decomposition in \eqref{eq:decomp-srct} serves as the crucial building block to decompose group mean differences in the following but might also be interesting in its own right depending on the application.

\begin{figure}[t]
    \centering
    \caption{Decomposition of Group Means in Job Corps Application} \label{fig:decomp-cm}
    \begin{subfigure}{1\textwidth}
        \centering
        \includegraphics[width=0.96\textwidth]{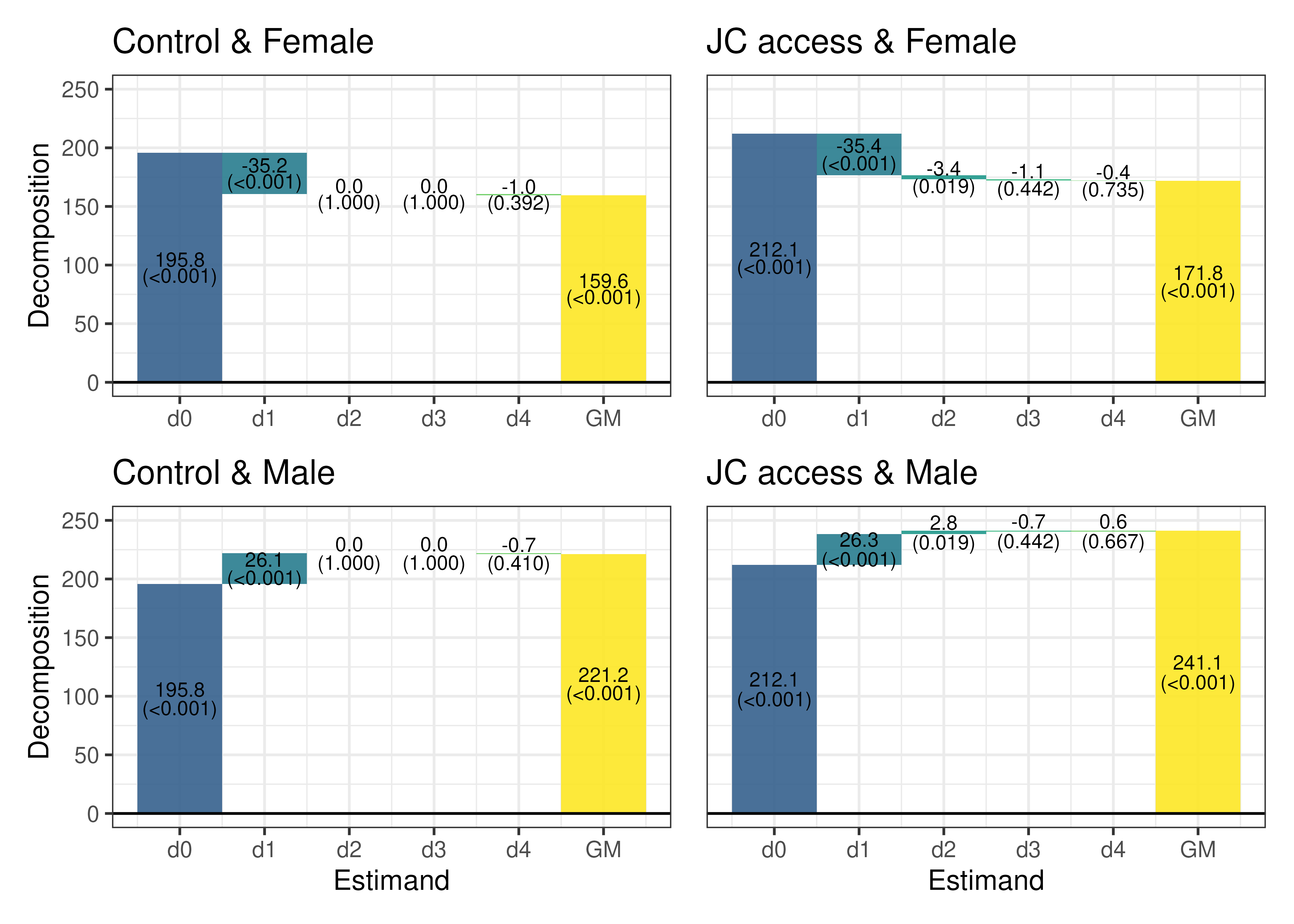}
    \end{subfigure}
    \subcaption*{\textit{Notes:} The graph shows the estimated decomposition introduced in Equation \eqref{eq:decomp-srct} for the group means of the subgroups described in the respective subfigure title. Point estimates and $p$-values (in brackets) are obtained using the estimation and inference procedures discussed in Section \ref{sec_estimation1}.}
\end{figure}

Figure \ref{fig:decomp-cm} decomposes the four group means entering the group differences-in-means in the JC application. The bars labeled ``GM'' correspond to the group mean and the waterfall bars to their left track the decomposition terms. The left column shows the group mean decompositions of the control group for females on top and males at the bottom. As there is only one control condition, $e_{ta'}(\xg) = e_{ta'} = 1$, which simplifies the interpretation. In particular, $d_{2} = d_{3} = 0$ by construction because no control versions can be targeted. $d_{0}$ shows that in the absence of any outcome heterogeneity the weekly earnings of untreated would be $\$196$. However, $d_{1}$ shows that female earnings are $\$35.2$ lower and male earnings are $\$26.1$ higher than this baseline. This documents significant outcome heterogeneity in form of the well-documented gender gap in earnings levels. Individual targeting captured by $d_{4}$ is negligible. This is expected as the control condition is randomly assigned and therefore $e_t(X)$ in the covariance of Equation $\eqref{eq:decomp-cm1}$ is approximately constant.

The right column of Figure \ref{fig:decomp-cm} provides the more interesting decomposition of the group that was randomly treated to receive access to JC, while receiving different vocational training. The baseline outcome, $d_{0}$, in the treated group is $\$212$ and therefore higher than in the control group. However, the outcome heterogeneities captured by $d_{1}$ are nearly identical to the control condition suggesting that the treatment did not change the gender earnings gap. Now we turn to the components that might be driven by differential targeting into vocational training. $d_{2}$ is significantly negative ($-\$3.4$, $p$ = 0.02) for women indicating that they predominantly receive training in vocations with relatively low average potential outcomes.
In contrast, men receive more often training in on average better paying vocations ($\$2.8$, $p$ = 0.02). In principle, women could achieve higher earnings when trained for service jobs than for craft jobs. Then the targeting of women towards service jobs would exploit group differences leading to positive $d_{3}$ that could even overcompensate the negative $d_{2}$.
However, we find no such group targeting of group outcomes or individualized targeting captured by $d_{3}$ and $d_{4}$.

\subsection{Decomposing a Single Group Difference-in-Means} \label{sec:decomp-diffs}

Following Section \ref{sec:decomp-single-mean} define $\delta_{0}(a,a') := d_{0}(a) - d_{0}(a')$ and $\delta_{j}(a,a',g) := d_{j}(a,g) - d_{j}(a',g) \text{ for } j \in \{1,2,3,4\}$ to decompose a single group DiM as
\begin{align}
E&[Y|T\in\mathcal{T}_a,X\in \xg] - E[Y|T\in\mathcal{T}_{a'},X\in \xg] \nonumber \\
& = \delta_{0}(a,a') + \delta_{1}(a,a',g) + \delta_{2}(a,a',g) + \delta_{3}(a,a',g) + \delta_{4}(a,a',g). \label{eq:delta-decomp1}
\end{align}

The interpretation of the decomposition parameters follows directly from the discussion in Section \ref{sec:decomp-single-mean}.
$\delta_{1}(a,a',g) = \sum_{t\in\ta} \sum_{t'\in\mathcal{T}_{a'}} e_{ta} e_{t'a'} [\tau_{t,t'}(\xg) - \tau_{t,t'}]$ indicates whether group effects systematically differ from average effects and captures real effect heterogeneity. Similarly, $\delta_{2}(a,a',g) = \sum_{t\in\ta} [e_{ta}(\xg) - e_{ta}] \mu_t - \sum_{t'\in\mathcal{T}_{a'}} [e_{t'a'}(\xg) - e_{t'a'}] \mu_{t'}$ captures pure treatment heterogeneity as it can only be non-zero if treatment probabilities are heterogeneous and average outcomes within an aggregation are non-constant. $\delta_2$ is positive (negative) if $g$ is better (worse) targeted on average outcomes in aggregation $a$ than in $a'$.

On the other hand $\delta_3$ and $\delta_4$ capture possible interactions between effect and treatment heterogeneity. Their most straightforward interpretation is that they indicate how the targeting and composition adjustments differ in aggregation $a$ compared to $a'$ for group $g$, see also Appendix \ref{app:decompstn} for the complete definitions and additional discussion. The difference between two groups $g$ and $g'$ within one aggregation $a$, $E[Y|T\in\mathcal{T}_a,X\in \xg] - E[Y|T\in\mathcal{T}_{a},X\in \mathcal{X}_{g'}]$, can be decomposed symmetrically. Most importantly, for the main result in the following section, it is irrelevant which difference is decomposed first, as it uses a path-independent difference in differences.

\begin{figure}[!htbp]
    \centering
    \caption{Decomposition of Gender Differences in Job Corps} \label{fig:decomp}
        \centering
        \includegraphics[width=0.96\textwidth]{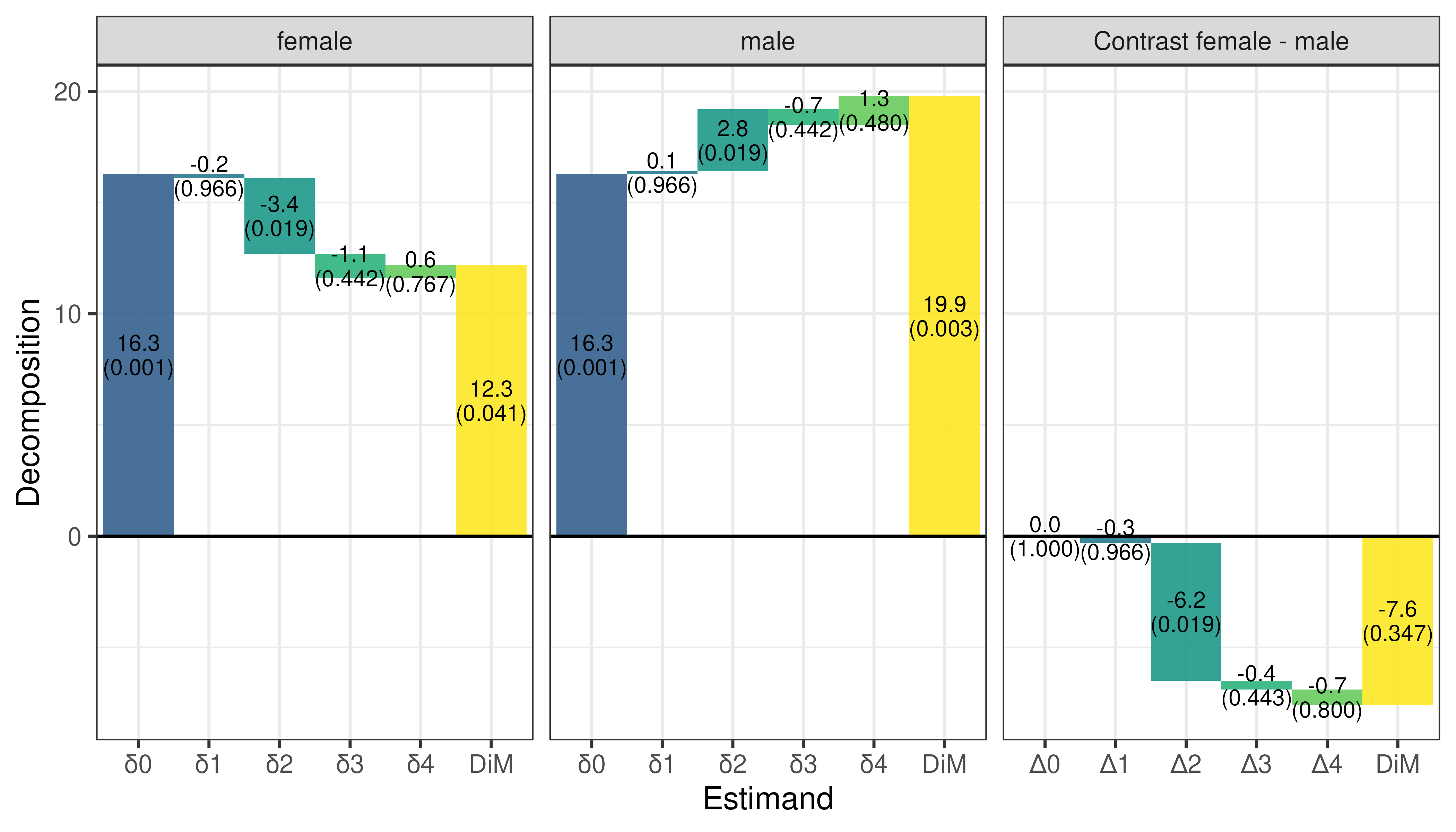}
    \subcaption*{\textit{Notes:} The numbers in the bars show the point estimate and the $p$-value in parentheses. They are obtained using the estimation and inference procedures discussed in Section \ref{sec_estimation1}.}
\end{figure}

The left two subfigures of Figure \ref{fig:decomp} show the decompositions for women and men in the JC application. They are obtained by taking the difference between the respective right and left subfigures of Figure \ref{fig:decomp-cm}. The right bars show the group difference-in-means documenting that women are estimated to benefit less from JC access than men. We note that $\delta_{0}+ \delta_{1}$ would be the expected effect if training assignment was fully random. In this hypothetical scenario, no gender differences would show up because $\delta_{1}$ is basically zero indicating that women and men do not react differently to the same training mix. However, the significantly negative $\delta_{2}$ for women shows that their actual treatment assignment is worse than random as they are targeted towards trainings with, on average, lower returns. In contrast, targeting is significantly better than random for men.

\subsection{Decomposing Group Differences-in-Means} \label{sec:decomp-diff-in-diffs}
We are now equipped to decompose the primary estimand. Define $\Delta_{j}(a,a',g,g') := \delta_{j}(a,g) - \delta_{j}(a',g') \text{ for } j \in \{1,2,3,4\}$. We obtain that the heterogeneity estimand group differences-in-means as defined in \eqref{eq_DiM} is composed of
    \begin{align*}
     &\Delta_{1}(a,a',g,g') & (\text{effect heterogeneity}) \\
     +& \Delta_{2}(a,a',g,g')  & (\text{average targeting heterogeneity}) \\
     +& \Delta_{3}(a,a',g,g')  & (\text{group targeting heterogeneity}) \\
     +& \Delta_{4}(a,a',g,g')  & (\text{individualized targeting heterogeneity})
\end{align*}

This shows that group heterogeneity may be driven by multiple components that both separately as well as in combination have specific economic interpretations. $\Delta_1$ is the only component that is necessarily driven by \textit{effect heterogeneity}. To see this note that 
\begin{equation} \label{eq:Delta1}
\Delta_{1}(a,a',g,g') = \sum_{t\in\ta} \sum_{t'\in\mathcal{T}_{a'}} e_{ta} e_{t'a'} [\tau_{t,t'}(\xg) - \tau_{t,t'}(\mathcal{X}_{g'})],  
\end{equation}
showing that it is a weighted average of group average treatment effect differences. $\Delta_1$ is higher (lower) when group $g$ has systematically higher (lower) effects than group $g'$. It can only differ from zero if there is at least some group-specific effect heterogeneity. Thus, it provides a clean test for true group effect heterogeneity that is not contaminated by treatment heterogeneity. Furthermore, it is the only decomposition parameter that can be non-zero when treatments are homogeneous and/or when the effective treatment is fully randomly assigned. The right panel of Figure \ref{fig:decomp} shows that pure effect heterogeneity does not contribute to the gender gap in effectiveness in the JC application.

$\Delta_2$, $\Delta_3$ and $\Delta_4$ are the consequence of \textit{treatment heterogeneity}. There are three levels of implicit \textit{targeting} parameters capturing separately to which extent treatment probabilities systematically vary with average, group or individual outcomes.
$\Delta_{2}(a,a',g,g') = \sum_{t\in\ta} [e_{ta}(\xg) - e_{ta}(\mathcal{X}_{g'})] \mu_{t} - \sum_{t'\in\mathcal{T}_{a'}} [e_{t'a'}(\xg) - e_{t'a'}(\mathcal{X}_{g'})] \mu_{t'}$ represents differences in \textit{group targeting} with respect to \textit{average outcomes}. It is purely driven by treatment heterogeneity. $\Delta_{2}$ is positive (negative) if group $g$ relative to $g'$ is better (worse) targeted on average outcomes in aggregation $a$ than in aggregation $a'$. If $a'$ represents a homogeneous control condition ($\mathcal{T}_{a'}=\{0\}$ like in the JC application), the expression simplifies to $\Delta_{2}(a,a',g,g') = \sum_{t\in\ta} [e_{ta}(\xg) - e_{ta}(\mathcal{X}_{g'})] \tau_{t,0}$. Then a more simple interpretation arises stating that $\Delta_{2}$ is positive (negative) if group $g$ is better (worse) targeted on average \textit{effects} compared to group $g'$.
The significantly negative $\Delta_{2}$ in Figure \ref{fig:decomp} shows therefore that women are targeted towards vocational trainings with on average lower returns to training.

$\Delta_3$ and $\Delta_4$ capture potential interactions between effect and treatment heterogeneity. Their complete formulas are in Appendix \ref{app:params-tests}; we focus on a high level discussion here. $\Delta_3$ represents differences in targeting beyond the group targeting of average outcomes captured by $\Delta_2$ and is positive (negative) if group $g$ relative to $g'$ is better (worse) targeted on group outcomes in aggregation $a$ than in aggregation $a'$. Similarly, $\Delta_4$ captures differences in individualized targeting based on confounders beyond group targeting covered by $\Delta_2$ and $\Delta_3$. The combination $\Delta_2 + \Delta_3 + \Delta_4$  therefore provides a summary of group differences stemming from targeting. Figure \ref{fig:decomp} documents that group and individualized targeting play a negligible role in explaining the group heterogeneity in the JC application.

\textit{Examples continued:} In Section \ref{sec:toy-exmpl1}, all decomposition parameters are zero except for $\Delta_{2} = \beta_3 = -1/3$. The non-zero interaction coefficient is therefore the result of group targeting of average outcomes as women are more likely to receive on average worse treatments like in the JC application.
In Section \ref{sec:toy-exmpl2}, all decomposition parameters are zero except $\Delta_{4} = \beta_3 = -1/3 \times Var(X_1 | A=1, G=1)$. The non-zero interaction coefficient is therefore the result of individualized targeting heterogeneity as targeting does not depend on $G$ but on $X_1$ creating a covariance even after conditioning on $G$. We do not find evidence for this mechanism in the JC application. % \hfill $\square$

\subsection{Generic Implications for Heterogeneity Analyses}

On a high level, the decompositions clarify how heterogeneous treatments complicate the causal content of seemingly simple heterogeneity estimands substantially. Group differences may be driven by effect heterogeneity ($\Delta_1$), treatment heterogeneity ($\Delta_2$), or interactions of both ($\Delta_3,\Delta_4$). It is important to note that the decomposition applies even if the effective treatment and its confounders are unobserved. Thus, interpretations of heterogeneity analyses should in general be linked to a discussion of treatment heterogeneity. If treatments are plausibly homogeneous, standard interpretations apply. If relevant treatment heterogeneity cannot be ruled out in a concrete application, the decomposition offers a principled framework for interpreting group differences and may prevent overly simplistic policy conclusions. In particular, it may prevent to prematurely conclude that ``treatment'' $a$ works better than $a'$ for group $g$ compared to $g'$, although no real effect heterogeneity exists. In cases where researchers expect substantial but unobservable variation within the analyzed treatment, they might even conclude that standard heterogeneity analyses are uninformative.

\subsection{Identification of Decomposition Parameters} \label{sec:identification}
All decomposition parameters are identified under a standard unconfoundedness assumption for the multi-valued treatment:

\begin{assump}[Identifying Assumptions] Assume that $X$ in Assumption \ref{ass_CIA} is observable, $T$ is observable, and effective treatment overlap holds  $P(T=t|X=x) > 0 \quad\forall~ x \in \mathcal{X}$.
\label{ass_observed}
\end{assump}
Under Assumption \ref{ass_observed}, $\mu_t(x) = E[Y|T=t,X=x]$ such that all decomposition components involving potential outcomes $\mu_t(X)$, $\mu_t(\xg)$, and $\mu_t$ are identified.

The plausibility of Assumption \ref{ass_observed} depends on the concrete application. In observational studies, unconfoundedness is arguably credible in the evaluation of active labor market policies (ALMPs) when rich background characteristics are measured.
% Socio-demographic information and pre-treatment outcomes (Mueser et al., 2007), regional information and labor market histories (Friedlander and Robins, 1995, Heckman et al., 1998b, Heckman and Smith, 1999, Dolton and Smith, 2010) are shown to be important. 
\citeA{lechner2013sensitivity} provide evidence that the common practice to control for socio-demographic information, pre-treatment outcomes, regional information and labor market histories remove most of the selection bias in ALMP studies.\footnote{\citeA{Caliendo2017UnobservablePolicies} show that often unobserved personality traits, attitudes, expectations, social networks and inter-generational information, do predict selection into ALMPs, but do not lead to relevant effect differences for wages or employment prospects suggesting commonly observable controls sufficiently capture the relevant confounding.} Consistent with this result, the meta-analysis by \citeA{Card2018WhatEvaluations} finds no evidence that non-experimental evaluations of ALMPs produce markedly different average impacts than experimental studies.

The 57 control variables applied in the JC application contain all categories of variables considered important by \citeA{lechner2013sensitivity} plus health, crime, and JC related variables. These control variables overlap mostly with those of \citeA{Flores2012EstimatingCorps} who also employ an unconfoundedness strategy in JC. Additionally, we show in Supplementary R Notebook S.2.5 that the covariate-adjusted difference between non-compliers and untreated group is insignificant (3.1, S.E.:~5.9) providing evidence that the control variables can successfully account for selection.\footnote{See also \citeA{Bodory2022EvaluatingLearning} for a similar discussion on the validity of unconfoundedness in the JC application.} It is therefore unlikely that selection bias drives the results in our labor economics application.

Assumption \ref{ass_observed} might be less credible in other applications. For example, \citeA{Bernard2024HowCompliance} provide meta-evidence for relevant variation in observational bias using an unconfoundedness based approach when considering a broad set of development economics applications.\footnote{Their primary observational method uses a partially linear model estimated via double debiased machine learning. When effects are heterogeneous, this model does not target the same estimand as experimental studies and, thus, their comparison is not necessarily an indication of bias.}
Nevertheless, \citeA{Avdeenko2025CostPakistan} employ an unconfoundedness based decomposition by \citeA{Heiler2021EffectTreatments} in the context of a development economics RCT.

\section{Extensions and Special Cases} \label{sec:extensions}
\subsection{Adjusted Group Differences-in-Means} \label{sec:adim-decomp}
The first natural extension is to consider the adjusted group differences-in-means 
\begin{align}
    & E[E[Y| T \in \mathcal{T}_a, X] | X \in \mathcal{X}_g] - E[E[Y| T \in \mathcal{T}_{a'}, X] | X \in \mathcal{X}_g] \notag  \\
     - \Big( &E[E[Y| T \in \mathcal{T}_a, X] | X \in \mathcal{X}_{g'}] - E[E[Y| T \in \mathcal{T}_{a'}, X] | X \in \mathcal{X}_{g'}] \Big) \label{eq_adjDiM}
\end{align}
that balances the distribution of confounding factors between the analyzed treatments in observational studies and not only the heterogeneity group indicator.\footnote{In principle, the adjusted DiM could be based on confounding variables that only assure unconfoundedness on the level of $\ta$ and $\tap$ respectively. These do not, in general, have to be identical to the $X$ used in Assumption \ref{ass_CIA}. In fully observational studies, it is usually hard to conceive different confounding variables between effective and aggregated treatments. Our approach can be extended at the expense of more involved derivations and additional selection bias components, see Appendix \ref{app:deriv-acm-xtilde} for more details.} This is the population quantity behind contrasting aggregated matching estimands between groups or, equivalently, projecting signals for conditional effects onto group indicators \cite{Semenova2021DebiasedFunctions}. Additionally, it covers a variety of estimands proposed in the literature as special cases; see Section \ref{sec:lit-rel} below.

A single adjusted group mean can be decomposed in a similar manner as the group mean in \eqref{eq:decomp-cm1} but takes into account the adjustment for covariates beyond the heterogeneity group (see derivation in Appendix \ref{app:deriv-acm}):
\begin{align} \label{eq:decomp-cm2}
    &E[E[Y|T\in\mathcal{T}_a,X]|X\in \xg] \nonumber \\ %= \sum_{t\in\ta} E[E[ \mathbbm{1}(T=t)Y(t)|T\in\mathcal{T}_a,X] |X\in \xg]  \nonumber \\
    & = \sum_{t\in\ta} \underbrace{e_{ta}(\xg) \mu_t(\xg)}_{\text{Stratified RCT}} + \underbrace{Cov[e_{ta}(X), \mu_{t}(X) |X\in \xg]}_{\text{Conditional individualized targeting}}  \nonumber \\ &\quad + \underbrace{[E[e_{ta}(X) |X\in \xg] - e_{ta}(\xg)] \mu_{t}(\xg)}_{\text{Aggregate composition adjustment}}  \nonumber \\
    % & = \sum_{t\in\ta} \underbrace{e_{ta}(\xg) \mu_t(\xg)}_{\text{Stratified RCT}} + \underbrace{Cov[e_{ta}(X), \mu_{t}(X) |X\in \xg]}_{\text{Adjusted targeting}} + \underbrace{E[e_{ta}(X) |X\in \xg] \mu_{t}(\xg) - d_{\mathcal{T}_a}(\xg)}_{\text{Adjusted treatment composition}}  \nonumber \\
    & =: d(a,g) + d_{4'}(a,g) + d_{5}(a,g).
\end{align}
$d(a,g)$ is the same synthetic stratified RCT as in \eqref{eq:decomp-cm1}. This also means that its decomposition into $d_1(a,g)$, $d_2(a,g)$ and $d_3(a,g)$ and the respective interpretations established in the previous section directly apply. 
$d_{4'}(a,g)$ captures again whether treatment assignment is individually targeted. However, it uses the covariance with the propensity score conditional on being in aggregation $a$, $e_{ta}(X)$, instead of the unconditional $e_{t}(X)$ in $d_{4}(a,g)$. It therefore takes into account that the effective treatment probabilities \textit{within} the analyzed treatment $a$ may vary beyond the heterogeneity group $g$.\footnote{\label{fn:d4s}To see this observe how the denominators differ when rewriting the individualized targeting components as $d_{4}(\mathcal{T}_a,\xg) = Cov\left(\frac{e_t(X)}{P(T\in\mathcal{T}_a | X\in \xg)},\mu_t(X) | X\in \xg\right)$ and $d_{4'}(\mathcal{T}_a,\xg) = Cov\left(\frac{e_t(X)}{P(T\in\mathcal{T}_a | X)},\mu_t(X) | X\in \xg\right)$.}
$d_{5}(a,g)$ is a consequence of adjusting the confounder distribution in $a$ to match the population distribution. It contrasts two hypothetical stratified RCTs that leave the potential outcome distribution untouched but use different assignment probabilities. 
Both assignment probabilities are averages over the conditional probability of treatment $t$ within aggregation $a$. The first averages $e_{ta}(X)$ over the whole group $g$. The second averages $e_{ta}(X)$ over group $g$ within aggregate treatment $a$ only, i.e.~it uses the actual group probabilities $e_{ta}(\xg) = E[e_{ta}(X)|X\in\xg,T\in\mathcal{T}_a]$. 
%The first one uses probabilities $E[e_{ta}(X) |X\in \xg]$ for assignment, i.e.~the conditional probability of treatment $t$ within aggregation $a$, $e_{ta}(X)$, averaged within subgroup $g$. The second one uses the actual group probabilities $e_{ta}(\xg) = E[e_{ta}(X)|X\in\xg,T\in\mathcal{T}_a]$. 
$d_{5}(a,g)$ is therefore driven by variation in propensity scores going beyond the group level and its interaction with treatment specific outcomes. %and has no clean targeting interpretation beyond that.

A single adjusted group DiM is then decomposed as
\begin{align}
E&[E[Y| T \in \mathcal{T}_a, X] | X \in \mathcal{X}_g] - E[E[Y| T \in \mathcal{T}_{a'}, X] | X \in \mathcal{X}_g] \label{eq:delta-decomp2} \\
& = \delta_{0}(a,a') + \delta_{1}(a,a',g) + \delta_{2}(a,a',g) + \delta_{3}(a,a',g) + \delta_{4'}(a,a',g) +  \delta_{5}(a,a',g).  \nonumber
\end{align}
with $\delta_{j}(a,a',g) := d_{j}(a,g) - d_{j}(a',g) \text{ for } j \in \{1,2,3,4',5\}$
and adjusted differences-in-means (ADiM) as defined in \eqref{eq_adjDiM} as
    \begin{align*}
&\Delta_{1}(a,a',g,g') & (\text{effect heterogeneity}) \\
     +& \Delta_{2}(a,a',g,g')  & (\text{average targeting heterogeneity}) \\
     +& \Delta_{3}(a,a',g,g')  & (\text{group targeting heterogeneity}) \\
     +& \Delta_{4'}(a,a',g,g')  & (\text{conditional individualized targeting heterogeneity}) \\
     +& \Delta_{5}(a,a',g,g')  & (\text{aggregate composition adjustment heterogeneity})
\end{align*}
with $\Delta_{j}(a,a',g,g') := \delta_{j}(a,g) - \delta_{j}(a',g') \text{ for } j \in \{1,2,3,4',5\}$. The interpretation of $\Delta_{1}$, $\Delta_{2}$ and $\Delta_{3}$ are already discussed in the previous section. The interpretation of $\Delta_{4'}$ is analogous to $\Delta_{4}$.
Finally, $\Delta_5$ becomes relevant if the confounder distributions in $a$ and/or $a'$ deviate from the population distribution. It tracks how treatment shares are implicitly changed by adjusting for covariate differences on the level of the aggregate treatments. $\Delta_5$ may be driven by treatment heterogeneity alone or its interaction with effect heterogeneity.\footnote{The exact mechanics leading to non-zero $\Delta_5$ are non-trivial and we do not elaborate on them here. However, progress can be made by recalling that it contrasts two different stratified RCTs that each can be decomposed following Equation \eqref{eq:decomp-srct}.} 

Appendix \ref{app:app} shows the decomposition for ADiM in the JC application. The differences to Figure \ref{fig:decomp} are negligible, which is expected as JC access is randomly assigned. Indeed, $\Delta_{4'}$ is small and insignificant similar to $\Delta_{4}$, and $\Delta_5$ is virtually zero (0.01, S.E.:~0.05).

\subsection{Special Cases and Relations to the Literature} \label{sec:lit-rel}

First, we note that under treatment homogeneity with $\mathcal{T}_a = \{1\}$ and $\mathcal{T}_{a'} = \{0\}$ all estimands collapse to familiar quantities and have the standard interpretation. For example, Equation \eqref{eq:delta-decomp1} simplifies to $E[Y(1)|T=1, X\in \xg] - E[Y(0)|T=0, X\in \xg]$ representing the group average treatment effect if unconfoundedness holds on the group level.\footnote{To see this note that then $Cov\left(\frac{e_t(X)}{e_t(\xg)},\mu_t(X) | X\in \xg\right) = E[Y(t)|T=t, X\in \xg] - \mu_t(\xg)$.} Otherwise, it can be decomposed into group effects and confounding bias following, e.g., \citeA{Heckman1998CharacterizingData}. Also, Equation \eqref{eq:delta-decomp2} collapses to the group average treatment effect $E[Y(1) - Y(0)| X\in \xg]$ under treatment homogeneity. 

Furthermore, DiM and ADiM decompositions in Sections \ref{sec_decompositionALL1} and \ref{sec:adim-decomp} are identical in two scenarios. First, if heterogeneity groups and confounders coincide like in an actually group stratified RCT as then $d_{4}(a,g) = d_{4'}(a,g) = d_{5}(a,g) = 0$. Second, if at least the analyzed treatment is (stratified) randomly assigned ensuring that $P(T\in\mathcal{T}_a | X) = P(T\in\mathcal{T}_a | X\in \xg)$ as then $d_{4}(a,g) = d_{4'}(a,g)$ (see footnote \ref{fn:d4s}) and $d_{5}(a,g) = 0$.\footnote{To see this note that then $E[e_{ta}(X)|X\in\xg] = e_{ta}(\xg) = \frac{e_t(\xg)}{P(T\in\mathcal{T}_a | X\in \xg)}$ for all $t\in\ta$.}

More importantly, the double aggregation framework and the decomposition of ADiM nests a variety of results in the literature on average estimands. First, consider unconditional/average estimands as special case with $g(X) = 1$ such that $\xg = \mathcal{X}$. Then, $\delta_0$ is a special case of the composite treatment effect in \citeA{Lechner2002ProgramPolicies}. Further, Proposition 8 in \citeA{VanderWeele2013CausalTreatment} is concerned with $E[E[Y| T \in \mathcal{T}_a, X]] - E[E[Y| T \in \mathcal{T}_{a'}, X]]$, which is the unconditional version of \eqref{eq:delta-decomp2}. They establish its interpretation as a contrast of two interventions where effective treatments are assigned according to their conditional population distribution in the different treatment aggregates. Their Proposition 8 is equivalent to the second equality in the derivation of the adjustment mean decomposition in Appendix \ref{app:deriv-acm} when setting $\xg = \mathcal{X}$. However, we go further in decomposing this contrast. Note that in the unconditional case $\delta_{1}=\delta_{2}=\delta_{3}=0$ and only $\delta_{0}$, $\delta_{4'}$ and $\delta_{5}$ are relevant. The same estimand is discussed by \citeA{Lee2024BridgingExposures} for continuous $T$. Additionally, they propose to consider contrasts of the form $E[E[Y| T \in \mathcal{T}_a, X]] - E[Y]$. As $E[Y] = d_0(a) + d_4(a,g)$ %= d_0(a) + d_{4'}(a,g)
is covered in our framework by setting $a(T) = 1$ and $g(X) = 1$, the decomposition could also be applied to their estimands. 
\citeA{vanderLaan2023NonparametricIntervention} also consider continuous $T$ and are interested in the adjusted group mean $E[E[Y|T \geq v,X]]$ for some threshold $v$. This corresponds to treatment aggregation function $a(T) = \mathbbm{1}(T \geq v)$. They interpret the estimand as partially stochastic intervention leaving all units with $T \geq v$ at their original treatment level but conditionally randomizing units with $T < v$ to values above the threshold. Our decomposition components $d_0$, $d_{4'}$ and $d_{5}$ could be applied to further decompose this estimand, while $d_{1}=d_{2}=d_{3}=0$ in their setting.

Second, the framework also relates to the causal decomposition of group disparities in
\citeA{Yu2025NonparametricDisparities} when grouping all treatments together ($a(T) = 1$). The paper considers a setting with $\ta = \mathcal{T} = \{0,1\}$ such that $e_{ta}(\cdot) = e_{t}(\cdot)$ and $P(T\in\mathcal{T}_a | X\in \xg) = 1$. The decomposed statistical estimand reads $E[Y|X \in \xg] - E[Y|X \in \xgp]$ in our notation. It would be decomposed as $\sum_{j=1}^4 d_j(a,g) - d_j(a,g')$ in our decomposition. $d_4(a,g) - d_4(a,g')$ is identical to the ``selection'' term in their equation (1). However, $d_j(a,g) - d_j(a,g')$ for $j \in \{1,2,3\}$ represent a different way to decompose the stratified experiment. We note that $d_1(a,g) - d_1(a,g')$ becomes the sum of their ``baseline'' component $\mu_0(\xg) - \mu_0(\xgp)$ and $e_{1} (\tau(\xg) - \tau(\xgp))$ in their setting, which is very similar to their ``effect'' component. The latter uses $e_{1}(\xg)$ instead of $e_{1}$ to weight the effect difference. While our decomposition does not separate the baseline term, their decomposition does not differentiate between average and group targeting. A hybrid decomposition-based on ours with additionally separated baseline term could therefore be a more detailed alternative to the decomposition in \citeA{Yu2025NonparametricDisparities}.

In summary, the double aggregation framework to analyze heterogeneity estimands in this paper can also be applied to simpler estimands discussed in the literature. This might be interesting on its own in the context of these papers. However, we focus here on its use to decompose heterogeneity estimands. Finally, we note that $\delta_0 + \delta_1$ recovers the random average treatment effect introduced by \citeA{Heiler2021EffectTreatments}, while their $\Delta$ is now decomposed into different components depending on the statistical estimand.

\subsection{Other research designs}
We note that estimands \eqref{eq_DiM} and \eqref{eq_adjDiM} do not cover a variety of important research designs. We provide an extension to multi-valued instrumental-variable-based designs where the treatment can be fully endogenous in Appendix  \ref{app:IV1}. We believe that many of the contributions of this paper can be extended to other settings such as difference-in-differences. However, even for the arguably simpler estimands \eqref{eq_DiM} and \eqref{eq_adjDiM}, the double aggregation setting has not been studied and reveals non-trivial complexities. Thus, we leave further extensions and discussions for future research.

\section{Multi-valued Treatment Effect Analysis versus Decomposition: Local Power Analysis}\label{sec_indirect1}

We now contrast the properties of testing group effect homogeneity in the decomposition framework with the conventional MVTE approach in the case of observed $X$ and $T$. We focus on finite-dimensional $T$ - a setting where both MVTE analysis and the decomposition are feasible.
The previous sections highlight that standard heterogeneity estimands \eqref{eq_DiM} and \eqref{eq_adjDiM} might not (only) capture effect heterogeneity in the double aggregation setting. Instead, they implicitly combine effect and treatment heterogeneity. On the other hand, $\Delta_1$ isolates effect heterogeneity at the scale of the (adjusted) DiM. Moreover, it allows us to assess an implication of the arguably empirically relevant null hypothesis of \textit{strong group effect homogeneity} 
\begin{align}
        H_0:~& \tau_{t,t'}(\xg) = \tau_{t,t'}(\xgp) \text{ for all } t \in \ta, t' \in \tap, \label{eq_strongeffecthomogeneity} \\ 
        %\text{ vs. } \\
        H_A:~& \tau_{t,t'}(\xg) \neq \tau_{t,t'}(\xgp) \text{ for some } t \in \ta, t' \in \tap. \label{eq_HAgheterogeneity}
\end{align}
In particular, rejecting \textit{weak group effect homogeneity} $\Delta_1 = 0$ is sufficient to reject strong group effect homogeneity \eqref{eq_strongeffecthomogeneity}. However, $H_A$ could hold while $\Delta_1 = 0$ if the heterogeneities offset each other when aggregating over all $t$-$t'$ combinations.\footnote{Appendix \ref{app:ih-th} provides similar discussions for testing treatment homogeneity.} This raises the question whether researchers interested in $H_0$ and access to the effective treatment could benefit from testing $\Delta_1 = 0$ instead of testing $H_0$ directly. The latter could, for example, be achieved by $\ell_p$-norm based tests such as Wald or supremum tests that rely on the (vector) of $t$-$t'$-specific MVTE estimates. Thus, the decomposition is not strictly required to assess the strong null. However, it can have significant power advantages in setups where the number of group effects to test $|\ta|\times|\tap|$ is large.

We now discuss the local power properties of statistical tests around the strong null hypotheses. For exposition, we focus on a simplified case where the aggregate control group defined by $\tap = \{0\}$ is homogeneous with potential outcomes $\mu_0(x) = 0$ for all $x \in \mathcal{X}$. Then, the strong null contains $J = |\ta|$ hypotheses. For MVTE, we consider conventional tests using a Wald ($\ell_2$) or supremum ($\ell_\infty$) statistic for the strong null with critical values that yield exact asymptotic size control. For the decomposition parameter, we test the weak null using a simple $t$-statistic. For all tests, we analyze their approximate, i.e.~first order, power properties along sequences of local alternatives around the strong null of the form \begin{align}
    \sqrt{n}(\tau(\xg) - \tau(\xgp)) \rightarrow \xi \in \mathbb{R}^J,
\end{align}
where $\tau(x) = (\tau_{1,0}(x),\dots,\tau_{J,0}(x))'$ for all $x \subseteq \mathcal{X}$ and $\xi = (\xi_1,\dots,\xi_J)'$. This means we consider $J$ differences in effective treatment means between groups $g$ and $g'$ that are in a root-$n$ neighborhood around zero (strong group effect homogeneity). 

\begin{table}[!h]
\centering
\begin{threeparttable}
  \caption{Local Power Comparison}
  \label{tab:power1}
    \begin{tabular}{l|c}
      Metric for Test   &  Approximate Local Power$^{\dagger}$  \\ \hline \\[-0.5ex]
      $\ell_2$ (Wald) &    $ 1 - \Phi\left( \frac{z_{1-\alpha}}{\sqrt{1 + {2||\xi||^2_2}/{J}}} - \frac{1}{\sqrt{2J}}\frac{{||\xi||_2^2}/{J}}{\sqrt{1 + {2||\xi||^2_2}/{J}}} \right)$ \\
      $\ell_\infty$ (Supremum) &  $1 - F_G\left(F^{-1}_{G,{1-\alpha}} - \sqrt{\frac{2\log J }{J}}||\xi||_{\infty}\right)$  \\
      $\Delta_1$ (Decomposition)  & $1 - \Phi\bigg(z_{1-\alpha/2} -\sum_{t\in \ta}e_{ta}\xi_t \bigg) + \Phi\bigg(z_{\alpha/2} - \sum_{t\in \ta}e_{ta}\xi_t \bigg)$  \\[1ex] \hline
  \end{tabular}
  \begin{tablenotes}[flushleft]
    \footnotesize
    \item \textit{Notes:} Approximate first-order local power comparisons between tests for sequences $\sqrt{n}(\tau(\xg) -\tau(\xgp)) \rightarrow \xi \in \mathbb{R}^{J}$, pointwise in $J$. All results are derived using the estimators and assumptions introduced in Section \ref{sec_estimation1}, a unit-variance assumption, and a fixed effective treatment dimension, see also Appendix \ref{sec_app_powerJ1} for more details. $\Phi(\cdot)$ and $z_{\alpha}$ are the cumulative distribution function and $\alpha$-quantile of the standard normal distribution. $F_G(\cdot)$ and $F^{-1}_{G,\alpha}$ are the cumulative distribution function and $\alpha$-quantile function of the $Gumbel(0,1)$ distribution.    \item[$^{\dagger}$] For $\ell_2$ and $\ell_\infty$, the power functions are tight upper bounds, for $\Delta_1$ they are exact (all up to second-order terms).
  \end{tablenotes}
\end{threeparttable}
\end{table}

Table \ref{tab:power1} summarizes the local power properties. Wald- and supremum-based tests have non-trivial power around the strong null. Wald has more power against dense alternatives and supremum against sparse alternatives as expected. However, power deteriorates quickly at an approximate $1/\sqrt{J}$ rate for both even when differences are relevant at each coordinate, i.e.~$||\xi||_2^2/J > 0$ (and thus also $||\xi||_\infty > 0$). For the supremum test, large $J$ leads to trivial power $\alpha$. For the Wald test, there can still be non-trivial power in high dimensions if the alternative is very dense in the sense that $||\xi||_2^2/J \nrightarrow 0$. However, when only a few (relative to $J$) group means are different, then the Wald test also deteriorates to trivial power. The decomposition-based test, on the other hand, has local power properties that are more robust to the dimensionality of the effective treatment. In particular, it has non-trivial power as $J\rightarrow \infty$ as long as the local propensity weighted differences are non-zero $\sum_{t\in\ta}e_{ta}\xi_t \neq 0$. Under sparse deviations from the strong null hypothesis, this metric can also be small. Under denser deviations, however, this can provide non-trivial power even when $J$ is large.
Thus, there is a trade-off between using conventional $\ell_p$-based tests and the decomposition analogues for rejecting the strong null: If the dimension of the effective treatment is large, these $\ell_p$ tests have low or trivial power. The decomposition-based test scales significantly better with $J$ but has little to trivial power against very sparse local alternatives and the direction where ($e_{ta}$-weighted) positive and negative deviations over the different treatment groups $\tau_{t,0}(\xg) - \tau_{t,0}(\xgp)$ aggregate to zero.

\begin{figure}[ht] 
    \centering
    \caption{Simulation of Finite Sample Power}
    \begin{subfigure}[b]{0.49\textwidth}
        \includegraphics[clip, width=\textwidth]{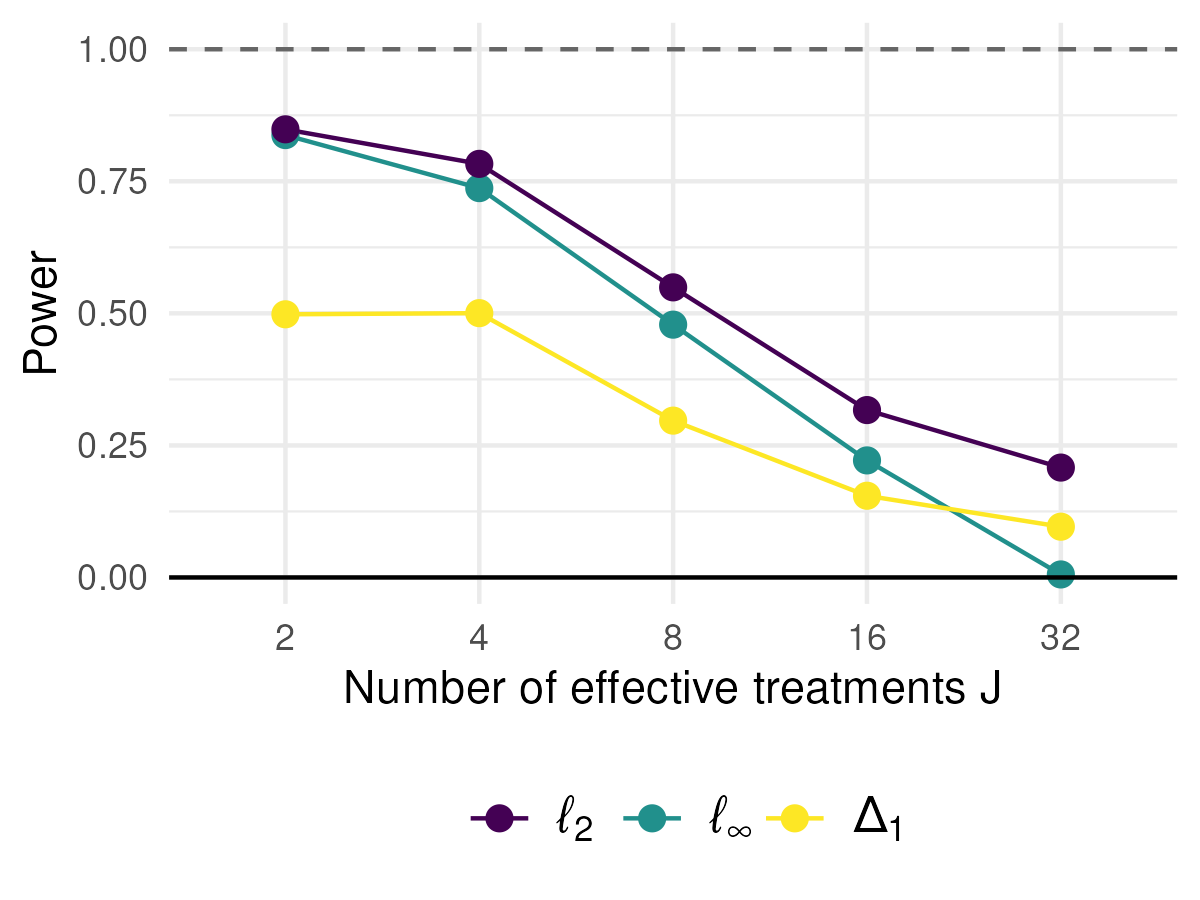}
        \caption{sparse}
        \label{fig:plot2}
    \end{subfigure}
    \begin{subfigure}[b]{0.49\textwidth}
        \includegraphics[clip, width=\textwidth]{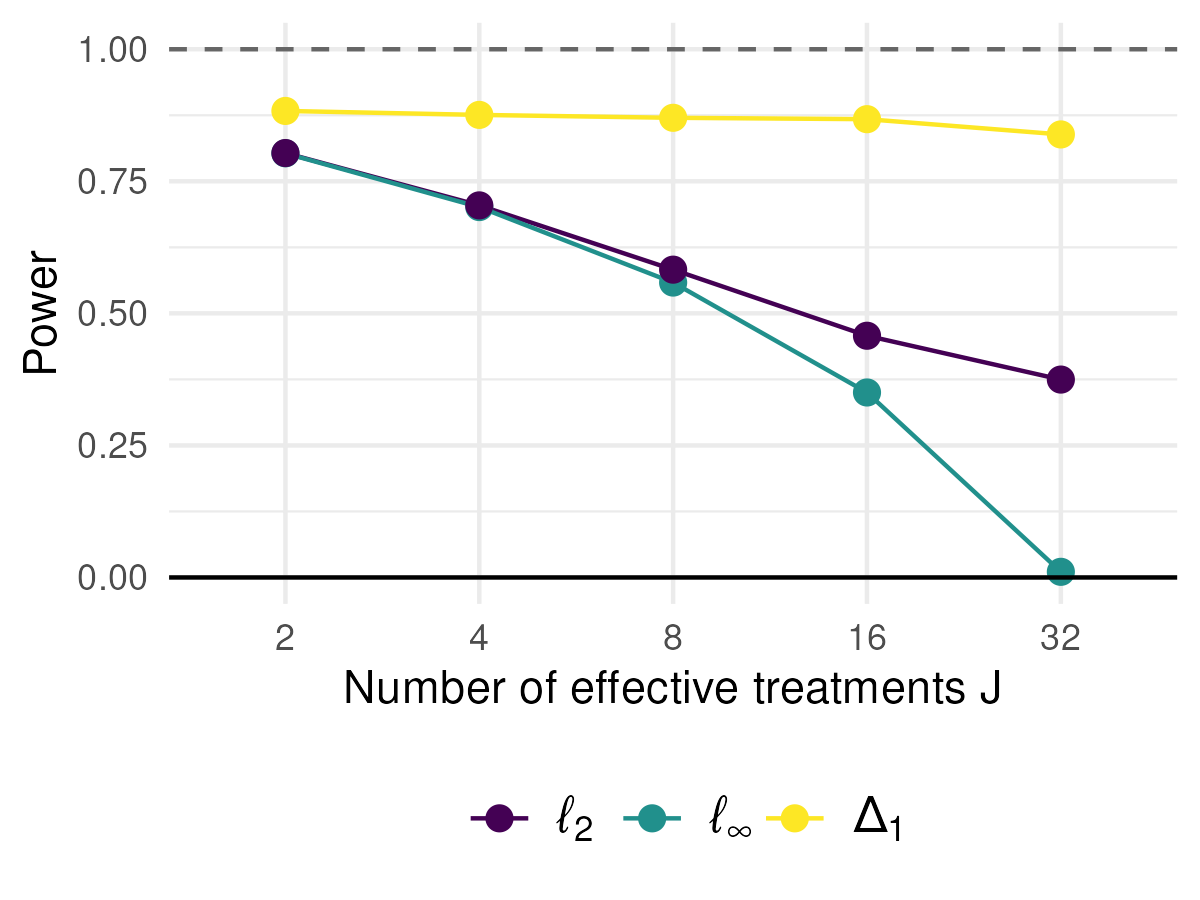}
        \caption{dense}
        \label{fig:plot1}
    \end{subfigure}
    \hfill
    \label{fig_powerplots1}
    \subcaption*{\textit{Notes:} The graphs display power for the DGP described in \ref{app:power-simul} for 10,000 replications with 1,000 observations. Panel (a) shows a sparse setting where only $\left[ J^{1/2} \right]$ entries are non-zero constants. Panel (b) shows a dense setting where all entries of $\xi$ are non-zero constants.}
\end{figure}

Figure \ref{fig_powerplots1} depicts the simulated finite sample power as a function of $J$ in a sparse and a dense design, see Appendix \ref{app:power-simul} for details. We can see that, in line with the theory, $\Delta_1$ is dominated by conventional tests for most $J$ under the sparse alternative by the $\ell_p$ based tests. However, it clearly dominates under the dense alternative and has near-constant power as $J$ increases.\footnote{
If the researcher's goal is limited to only having large power for strong null \eqref{eq_strongeffecthomogeneity}, other tests should also be considered. We conjecture that suitably propensity \textit{weighted} versions of $\ell_p$ based differences, analogously to the decomposition, could have qualitatively similar power curves as the test based on $\Delta_1$ but with $\sum_t e_{ta}\xi_t$ replaced by functions like $\sum_te_{ta}^p|\xi_t|^p$ (which might well be small for large $J$).  Moreover, \citeA{Kock2019PowerProblems}, Theorem 5.2 suggests that any of these tests are asymptotically enhanceable in the sense of \citeA{Fan2015PowerTests} over some part of the parameter space. Thus, we do not claim that $\Delta_1$ provides an ``optimal'' test for strong effect homogeneity hypothesis over subspaces under growing effective treatment dimensions. However, as proxy it can have advantages over common tests based on version-by-version MVTE estimates.
}

% Rejection of the implied hypothesis is sufficient for the rejection of strong effect homogeneity. Furthermore by definition, as argued above, it is a direct test for the synthetic estimand $\Delta_1(a,a',g,g')$. Hence it tests whether in a hypothetical experiment that has no treatment heterogeneity at given unconditional propensities, there is aggregate effect homogeneity. 

From a substantive perspective, the decomposition parameter $\Delta_1$ weights group differences by unconditional propensities and is therefore on the scale of the aggregate effect from the stratified experiment. This avoids rejecting effect homogeneity in setups where there is statistical variation in a few versions that are observed for a very small minority in the population. Thus, the particular form of $\Delta_1$ in \eqref{eq:Delta1} aligns the statistical with the economic significance of its components. 

In JC we find no evidence against strong group effect homogeneity. $\Delta_{1}$ representing $e_{ta}$ weighted effect differences has a $p$-value of 0.96. The MVTE $\ell_2$ ($p$ = 0.69) and $\ell_\infty$ ($p$ = 0.87) tests do not reject the strong null as well.

\section{Estimation and Inference} \label{sec_estimation1}
\subsection{Orthogonal Moments Building Blocks and Estimators}
We now provide a brief overview of the components used for constructing our orthogonal moments/influence functions. Some of them have a typical augmented inverse probability weighting structure. However, since most decomposition estimands are novel and extend beyond averages, they require correspondingly novel orthogonal moments, see Appendix \ref{sec_app_implement1} for all details. First note that the decomposition estimands consist of many parameters that can vary over covariate and group aggregation. Each of these are obtained by summing over $t$-specific components in the respective aggregate treatment. In particular, for given aggregations $a$ and $g$, each estimand is a sum over linear combinations of eight $t$-specific primitive parameters $\theta_{a,g,t,p}$ for $p=1,\dots,8$. Their influence functions all have a convenient weighted linear structure
\begin{align}
    \Psi_{a,t,p} = \Psi_{X,a,g,p}(\Psi_{Y,a,g,t,p} - \theta_{a,g,t,p} \Psi_{T,a,g}). \label{eq_linearMoment1}
\end{align}
We compactly summarize their functional forms in Table \ref{tab_IF_text1}. 
The specific entries do also depend on nuisance parameters and components of their respective influence functions $\Psi_{Y,\cdot}$. These influence functions have the same linear structure \eqref{eq_linearMoment1} and all of their components are contained in Table \ref{tab:agg-np_text1}.

\begin{table}[!h] 
\caption{IFs of components for decomposition parameters}\label{tab_IF_text1}
\centering \tiny %>{$}l<{$} %@{}c@{}
\begin{tabular}{|>{$}l<{$}|>{$}c<{$}|>{$}c<{$}|>{$}c<{$}|}
\hline
 & \Psi_X & \Psi_Y & \Psi_T \\ \hline
e_{ta}(\xg) \mu_t(\xg)  & \frac{\mathbbm{1}(X\in\xg)}{P(X\in\xg)} & \frac{\mathbbm{1}(T\in\mathcal{T}_a) \mathbbm{1}(T = t)}{P(T\in\mathcal{T}_a|X\in\xg)} \mu_t(\xg) + e_{ta}(\xg) \Psi_{Y,\mu_t} &  \left(1 + \frac{\mathbbm{1}(T\in\mathcal{T}_a)}{P(T\in\mathcal{T}_a|X\in\xg)} \right) \\ \hline
E[e_{ta}(X)|X\in\xg]\mu_t(\xg)  & \frac{\mathbbm{1}(X\in\xg)}{P(X\in\xg)} & \Psi_{Y,E[e_{ta}(X)|X\in\xg]} \mu_t(\xg) + E[e_{ta}(X)|X\in\xg] \Psi_{Y,\mu_t} &  2 \\ \hline
E[e_{ta}(X)\mu_t(X)|X\in \xg]  & \frac{\mathbbm{1}(X\in\xg)}{P(X\in\xg)} & \frac{\mathbbm{1}(T \in \ta) (\mathbbm{1}(T = t) - e_{ta}(X))}{P(T \in \ta|X)} \mu_t(X) + e_{ta}(X) \Psi_{Y,\mu_t} &  1 \\ \hline
e_{ta}\mu_t(\xg)  & 1 & \frac{\mathbbm{1}(T \in \ta)}{P(T \in \ta)}\mathbbm{1}(T = t) \mu_t(\xg) + e_{ta} \frac{\mathbbm{1}(X\in\xg)}{P(X\in\xg)} \Psi_{Y,\mu_t} &  \left(\frac{\mathbbm{1}(T \in \ta)}{P(T \in \ta)} + \frac{\mathbbm{1}(X\in\xg)}{P(X\in\xg)} \right)\\ \hline
e_{ta}(\xg)\mu_t  & 1 & \frac{\mathbbm{1}(T\in\mathcal{T}_a,X\in\xg)}{P(T\in\mathcal{T}_a,X\in\xg)} \mathbbm{1}(T = t) \mu_t + e_{ta}(\xg) \Psi_{Y,\mu_t} &  \left(\frac{\mathbbm{1}(T\in\mathcal{T}_a,X\in\xg)}{P(T\in\mathcal{T}_a,X\in\xg)} + 1\right) \\ \hline
%Cov(e_{ta}(X),\mu_t(X)|X\in \xg)  & \multicolumn{3}{c|}{$\Psi_{Cov(e_{ta}(X),\mu_t(X)|X\in \xg)} = \Psi_{E[e_{ta}(X)\mu_t(X)|X\in \xg]} - \Psi_{E[e_{ta}(X)|X\in\xg]\mu_t(\xg)}$} \\ \hline
e_{t}(\xg) \mu_t(\xg)  & \frac{\mathbbm{1}(X\in\xg)}{P(X\in\xg)} & \mathbbm{1}(T = t) \mu_t(\xg) + e_{t}(\xg) \Psi_{Y,\mu_t} &  2 \\ 
\hline
E[e_t(X)\mu_t(X)|X\in \xg]  & \frac{\mathbbm{1}(X\in\xg)}{P(X\in\xg)} & \mathbbm{1}(T=t)Y &  1 \\ \hline
%Cov(e_t(X),\mu_t(X)|X\in \xg) & \multicolumn{3}{c|}{$\Psi_{Cov(e_t(X),\mu_t(X)|X\in \xg)} = \Psi_{E[e_t(X)\mu_t(X)|X\in \xg]} - \Psi_{e_{t}(\xg) \mu_t(\xg)}$} \\ \hline
e_{ta}\mu_t  & 1 & \frac{\mathbbm{1}(T \in \ta)}{P(T \in \ta)}\mathbbm{1}(T = t) \mu_t + e_{ta} \Psi_{Y,\mu_t} &  \left(\frac{\mathbbm{1}(T \in \ta)}{P(T \in \ta)} + 1 \right)\\ \hline
\end{tabular} \label{tab:agg-tc2}
\end{table}

\begin{table}[ht] 
\caption{IF components of aggregated nuisance parameters}\label{tab:agg-np_text1}
\centering \tiny
\begin{tabular}{|>{$}l<{$}|>{$}c<{$}|>{$}c<{$}|>{$}c<{$}|}
\hline
& \Psi_X & \Psi_Y & \Psi_T \\ \hline
e_{a} &  1 & \mathbbm{1}(T \in \ta) & 1 \\ \hline
e_{g} &  1 & \mathbbm{1}(X\in\xg) & 1 \\ \hline
e_{t} &  1 & \mathbbm{1}(T = t) & 1 \\ \hline
\mu_t &  1 & \frac{\mathbbm{1}(T=t) (Y - \mu_t(X))}{P(T=t|X)} + \mu_t(X) & 1 \\ \hline
e_{ta} &  \frac{\mathbbm{1}(T \in \ta)}{P(T \in \ta)} & \mathbbm{1}(T = t) & 1 \\ \hline
e_t(\xg) &  \frac{\mathbbm{1}(X\in\xg)}{P(X\in\xg)} & \mathbbm{1}(T=t) & 1 \\ \hline
\mu_t(\xg) &  \frac{\mathbbm{1}(X\in\xg)}{P(X\in\xg)} & \frac{\mathbbm{1}(T=t) (Y - \mu_t(X))}{P(T=t|X)} + \mu_t(X) & 1 \\ \hline
e_{ag} &  1 & \mathbbm{1}(T\in\mathcal{T}_a,X\in\xg) & 1 \\
\hline
e_a(\xg) &  \frac{\mathbbm{1}(X\in\xg)}{P(X\in\xg)} & \mathbbm{1}(T \in \ta) & 1 \\ \hline
\mu_a (\xg) & \frac{\mathbbm{1}(T\in\mathcal{T}_a,X\in\xg)}{P(T\in\mathcal{T}_a,X\in\xg)} & Y & 1 \\ \hline
e_{ta}(\xg) &  \frac{\mathbbm{1}(T\in\mathcal{T}_a,X\in\xg)}{P(T\in\mathcal{T}_a,X\in\xg)} & \mathbbm{1}(T = t) & 1 \\ \hline
E[e_{ta}(X)|X\in\xg] &  \frac{\mathbbm{1}(X\in\xg)}{P(X\in\xg)} & \frac{\mathbbm{1}(T \in \ta) (\mathbbm{1}(T = t) - e_{ta}(X))}{P(T \in \ta|X)} + e_{ta}(X) & 1 \\ \hline
\end{tabular} 
\end{table}

The estimators for all decomposition estimands are then obtained by solving the estimating equations for their respective linear combinations based on Table \ref{tab_IF_text1} using estimated nuisances.
All unknown \textit{aggregate} nuisances for this step are obtained via the influence functions in Table \ref{tab:agg-np_text1}. All \textit{granular} nuisances required for both steps are obtained via $K$-fold cross-fitting \cite{Chernozhukov2018}.

In particular, for a given $a$ and $g$, we denote as $\theta(a,g) \in \mathbb{R}^{8}$ all the estimands corresponding to the $\ta$-aggregated (i.e.~summed over $t \in \ta)$ influence functions components from Table \ref{tab_IF_text1}.  $\hat{\theta}(a,g)$ are their respective estimators. Again, all these decomposition estimands and estimators can be obtained via linear combinations of their respective components. For example, for decomposition parameter $d_0(g) = \sum_{t\in\ta}e_{ta}\mu_t$, the implied estimator is given by $\hat{d}_0(a) = \sum_{t\in\ta}\hat{e}_{ta}\hat{\mu}_t$
where all components are obtained by the solution to their influence function evaluated at estimated nuisances
\begin{align}
    \frac{1}{n}\sum_{i=1}^n\left[\frac{\mathbbm{1}(T_i=t)(Y_i-\hat{\mu}_t(X_i))}{\hat{e}_t(X_i)} + \hat{\mu}_t(X_i) - \hat{\mu}_t\right] = 0, \\
    \frac{1}{n}\sum_{i=1}^n\left[\frac{\mathbbm{1}(T_i \in \ta)}{\hat{e}_a}(\mathbbm{1}(T_i=t) - \hat{e}_{ta})\right] = 0, 
\end{align}
with additional aggregate nuisance $e_a$ estimated via $n^{-1}\sum_{i=1}^n\left[\mathbbm{1}(T_i \in \ta) - \hat{e}_a\right] = 0$.
The granular nuisances $\hat{\mu}_t(X)$ and $\hat{e}_{t}(X)$ are obtained via cross-fitting using suitable machine learning or other non/semiparametric prediction methods.

\subsection{Large Sample Properties}
We now present the main technical assumptions and large sample properties of the influence function based decomposition estimators and discuss how they extend and nest existing results in the semiparametric debiased machine learning literature.
Denote the variance-covariance matrix of the full vector of decomposition component influence functions as $\Sigma$, see Appendix \ref{app_proof_multiCLT} for a formal definition. Let $J = J_n = |\mathcal{T}_a|$ be the number of treatments in aggregation $a$.
Nuisances are $\eta \in \mathcal{H}$ where $\mathcal{H}$ is a convex subset of some normed vector space. Denote the realization set of the estimated nuisance quantities by $\mathcal{H}_n = \mathcal{E}_n \times \mathcal{M}_n \subset \mathcal{H}$ with $
	\mathcal{E}_n = E_{0,n}\times E_{1,n}\times\dots\times E_{J,n}$ and $
	\mathcal{M}_n = M_{0,n}\times M_{1,n}\times\dots\times M_{J,n}$,
where $E_{t,n}$ and $M_{t,n}$ are the realization sets containing estimates $\hat{e}_t(X)$ and $\hat{\mu}_t(X)$ with probability $1-u_n$. Define their slowest $L_q$ error rates as 
\begin{align}
	r_{e,n,q} &= \sup_{t \in \ta}\sup_{\hat{e}_t \in E_{t,n}} E[(\hat{e}_t(X) - e_t(X))^q]^{1/q},  \\
	r_{\mu,n,q} &= \sup_{t \in \ta}\sup_{\hat{\mu}_t \in M_{t,n}} E[(\hat{\mu}_t(X) - \mu_t(X))^q]^{1/q}.
\end{align}
We write $a_n \lesssim b_n$ and $a_n \lesssim_P b_n$, whenever $a_n = O(b_n)$ or $a_n = O_p(b_n)$ respectively. If not stated differently, the following assumptions are uniformly over $n$. 

\noindent {\textbf{Assumption A (Regularity, Treatment, and Learning Rates)}}
%\begin{assump}[Regularity, Treatment, and Learning Rates]
\textit{Let $\underline{c} > 0$ be a universal constant. For given $a$ and $g$, we assume that \begin{enumerate}[itemsep=0pt] \singlespacing
\item[A.1)] (Moments) Conditional potential outcome moments are bounded, i.e.~for some $m > 0$ \begin{align*}\sup_{x\in\mathcal{X},t\in\ta}E[|Y(t)|^{2+m}|X=x] \lesssim 1.\end{align*}
\item[A.2)] (Eigenvalue) The variance of the influence function is non-degenerate, i.e.~the smallest eigenvalue of $\Sigma$ is bounded away from zero.
\item[A.3)] (Aggregate Overlap) The shares of aggregate treatment and heterogeneity groups are bounded away from zero 
\begin{align*} \min\{P(\xg),P(\ta),P(\ta|\xg)\} > \underline{c}.\end{align*}
\item[A.4)] (Weak Effective Treatment Overlap) Each treatment group contains a non-trivial share of observations \textbf{relative} to the total number of treatments \begin{align*}
    Je_t \lesssim 1 \textit{ or } e_t > \underline{c} \textit{ for any } t \in \ta,
\end{align*}
and it does so for all covariates \begin{align*}
    \sup_{t\in\ta,x\in\mathcal{X}}\frac{e_t}{e_t(x)} \lesssim 1.
\end{align*}
\item[A.5)] (Machine Learning Rates) The cross-fitted nuisance learners are consistent at sufficiently fast rates \begin{align*}
    r_{\mu,n,2} + J r_{e,n,2}  &= o(1), \\
    \sqrt{n}J r_{\mu,n,2} r_{e,n,2} &= o(1).
\end{align*}
\item[A.6)] (Bounded Relative Prediction Error) On the realization set with probability $1-u_n = 1- o(1)$, the worst relative prediction error for the cross-fitted propensities are bounded \begin{align*}
    \sup_{t\in\ta,x\in\mathcal{X}}\sup_{\hat{e}_t\in E_{t,n}}\frac{e_t(x)}{\hat{e}_t(x)} \lesssim 1.
\end{align*}
\end{enumerate}}
%\end{assump}

Assumption A.1 and A.2 are standard regularity conditions. Assumption A.3 is an overlap condition with respect to both treatment and heterogeneity groups. This means that both aggregates are comprised of a sufficiently large share of observations in the population and that their intersection is non-empty. A.4 has two components. First, the propensities of the effective treatment are either bounded away from zero or vanish at a rate of mostly $J^{-1}$. Thus, along sequences of DGPs, the $t$-specific propensities are allowed to converge to zero, i.e.~we do not impose strong overlap on the level of the effective treatment. It also accommodates the aggregation of both small and large effective treatment groups. Second, we require that the conditional propensities of the effective treatment are on similar scale as the unconditional propensities. This effectively means that, as the dimensionality of the treatment grows with most groups containing smaller and smaller shares at a particular rate, the propensities for each group vanish comparably fast/slow. A.4 can be inspected visually via the $1/e_t$ rescaled effective treatment propensity score densities close to the boundary. If any of these diverge at zero, weak overlap is likely violated, equivalently to generic propensities under limited overlap \cite{Heiler2021ValidScores}. The \href{https://mcknaus.github.io/assets/code/RNB_HK2_decomposition.nb.html}{Supplementary R Notebook} S.2.2 provides an example based on the Job Corps data where we do not detect violations of Assumption A.4. 

Assumption A.5 is a requirement on the $L_2$ approximation qualities of the nuisance functions. In the case of a fixed $J$, the assumption collapses to the standard debiased machine learning requirements as in \citeA{Chernozhukov2018}. They are stronger than the typical debiased machine learning requirements when $J$ diverges. This is the price of the growing nuisance parameter space. Additionally, one might be worried about the approximation qualities of $r_{\mu,n,2}$ and $r_{e,n,2}$ themselves also depending on $J$. For many learners, however, they tend to counterbalance each other. In particular, conditional means become harder to estimate with vanishing sample frequencies, while propensities actually become easier. The latter is due to the fact that the smaller the frequency, the lower the (conditional) variance of the corresponding random indicator. Thus, estimators of conditional selection probabilities often exhibit \textit{superefficiency} properties in asymptotic regimes with $J\rightarrow \infty$, which weakens learning rate requirements. For example, for regular parametric nuisance models, we have that $r_{\mu,n,2} \lesssim \sqrt{{J/n}}$ and $r_{e,n,2} \lesssim \sqrt{1/nJ}$. Thus, the learning requirements in A.5 reduce to $ J = o(n^{1/2})$, which is a moderate but relevant constraint on the dimensionality of the effective treatment. In more nonparametric setups, the growth requirements can become stronger but still admit a reasonably growing number of treatments. For example, for $q$-dimensional kernel regression for a regression function with $s$ H\"older continuous derivatives and optimal bandwidth choice, the requirement in A.5 is $ J = o(n^{\frac{1}{2}\left[\frac{2s - q}{2s + q}\right]})$, see also \citeA{Heiler2021EffectTreatments} for additional examples.\footnote{Condition A.5 is reminiscent of, but different from non-parametric projection methods for estimation of heterogeneous treatment effects \cite{Semenova2021DebiasedFunctions,Heiler2024HeterogeneousPolarization}. There, growing basis functions yield additional bias and learning rates have to be faster compared to unconditional estimands. In our setup, additional bias arises from the effective treatment dimension that yields both an expanding nuisance parameter space and limited overlap.} 

A.6 is another assumption related to the approximation quality of the propensity scores. It requires that, uniformly over the covariate space, the ratio of true to estimated propensities should be bounded with high probability. This holds trivially when the propensity scores are uniformly consistent. However, this is not necessary. We expect this to hold for most frequency based methods with conventional loss functions.\footnote{Uniform consistency applies to various nonparametric estimators such as kernel or series regression, see also \citeA{Ma2024TestingOverlap} for semiparametric single index models and \citeA{Hardle2025UniformForests} for generalized random forests.} 
We obtain the following Theorem:

\begin{thm}[Large Sample Inference] \label{thm_asyN1}
    For any aggregation $a$, $g$, and bounded sequence vector $c_n \lesssim 1$ with $c_n \neq 0$, given Assumptions A.1 - A.6, we have that all decomposition parameters are asymptotically normal \begin{align*}
        \left[c_n'\Sigma c_n\right]^{-\frac{1}{2}}c_n'\left(\hat{\theta}(a,g) - \theta(a,g)\right) \overset{d}{\rightarrow} \mathcal{N}(0,1).
    \end{align*}
    Moreover when $J \left(\frac{J}{n}\right)^{\left[\frac{1}{2}\wedge \frac{m}{2+m}\right]} = o(1)$, then asymptotic normality applies with $\Sigma$ replaced by its sample equivalent with estimated nuisances. 
\end{thm}

Theorem \ref{thm_asyN1} implies the asymptotic validity of using the suggested influence-function based estimators for estimation and inference on any decomposition parameters. As the treatment and covariate aggregations are mutually exclusive, this directly implies an analogous result for all decomposition estimands. We note that, for a fixed $J$, our estimators reach their respective semiparametric efficiency bound as they are based on efficient influence functions.  The variance assumption relating potential outcome moments to the estimation of $\Sigma$ is sufficient, but not necessary, see Appendix \ref{sec_app_variance1} for more details. Critical values and quantiles for tests and confidence intervals can be obtained using the standard normal distribution. The dimensionality of the effective treatment $J$ does not affect the rate of convergence. This is due to the fact that, while the number of nuisance parameters and potential outcomes grows, the decomposition is always made up of the same 8 components regardless of $J$.

\section{Generalized Treatment Spaces} \label{sec:generalized1}
\subsection{Generalized Treatment Spaces and Aggregations}
Identification, estimation, and inference generalize beyond the discussed multi-valued effective treatment case.
We now present the case where treatment and aggregation spaces are allowed to be (union of) uncountable as well as countable or finite sets $\{t_r\} = \{t_1,t_2,\dots\}$ nesting all results in Sections \ref{sec_decompositionALL1} and \ref{sec_estimation1}. Proofs and further derivations are in Appendix \ref{sec_app_generalizedT}.
For any $x \subseteq \mathcal{X}$, we define a (conditional) probability measure of treatment $T$ according to the Lebesgue decomposition theorem as $\nu_{t|x} = \nu_{t|x}^c + \nu_{t|x}^d$. Thus, for any measurable set $S$, we have \begin{align}
    \nu_{t|x}^c(S) = \int_{S\backslash\{t_r\}}f(t|x)d\lambda( t), \quad 
    \nu_{t|x}^d(S) = \sum_{t\in\{t_r\}}P(T=t|x)\delta_{t}(S).
 \end{align}
where $f(t|x)$ is a (conditional) density function, $\lambda(\cdot)$ the Lebesgue measure, and $\delta_{t}(S) = \mathbbm{1}(t \in S)$ the Dirac function. We further define the following spaces \begin{align}
    L^2(\ta\backslash \{t_r\},\lambda) &= \{ g : \int_{\ta\backslash \{t_r\}}|g(t)|^2d\lambda(t) < \infty \}, \\
    \ell^2(\{t_r\}) &= \{ g : \sum_{t\in\{t_r\}}|g(t)|^2 < \infty \}. 
\end{align}
$L^2(\ta\backslash \{t_r\},\lambda)$ is a function space over measure space $(\ta\backslash \{t_r\},\mathcal{B},\lambda)$. $\ell^2(\{t_r\})$ is a space of square-summable sequences over $\{t_r\}$. As composite they yield a hybrid space of ordered pairs $\la = L^2(\ta\backslash \{t_r\},\lambda) \oplus \ell^2(\{t_r\})$. 
Lastly, for generic functions (possibly indexed by $x$) $g_1,g_2:\mathcal{T}\rightarrow \mathbb{R}$, we define the following inner products relative to the Lebesgue and counting measure as  \begin{align}
    \ip{g_1,g_2}_{L^2(\ta\backslash \{t_r\},\lambda)}
    &= \int_{\ta\backslash \{t_r\}}g_1(t)g_2(t)d\lambda(t), \\
    \ip{g_1,g_2}_{\ell^2(\{t_r\})} &= \sum_{t\in\{t_r\}} g_1(t)g_2(t),
\end{align}
and, consequently, the composite inner product \begin{align}
    \ip{g_1,g_2}_{\la}
    &= \ip{g_1,g_2}_{L^2(\ta\backslash \{t_r\},\lambda)} + \ip{g_1,g_2}_{\ell^2(\{t_r\})}.
\end{align}
Under abuse of notation, when referring to elements in $\mathcal{L}^2_a$, we subsequently denote ordered pairs
% \begin{align}
    $\mu = (\{\mu(t): t \in \ta \backslash \{t_r\}\}, \{\mu({t_r})\})$ and %\label{eq_abuse1}\\
    $e_a = (\{f_a(t) : t \in \ta \backslash \{t_r\}\}, \{e_{t_ra}\})$, %\label{eq_abuse2}
%\end{align} 
where $\mu(t) = \mu_t$ and same for $\mu(x), e_a(x)$, and $e(x)$ for all $x \subseteq \mathcal{X}$. 

\subsection{Generalized Decompositions}
We now decompose a the group mean and its adjusted counterpart as well as the synthetic stratified experiment. Paralleling the results in Section \ref{sec_decompositionALL1}, we obtain 
\begin{align}
     E[Y|T\in\mathcal{T}_a,X\in \xg]
     &=\ip{e_{a}(\xg),\mu(\xg)}_{\la} +\frac{\ip{Cov(e_{a}(X),\mu(X)|X\in\xg),1}_{\la}}{P(T\in\ta|X\in\xg)}, \end{align}
     as well as
     \begin{align}
     E\left[E[Y|T\in\ta,X]|X\in\xg\right]
     &= \ip{e_{a}(\xg),\mu(\xg)}_{\la} + \ip{Cov(e_{a}(X),\mu(X)|X\in\xg),1}_{\la} \notag \\ &\quad +  \ip{E[e_{a}(X)|X\in\xg] - e_{a}(\xg),\mu(\xg)}_{\la}. 
\end{align}
% Now define unconditional estimands over the composite space as ordered pairs \begin{align}
%     \mu &= (\{\mu_t : t \in \ta \backslash \{t_r\}\}, \{\mu_{t_r}\}), \\
%     e_a &= (\{f_a(t) : t \in \ta \backslash \{t_r\}\}, \{e_{t_r}\}).
% \end{align} 
We further decompose the synthetic stratified experiment as
\begin{align}
    \ip{e_{a}(\xg),\mu(\xg)}_{\la} &= \ip{e_{a},\mu}_{\la}
    + \ip{e_{a},\mu(\xg)-\mu}_{\la} 
    + \ip{e_{a}(\xg) - e_a,\mu}_{\la} \notag \\
    &\quad + \ip{e_{a}(\xg) - e_a,\mu(\xg) - \mu}_{\la}. \label{eq_general_decomp_sexp1}
\end{align} 
All inner products have the same substantive interpretations as their discrete analogues in Section \ref{sec_decompositionALL1}.
The decomposition for the differences in terms of $a$ and $g$ can be obtained from these components as in Section \ref{sec:decomp-diffs} and Section \ref{sec:decomp-diff-in-diffs}. 
\subsection{Estimation and Inference}
We now extend the results of Section \ref{sec_estimation1} to the generalized treatment and aggregation spaces. We consider parameter $d_{0}(a)$ in what follows for exposition, but the theory readily extends to all decomposition parameters as well as their linear combinations as in Theorem \ref{thm_asyN1}. 
The target parameters in the decomposition now live in the general treatment and aggregation spaces $\mathcal{T}$ and $\ta$ respectively. We still employ the same discretized estimator as in Section \ref{sec_estimation1}. 
However, we now use $\tilde{J} = J^* + J$ treatments for estimation where $J = J_n$ is used for $\{t_r\} = \{t_r\}_{r=1}^J$ discrete treatments with non-zero probability mass as in Section \ref{sec_estimation1} and $J^* = J^*_n$ uses the same estimator but for $J^*$ user-defined partitions of the treatment aggregation space minus the $J$ components $\{t_r\}$. In particular, we define the partitioning $\{v_j\}_{j=1}^{J^*}$ such that $v_j \cap v_k = \emptyset$ if $j\neq k$ and $\bigcup_{j=1}^{J^*} v_j = \ta \backslash \{t_r\}$. The implied target estimand for the decomposition using a given partitioning is then given by pseudo-true parameter \begin{align}
    d^*_{0}(a) = \sum_{j=1}^{J^*}E[E[Y|T\in v_j,X]]P(T\in v_j|T \in \ta) + \sum_{t\in\{t_r\}}E[E[Y|T=t,X]]e_{ta}.
\end{align}
Theorem \ref{thm_asyN1} directly implies that the proposed estimator is asymptotically normal around this $d^*_{0}(a)$.
However, the true parameter of interest according to population decomposition \eqref{eq_general_decomp_sexp1} equals 
\begin{align}
    d_{0}(a) 
    &= \ip{e_{a},\mu}_{\la} \notag \\
    &= \int_{\ta \backslash \{t_r\}} E[E[Y|T=t,X]]f(t|T\in \ta)d\lambda(t) + \sum_{t\in\{t_r\}}E[E[Y|T=t,X]]e_{ta}.
\end{align}
$d^*_{0}(a)$ can be seen as an approximation to $d_{0}(a)$ where propensities are approximated by shares and conditional means by density weighted shares within discrete partitions. We can thus bound their difference as 
% \begin{align*}
%     d^*_{0}(a) - d_{0}(a) 
%     &= \frac{1}{P(T\in \ta)}\left(\int_{\ta \backslash \{t_r\}} P_{f,J^*}\mu(t)P_{J^*}f(t)d\lambda(t) - \int_{\ta \backslash \{t_r\}} \mu(t) f(t) d\lambda(t) \right)
% \end{align*}
%
%Now note that both projections errors must also be in the respective $L^2$ spaces. Thus, we obtain 
\begin{align}
    d^*_{0}(a) - d_{0}(a) 
 %   &\lesssim \frac{1}{P(T\in \ta)}\left(\ip{P_{f,J^*}\mu - \mu ,f}_{L^2(\ta \backslash \{t_r\},\lambda)} + \ip{P_{f,J^*}\mu,P_{J^*}f - f}_{L^2(\ta \backslash \{t_r\},\lambda)} \right) \\
%    &\lesssim \ip{P_{f,J^*}\mu - \mu ,1}_{L^2(\ta \backslash \{t_r\},f,\lambda)} + ||\mu||_{L^2(\ta \backslash \{t_r\},\lambda)}||f - P_{J^*}f||_{L^2(\ta \backslash \{t_r\},\lambda)} \\
    &\lesssim ||  \mu - P_{f,J^*}\mu||_{L^2(\ta \backslash \{t_r\},f,\lambda)} + ||f - P_{J^*}f||_{L^2(\ta \backslash \{t_r\},\lambda)},
\end{align}
with induced norm $||g||_{L^2(T,f,\lambda)} = \ip{g,g}_{L^2(T,f,\lambda)}^{1/2}$ defined by the $f$-weighted inner product $\ip{g_1,g_2}_{L^2(T,f,\lambda)} = \int_Tg_1(t)g_2(t)f(t)d\lambda(t)$. $P_{J^*,f}\mu$ and $P_{J^*}f$ are the orthogonal projections of $\mu(t) = E[E[Y|T=t,X]]$ and $f(t)$ respectively onto the subspace of piecewise constant functions in an ($f$-weighted) $L^2(\ta \backslash \{t_r\},\lambda)$ sense \begin{align}
P_{f,J^*}\mu &= \arg \min_{\tilde{\mu} \in V_{J^*}}\int_{\ta \backslash \{t_r\}} |\mu(t) - \tilde{\mu}(t)|^2f(t)d\lambda(t), \\
P_{J^*}f &= \arg \min_{\tilde{f} \in V_{J^*}}\int_{\ta \backslash \{t_r\}} |f(t) - \tilde{f}(t)|^2d\lambda(t), \\
\mathcal{V}_{J^*} &= {\bigg\{ } \sum_{j=1}^{J^*}c_j\mathbbm{1}(t \in v_j) \bigg| c_j \in \mathbb{R} {\bigg \} }.
\end{align}
We obtain the following Theorem:
\begin{thm} \label{thm_general_asyN1} Under the assumptions of Theorem \ref{thm_asyN1} with all statements about $J$ and treatment propensities also applying to $J^*$ and corresponding partition propensity scores, we have that
\begin{align*}
      \Sigma_{11}^{-1/2}\sqrt{n}(\hat{d}_{0}(a) - d_{0}(a)) &= Z + B_n^* + o_p(1), \end{align*} 
      where $Z \overset{d}{=} \mathcal{N}(0,1)$ and the discretization bias obeys
      \begin{align*}
      B_n^* &\lesssim \sqrt{n}\left(||\mu - P_{f,J^*}||_{L^2(\ta \backslash \{t_r\},f,\lambda)} + ||f - P_{J^*}f||_{L^2(\ta \backslash \{t_r\},\lambda)}\right). 
      \end{align*}
\end{thm}
Thus, there is a non-random discretization bias that is determined by the properties of the local partitioning projection of the respective treatment distribution over $\ta$. These can be controlled if the class of potential outcomes and treatment density on $\ta$ is not too complex. For example, for smooth function classes, a simple requirement for $J^*$ relative to the degree of smoothness assures asymptotic normality around the true decomposition target with asymptotically negligible discretization bias: 

\begin{corr} \label{corr_asyN_smooth1}
Let $\ta$ be compact. If $\mu,f \in C^q(\ta \backslash \{t_r\})$, then $
        B_n^* \lesssim \sqrt{{n}\big/{{J^*}^{2q}}} $.
    Moreover if $n/({J^*}^{2q}) = o(1)$, then \begin{align*}
      \Sigma_{11}^{-1/2}\sqrt{n}(\hat{d}_{0}(a) - d_{0}(a)) \overset{d}{\rightarrow} \mathcal{N}(0,1).
    \end{align*}
\end{corr}

Given the constraints on $J$ in the previous section, the overall number of partitions to approximate aggregates for general treatment spaces has to be in a balanced range for the discretization bias $B_n^*$ to be asymptotically negligible. The smoother the functions, the less constraining the lower bound. The upper bounds used in the Assumptions for Theorem \ref{thm_asyN1} must also apply to both $J$ and $J^*$. For example, when we have twice continuously differentiable potential outcomes and treatment density, $q=2$, $J^*$ must grow somewhere strictly in between rates of  $n^{1/4}$ and $n^{1/2}$ for valid inference. $J$ has no lower bound by construction as in Section \ref{sec_estimation1}. Thus, due to its fixed dimension, the decomposition can achieve faster rates of convergence compared to nonparametric estimation of the average structural function with a continuous treatment under comparable smoothness assumptions \cite{Colangelo2026DoubleTreatments}.

\section{Concluding Remarks} \label{sec_conclusion1}
This paper provides a comprehensive package how to interpret group heterogeneity when treatments are heterogeneous in general as well as how distinct drivers of heterogeneity can be estimated and tested for if effective treatment and confounders are observed in particular.

From a practical point of view, our results also serve as a cautionary tale regarding the policy relevance of conventional heterogeneity analysis when treatments are heterogeneous. In addition, collecting data on treatment versions as well as documentation and discussion of assignment mechanisms is likely to improve on policy relevance of the analysis. 

From a methodological perspective, the presented estimation approach is limited to setups where $J = o(n^{1/2})$. In setups with even more high-dimensional treatments, the method could be supplemented with additional, data-driven steps that first extract relevant treatment heterogeneity or versions, e.g.~based on maximizing within-aggregate heterogeneity. Interesting extensions could also lie in the adaption of the methodology to other target parameters such as quantile treatment effects and/or alternative research designs.

{\setstretch{1.12} 
\bibliographystyle{apacite}
\renewcommand{\APACrefYearMonthDay}[3]{\APACrefYear{#1}}
\bibliography{references}
}

\newpage
%---------------------------------Appendix
\begin{appendices}
\counterwithin{figure}{section}
\counterwithin{table}{section}

\huge \noindent \textbf{Supplemental Appendix}

%\large \noindent \textit{Not meant for publication but to be provided as supplementary material in the online repositories of the Journal and the homepages of the authors.}

\normalsize

\section{Supplementary Material for Sections \ref{sec_setting1} -- \ref{sec:decomp-diff-in-diffs}} \label{app:deriv-both}

\subsection{Derivation of Mean Decomposition} \label{app:deriv-rct-prof}

Under Assumption \ref{ass_CIA}, we can write

{\footnotesize \begin{align*}
    E[Y|T\in\mathcal{T}_a,X\in \xg] 
    % &= E\left[\sum_{t\in\mathcal{T}} \mathbbm{1}(T=t)Y(t)|T\in\mathcal{T}_a,X\in \xg\right] \\
    % &= E\left[\sum_{t\in\ta} \mathbbm{1}(T=t)Y(t)|T\in\mathcal{T}_a,X\in \xg\right] \\
    &= \sum_{t\in\ta} E[\mathbbm{1}(T=t)Y(t)|T\in\mathcal{T}_a,X\in \xg] \\
    % &= \sum_{t\in\ta} P(T=t | T\in\mathcal{T}_a,X\in \xg) E[Y(t)|T=t,T\in\mathcal{T}_a,X\in \xg] \\
    % &= \sum_{t\in\ta} P(T=t | T\in\mathcal{T}_a,X\in \xg) E[Y(t)|T=t,X\in \xg] \\
    &= \sum_{t\in\ta} P(T=t | T\in\mathcal{T}_a,X\in \xg) E[E[Y(t)|T=t,X]|T=t,X\in \xg] \\
    % &= \sum_{t\in\ta} P(T=t | T\in\mathcal{T}_a,X\in \xg) E[E[Y(t)|X]|T=t,X\in \xg] & (CIA) \\
    &= \sum_{t\in\ta} e_{ta}(\xg) E[\mu_t(X)|T=t,X\in \xg] & \\
    % &= \sum_{t\in\ta} e_{ta}(\xg) \frac{E[\mathbbm{1}(T=t) \mu_t(X) | X\in \xg]}{P(T=t|X\in \xg)} \\
    % &= \sum_{t\in\ta} e_{ta}(\xg) \frac{E[E[\mathbbm{1}(T=t) \mu_t(X) |X] | X\in \xg]}{P(T=t|X\in \xg)} \\
    &= \sum_{t\in\ta} e_{ta}(\xg) \frac{E[e_t(X) \mu_t(X) | X\in \xg]}{e_t(\xg)} \\
    % &= \sum_{t\in\ta} e_{ta}(\xg) \frac{E[e_t(X) | X\in \xg] E[\mu_t(X) | X\in \xg] + Cov(e_t(X),\mu_t(X) | X\in \xg)}{e_t(\xg)} \\
    % &= \sum_{t\in\ta} e_{ta}(\xg) E[\mu_t(X) | X\in \xg]  + e_{ta}(\xg) \frac{Cov(e_t(X),\mu_t(X) | X\in \xg)}{e_t(\xg)} \\
    &= \sum_{t\in\ta} e_{ta}(\xg) \mu_t(\xg)  + \frac{Cov(e_t(X),\mu_t(X) | X\in \xg)}{P(T\in\mathcal{T}_a | X\in \xg)}.
\end{align*}}
% because
% {\footnotesize \begin{align*}
%     e_{ta}(\xg) &=  P(T=t|X\in \xg,T\in\mathcal{T}_a) 
%     = \frac{P(T=t,X\in \xg,T\in\mathcal{T}_a)}{P(X\in \xg,T\in\mathcal{T}_a)} \\
%     & = \frac{P(T\in\mathcal{T}_a|T=t,X\in \xg)P(T=t|X\in \xg) P(X\in \xg)}{P(T\in\mathcal{T}_a|X\in \xg) P(X\in \xg)} \\
%     & = \frac{P(T\in\mathcal{T}_a|T=t,X\in \xg)P(T=t|X\in \xg)}{P(T\in\mathcal{T}_a|X\in \xg)} \\
%     & = \frac{P(t\in\mathcal{T}_a)P(T=t|X\in \xg)}{P(T\in\mathcal{T}_a|X\in \xg)} 
%     = \frac{\mathbbm{1}(t\in\mathcal{T}_a)e_t(\xg)}{P(T\in\mathcal{T}_a|X\in \xg)} \\
%     \Rightarrow & \quad \frac{e_{ta}(\xg)}{e_t(\xg)} = \frac{\mathbbm{1}(t\in\mathcal{T}_a)}{P(T\in\mathcal{T}_a|X\in \xg)}
% \end{align*}}
% and the sum is running over $t\in\ta$ such that $\mathbbm{1}(t\in\mathcal{T}_a) = 1$ in all summands.

\subsection{Derivation of Adjusted Mean Decomposition}  \label{app:deriv-acm}

The adjusted differences-in-means estimand explicitly relies on confounding variables $X$. Thus, $X$ must naturally be observed for the estimand to feasible. $T$ might still be unobserved.

\begin{assump}[Decomposition Assumptions ADiM] \label{ass_adim} In addition to Assumption \ref{ass_CIA}, assume that $X$ is observable, and $P(T \in \mathcal{T}_{a},X=x) > 0$ for at least some $x \in \xg$.
\end{assump}
Under Assumption \ref{ass_adim}, we can write
{\footnotesize \begin{align*}
    E&[E[Y|T\in\mathcal{T}_a,X]|X\in \xg] \\
    % &= E\left[E\left[\sum_{t\in\mathcal{T}} \mathbbm{1}(T=t)Y(t)|T\in\mathcal{T}_a,X\right] \middle|X\in \xg\right] \\
    % &= E\left[E\left[\sum_{t\in\ta} \mathbbm{1}(T=t)Y(t)|T\in\mathcal{T}_a,X\right] \middle|X\in \xg\right] \\
    &= \sum_{t\in\ta} E[E[ \mathbbm{1}(T=t)Y(t)|T\in\mathcal{T}_a,X] |X\in \xg] \\
    % &= \sum_{t\in\ta} E[P(T=t | T\in\mathcal{T}_a,X) E[Y(t)|T=t,T\in\mathcal{T}_a,X] |X\in \xg] \\
    % &= \sum_{t\in\ta} E[P(T=t | T\in\mathcal{T}_a,X) E[Y(t)|T=t,X] |X\in \xg] & \\
    % &= \sum_{t\in\ta} E[P(T=t | T\in\mathcal{T}_a,X) E[Y(t)|X] |X\in \xg] & (CIA) \\
    &= \sum_{t\in\ta} E[e_{ta}(X) \mu_{t}(X) |X\in \xg]\\
    % &= \sum_{t\in\ta} E[e_{ta}(X) |X\in \xg] E[\mu_{t}(X) |X\in \xg] + Cov[e_{ta}(X), \mu_{t}(X) |X\in \xg]\\
    &= \sum_{t\in\ta} E[e_{ta}(X) |X\in \xg] \mu_{t}(\xg) + Cov[e_{ta}(X), \mu_{t}(X) |X\in \xg]\\
    &= \sum_{t\in\ta} e_{ta}(\xg) \mu_{t}(\xg) + Cov[e_{ta}(X), \mu_{t}(X) |X\in \xg] + (E[e_{ta}(X) |X\in \xg] - e_{ta}(\xg)) \mu_{t}(\xg).
\end{align*}}

\subsection{Derivation of Adjusted Mean Decomposition with Covariate Adjustment for Aggregate Treatment Only}  \label{app:deriv-acm-xtilde}
Define $\tilde{X}$ as some subset of $X$ such that $\sum_{t\in\ta}\mathbbm{1}(T=t)Y(t) \bigCI \mathbbm{1}(T \in \ta)~|\tilde{X}$, i.e.~``binary unconfoundedness'' applies. Denote $\mu_{t,t}(\tilde{X}) := E[Y(t)|T=t,\tilde{X}]$ as the conditional average potential outcome of the treated. The decomposition for the adjusted component in Equation \eqref{eq:decomp-cm2} then changes to

{\footnotesize \begin{align*}
    E&[E[Y|T\in\mathcal{T}_a,\tilde{X}]|\tilde{X}\in \xg] \\
    % &= E\left[E\left[\sum_{t\in\ta} \mathbbm{1}(T=t)Y(t)|T\in\mathcal{T}_a,\tilde{X}\right] \middle|\tilde{X}\in \xg\right] \\
    % &= \sum_{t\in\ta} E[E[ \mathbbm{1}(T=t)Y(t)|T\in\mathcal{T}_a,\tilde{X}] |\tilde{X}\in \xg] \\
    % &= \sum_{t\in\ta} E[P(T=t | T\in\mathcal{T}_a,\tilde{X}) E[Y(t)|T=t,T\in\mathcal{T}_a,\tilde{X}] |\tilde{X}\in \xg] \\
    % &= \sum_{t\in\ta} E[P(T=t | T\in\mathcal{T}_a,\tilde{X}) E[Y(t)|T=t,\tilde{X}] |\tilde{X}\in \xg] & \\
    &= \sum_{t\in\ta} E[e_{ta}(\tilde{X}) \mu_{t,t}(\tilde{X}) |\tilde{X}\in \xg] \\
    % &= \sum_{t\in\ta} E[e_{ta}(\tilde{X}) |\tilde{X}\in \xg] E[\mu_{t,t}(\tilde{X}) |\tilde{X}\in \xg] + Cov[e_{ta}(\tilde{X}), \mu_{t,t}(\tilde{X}) |\tilde{X}\in \xg]\\
    &= \sum_{t\in\ta} e_{ta}(\xg) E[\mu_{t,t}(\tilde{X}) |\tilde{X}\in \xg] + Cov[e_{ta}(\tilde{X}), \mu_{t,t}(\tilde{X}) |\tilde{X}\in \xg] \\
    & \quad  + (E[e_{ta}(\tilde{X}) |\tilde{X}\in \xg] - e_{ta}(\xg)) E[\mu_{t,t}(\tilde{X}) |\tilde{X}\in \xg] \\
    &= \sum_{t\in\ta} e_{ta}(\xg) \mu_{t}(\xg) + Cov[e_{ta}(\tilde{X}), \mu_{t}(\tilde{X}) |\tilde{X}\in \xg] + (E[e_{ta}(\tilde{X}) |\tilde{X}\in \xg] - e_{ta}(\xg)) \mu_{t}(\xg) \\
    & \quad + e_{ta}(\xg) (E[\mu_{t,t}(\tilde{X}) |\tilde{X}\in \xg] - \mu_{t}(\xg)) \\
    & \quad  + Cov[e_{ta}(\tilde{X}), \mu_{t,t}(\tilde{X}) - \mu_{t}(\tilde{X}) |\tilde{X}\in \xg] \\
    & \quad + (E[e_{ta}(\tilde{X}) |\tilde{X}\in \xg] - e_{ta}(\xg)) (E[\mu_{t,t}(\tilde{X}) |\tilde{X}\in \xg] - \mu_{t}(\xg)).    
\end{align*}}
% where we could replace 
% $$\mu_{t,t}(\tilde{X}) - \mu_{t}(\tilde{X}) = Cov\left(\frac{e_t(X)}{e_t(\tilde{X})},\mu_t(X) | \tilde{X} \right)$$ 

The first line resembles the decomposition in main text with $X$ replaced by $\tilde{X}$. However, three additional terms appear because usually $\mu_{t,t}(\tilde{X}) \neq \mu_{t}(\tilde{X})$, i.e.~there is a form of conditional selection bias on the level of the effective treatment. Thus, these three terms can be interpreted as (i) group targeting based on group selection bias (ii) individualized target based on individual selection bias given group membership, and (iii) interaction between composition adjustment for aggregate treatment balance and group selection bias.

\subsection{Decomposition Details} \label{app:decompstn}

\subsubsection{All Decomposition Parameters} \label{app:params-tests}

Here we spell out the $\delta$ and $\Delta$ decomposition parameters. Occasionally they are rearranged in terms of effects to indicate how they implicitly test forms of effect heterogeneity:

% {\footnotesize
% {\footnotesize \begin{align*}
% d_{0}(a) & = \sum_{t\in\ta} e_{ta} \mu_t \\
% d_{1}(a,g) & = \sum_{t\in\ta} e_{ta} [\mu_t(\xg) - \mu_t] & (OH)\\
% d_{2}(a,g) & = \sum_{t\in\ta} [e_{ta}(\xg) - e_{ta}] \mu_t & (TH1\&TH2) \\
% d_{3}(a,g) & = \sum_{t\in\ta} [e_{ta}(\xg) - e_{ta}] [\mu_t(\xg) - \mu_t] & (TH1\&TH2\& TH3\&OH) \\
% d_{4}(a,g) & = \sum_{t\in\ta} \frac{Cov(e_t(X),\mu_t(X) | X\in \xg)}{P(T\in\mathcal{T}_a | X\in \xg)}  & (TH1\&TH-G\& OH) \\
% d_{4'}(a,g) & = \sum_{t\in\ta} Cov[e_{ta}(X), \mu_{t}(X) |X\in \xg]   & (TH-G\& OH)\\
% d_{5}(a,g) & = \sum_{t\in\ta} [E[e_{ta}(X) |X\in \xg] - e_{ta}(\xg)] \mu_{t}(\xg)  & (TH) 
% \end{align*}}
% }

 {\scriptsize \begin{align*}
\delta_{0}(a,a') & = \sum_{t\in\ta} e_{ta} \mu_t - \sum_{t'\in\mathcal{T}_{a'}} e_{t'a'} \mu_{t'} =  \sum_{t\in\ta} \sum_{t'\in\mathcal{T}_{a'}} e_{ta} e_{t'a'}  \tau_{t,t'} \\
\delta_{1}(a,a',g) & = \sum_{t\in\ta} e_{ta} [\mu_t(\xg) - \mu_t] - \sum_{t'\in\mathcal{T}_{a'}} e_{t'a'} [\mu_{t'}(\xg) - \mu_{t'}]  = \sum_{t\in\ta} \sum_{t'\in\mathcal{T}_{a'}} e_{ta} e_{t'a'} [\tau_{t,t'}(\xg) - \tau_{t,t'}] \\
% d_{1}(a,g) - d_{1}(a,g') & = \sum_{t\in\ta} e_{ta} [\mu_t(\xg) - \mu_t] - e_{ta} [\mu_t(\mathcal{X}_{g'}) - \mu_t]  & (OH) \\
% & = \sum_{t\in\ta} e_{ta} [\mu_t(\xg) - \mu_t(\mathcal{X}_{g'})] \\
\delta_{2}(a,a',g) & = \sum_{t\in\ta} [e_{ta}(\xg) - e_{ta}] \mu_t - \sum_{t'\in\mathcal{T}_{a'}} [e_{t'a'}(\xg) - e_{t'a'}] \mu_{t'}  
 % & = \sum_{t\in\ta} e_{ta}(\xg)\mu_t - \sum_{t\in\ta} e_{ta} \mu_t \\
 % & - \sum_{t'\in\mathcal{T}_{a'}} e_{t'a'}(\xg) \mu_{t'} + \sum_{t'\in\mathcal{T}_{a'}} e_{t'a'} \mu_{t'} \\ 
 % & = \sum_{t'\in\mathcal{T}_{a'}} e_{t'a'}(\xg) \sum_{t\in\ta} e_{ta}(\xg)\mu_t - \sum_{t'\in\mathcal{T}_{a'}} e_{t'a'}\sum_{t\in\ta} e_{ta} \mu_t \\
 % & - \sum_{t\in\ta} e_{ta}(\xg) \sum_{t'\in\mathcal{T}_{a'}} e_{t'a'}(\xg) \mu_{t'} + \sum_{t\in\ta} e_{ta} \sum_{t'\in\mathcal{T}_{a'}} e_{t'a'} \mu_{t'} \\ 
 % & = \sum_{t'\in\mathcal{T}_{a'}} \sum_{t\in\ta} e_{ta}(\xg)  e_{t'a'} (\xg) \mu_t - \sum_{t'\in\mathcal{T}_{a'}} \sum_{t\in\ta} e_{ta} e_{t'a'} \mu_t \\
 % & - \sum_{t\in\ta} \sum_{t'\in\mathcal{T}_{a'}} e_{ta}(\xg)  e_{t'a'}(\xg) \mu_{t'} + \sum_{t\in\ta} \sum_{t'\in\mathcal{T}_{a'}}  e_{ta} e_{t'a'} \mu_{t'} \\ 
 % & = \sum_{t'\in\mathcal{T}_{a'}} \sum_{t\in\ta} [e_{ta}(\xg)  e_{t'a'}(\xg) - e_{ta} e_{t'a'}] \mu_t \\
 % & - \sum_{t\in\ta} \sum_{t'\in\mathcal{T}_{a'}} [e_{ta}(\xg)  e_{t'a'}(\xg) - e_{ta} e_{t'a'}]  \mu_{t'} \\ 
  = \sum_{t\in\ta} \sum_{t'\in\mathcal{T}_{a'}} [e_{ta}(\xg)  e_{t'a'}(\xg) - e_{ta} e_{t'a'}] \tau_{t,t'} \\
% d_{2}(a,g) - d_{2}(a,g') & = \sum_{t\in\ta} [e_{ta}(\xg) - e_{ta}] \mu_t - \sum_{t\in\ta} [e_{ta}(\mathcal{X}_{g'}) - e_{ta}] \mu_t  & (TH) \\
%  & = \sum_{t\in\ta} [e_{ta}(\xg) - e_{ta}(\mathcal{X}_{g'})] \mu_t \\
\delta_{3}(a,a',g) & = \sum_{t\in\ta} [e_{ta}(\xg) - e_{ta}] [\mu_t(\xg) - \mu_t] - \sum_{t'\in\mathcal{T}_{a'}} [e_{t'a'}(\xg) - e_{t'a'}] [\mu_{t'}(\xg) - \mu_{t'}] \\
& = \sum_{t\in\ta} \sum_{t'\in\mathcal{T}_{a'}} [e_{ta}(\xg)  e_{t'a'}(\xg) - e_{ta} e_{t'a'}] [\tau_{t,t'}(\xg) - \tau_{t,t'}] \\
% d_{3}(a,g) - d_{3}(a,g') & = \sum_{t\in\ta} [e_{ta}(\xg) - e_{ta}] [\mu_t(\xg) - \mu_t] - [e_{ta}(\mathcal{X}_{g'}) - e_{ta}] [\mu_t(\mathcal{X}_{g'}) - \mu_t]  & (TH\& OH) 
% & = \sum_{t\in\ta} e_{ta}(\xg) \mu_t(\xg) - e_{ta}(\xg) \mu_t - e_{ta} \mu_t(\xg) + e_{ta} \mu_t \\
% & - e_{ta}(\mathcal{X}_{g'}) \mu_t(\mathcal{X}_{g'}) + e_{ta}(\mathcal{X}_{g'}) \mu_t + e_{ta} \mu_t(\mathcal{X}_{g'}) - e_{ta} \mu_t \\
% & = \sum_{t\in\ta} e_{ta}(\xg) \mu_t(\xg) - e_{ta}(\mathcal{X}_{g'}) \mu_t(\mathcal{X}_{g'})  \\
% & - [e_{ta}(\xg) - e_{ta}(\mathcal{X}_{g'})] \mu_t - e_{ta} [ \mu_t(\xg) - \mu_t(\mathcal{X}_{g'}) ]  \\
\delta_{4}(a,a',g) & = \sum_{t\in\ta} \frac{Cov(e_t(X),\mu_t(X) | X\in \xg)}{P(T\in\mathcal{T}_a | X\in \xg)} - \sum_{t'\in\tap} \frac{Cov(e_{t'}(X),\mu_{t'}(X) | X\in \xg)}{P(T\in\tap | X\in \xg)}  \\
\delta_{4'}(a,a',g) & = \sum_{t\in\ta} Cov[e_{ta}(X), \mu_{t}(X) |X\in \xg] - \sum_{t'\in\tap} Cov[e_{t'a'}(X), \mu_{t'}(X) |X\in \xg]     \\
%  & = \sum_{t\in\ta} \sum_{t'\in\mathcal{T}_{a'}} Cov[e_{ta}(X) e_{t'a'}(X), \tau_{t,t'}(X) |X\in \xg]
\delta_{5}(a,a',g) & = \sum_{t\in\ta} [E[e_{ta}(X) |X\in \xg] - e_{ta}(\xg)] \mu_{t}(\xg) - \sum_{t'\in\tap} [E[e_{t'a'}(X) |X\in \xg] - e_{t'a'}(\xg)] \mu_{t'}(\xg) \\
& = \sum_{t\in\ta} \sum_{t'\in\tap} [E[e_{ta}(X) |X\in \xg] - e_{ta}(\xg)] [E[e_{t'a'}(X) |X\in \xg] - e_{t'a'}(\xg)] \tau_{t,t'}(\xg) \\
\Delta_{1}(a,a',g,g') & = \sum_{t\in\ta} \sum_{t'\in\mathcal{T}_{a'}} e_{ta} e_{t'a'} [\tau_{t,t'}(\xg) - \tau_{t,t'}(\mathcal{X}_{g'})] \\
\Delta_{2}(a,a',g,g') & = \sum_{t\in\ta} [e_{ta}(\xg) - e_{ta}(\mathcal{X}_{g'})] \mu_{t} - \sum_{t'\in\mathcal{T}_{a'}} [e_{t'a'}(\xg) - e_{t'a'}(\mathcal{X}_{g'})] \mu_{t'} \\
% &  = \sum_{t\in\ta} e_{ta}(\xg)  \mu_{t} - \sum_{t\in\ta} e_{ta}(\mathcal{X}_{g'}) \mu_{t}  \\
% & - \sum_{t'\in\mathcal{T}_{a'}} e_{t'a'}(\xg)  \mu_{t'} + \sum_{t'\in\mathcal{T}_{a'}} e_{t'a'}(\mathcal{X}_{g'}) \mu_{t'} \\
% &  = \sum_{t'\in\mathcal{T}_{a'}} e_{t'a'}(\xg) \sum_{t\in\ta} e_{ta}(\xg)  \mu_{t} - \sum_{t'\in\mathcal{T}_{a'}} e_{t'a'}(\mathcal{X}_{g'}) \sum_{t\in\ta} e_{ta}(\mathcal{X}_{g'}) \mu_{t}  \\
% & -  \sum_{t\in\ta} e_{ta}(\xg) \sum_{t'\in\mathcal{T}_{a'}} e_{t'a'}(\xg)  \mu_{t'} + \sum_{t\in\ta} e_{ta}(\mathcal{X}_{g'}) \sum_{t'\in\mathcal{T}_{a'}} e_{t'a'}(\mathcal{X}_{g'}) \mu_{t'} \\
% &  = \sum_{t'\in\mathcal{T}_{a'}} \sum_{t\in\ta} e_{ta}(\xg) e_{t'a'}(\xg) \mu_{t} - \sum_{t'\in\mathcal{T}_{a'}}  \sum_{t\in\ta} e_{ta}(\mathcal{X}_{g'}) e_{t'a'}(\mathcal{X}_{g'}) \mu_{t}  \\
% & -  \sum_{t\in\ta} \sum_{t'\in\mathcal{T}_{a'}}  e_{ta}(\xg) e_{t'a'}(\xg)  \mu_{t'} + \sum_{t\in\ta}  \sum_{t'\in\mathcal{T}_{a'}} e_{ta}(\mathcal{X}_{g'}) e_{t'a'}(\mathcal{X}_{g'}) \mu_{t'} \\
% &  = \sum_{t'\in\mathcal{T}_{a'}} \sum_{t\in\ta} e_{ta}(\xg) e_{t'a'}(\xg) (\mu_{t} - \mu_{t'})  - \sum_{t'\in\mathcal{T}_{a'}}  \sum_{t\in\ta} e_{ta}(\mathcal{X}_{g'}) e_{t'a'}(\mathcal{X}_{g'})(\mu_{t} - \mu_{t'})  \\
% &  = \sum_{t'\in\mathcal{T}_{a'}} \sum_{t\in\ta} [e_{ta}(\xg) e_{t'a'}(\xg) - e_{ta}(\mathcal{X}_{g'}) e_{t'a'}(\mathcal{X}_{g'})] (\mu_{t} - \mu_{t'}) \\
&  = \sum_{t\in\ta} \sum_{t'\in\mathcal{T}_{a'}} [e_{ta}(\xg) e_{t'a'}(\xg) - e_{ta}(\mathcal{X}_{g'}) e_{t'a'}(\mathcal{X}_{g'})] \tau_{t,t'} \\
\Delta_{3}(a,a',g,g') & = \sum_{t\in\ta} [e_{ta}(\xg) - e_{ta}] [\mu_t(\xg) - \mu_t] - [e_{ta}(\mathcal{X}_{g'}) - e_{ta}] [\mu_t(\mathcal{X}_{g'}) - \mu_t] \\
& -  \sum_{t'\in\mathcal{T}_{a'}} [e_{t'a}(\xg) - e_{t'a}] [\mu_{t'}(\xg) - \mu_{t'}] - [e_{t'a}(\mathcal{X}_{g'}) - e_{t'a}] [\mu_{t'}(\mathcal{X}_{g'}) - \mu_{t'}] \\
& = \sum_{t\in\ta} \sum_{t'\in\mathcal{T}_{a'}} [e_{ta}(\xg)  e_{t'a'}(\xg) - e_{ta} e_{t'a'}] [\tau_{t,t'}(\xg) - \tau_{t,t'}] \\
& \quad - [e_{ta}(\mathcal{X}_{g'})  e_{t'a'}(\mathcal{X}_{g'}) - e_{ta} e_{t'a'}] [\tau_{t,t'}(\mathcal{X}_{g'}) - \tau_{t,t'}] \\
\Delta_{4}(a,a',g,g') & = \sum_{t\in\ta} \frac{Cov(e_t(X),\mu_t(X) | X\in \xg)}{P(T\in\mathcal{T}_a | X\in \xg)} - \sum_{t'\in\tap} \frac{Cov(e_{t'}(X),\mu_{t'}(X) | X\in \xg)}{P(T\in\tap | X\in \xg)}  \\
& - \sum_{t\in\ta} \frac{Cov(e_t(X),\mu_t(X) | X\in \xgp)}{P(T\in\mathcal{T}_a | X\in \xgp)} + \sum_{t'\in\tap} \frac{Cov(e_{t'}(X),\mu_{t'}(X) | X\in \xgp)}{P(T\in\tap | X\in \xgp)} \\
\Delta_{4'}(a,a',g,g') & = \sum_{t\in\ta} Cov[e_{ta}(X), \mu_{t}(X) |X\in \xg] - \sum_{t'\in\tap} Cov[e_{t'a'}(X), \mu_{t'}(X) |X\in \xg]  \\
& - \sum_{t\in\ta} Cov[e_{ta}(X), \mu_{t}(X) |X\in \xgp] + \sum_{t'\in\tap} Cov[e_{t'a'}(X), \mu_{t'}(X) |X\in \xgp]  \\
%  & = \sum_{t\in\ta} \sum_{t'\in\mathcal{T}_{a'}} Cov[e_{ta}(X) e_{t'a'}(X), \tau_{t,t'}(X) |X\in \xg]
\Delta_{5}(a,a',g,g') & = \sum_{t\in\ta} \sum_{t'\in\tap} [E[e_{ta}(X) |X\in \xg] - e_{ta}(\xg)] [E[e_{t'a'}(X) |X\in \xg] - e_{t'a'}(\xg)] \tau_{t,t'}(\xg) \\
& - \sum_{t\in\ta} \sum_{t'\in\tap} [E[e_{ta}(X) |X\in \xgp] - e_{ta}(\xgp)] [E[e_{t'a'}(X) |X\in \xgp] - e_{t'a'}(\xgp)] \tau_{t,t'}(\xgp)
\end{align*}}

\subsubsection{Implied Hypothesis Regarding Treatment Heterogeneity} \label{app:ih-th}

This appendix extends the discussion on implied hypotheses in Section \ref{sec_indirect1} to \textit{testing treatment heterogeneity} on the group level. In the spirit of strong group effect homogeneity in Equation \eqref{eq_strongeffecthomogeneity}, we can articulate 
% \item \textit{strong group outcome homogeneity} (OH)
% {\footnotesize \begin{align}
%         H_0:~& \mu_{t}(\xg) = \mu_{t}(\xgp) \text{ for all } t \in \ta \label{eq:h0-oh} \\ 
%         %\text{ vs. } \\
%         H_A:~& \mu_{t}(\xg) \neq \mu_{t}(\xgp) \text{ for some } t \in \ta \nonumber
% \end{align}}
\begin{itemize}[left=0pt]
    \item \textit{group treatment assignment homogeneity}:
     {\footnotesize \begin{align}
        H_0:~& e_{ta}(\xg) = e_{ta}(\xgp) \text{ for all } t \in \ta \text{, and } e_{t'a'}(\xg) = e_{t'a'}(\xgp) \text{ for all } t' \in \tap, \label{eq:h0-tah} \\
        H_A:~& e_{ta}(\xg) \neq e_{ta}(\xgp) \text{ for some } t \in \ta \text{, and/or } e_{t'a'}(\xg) \neq e_{t'a'}(\xgp) \text{ for some } t' \in \tap, 
    \end{align}}
    % \item with either \textit{average treatment version irrelevance} (TH2):
    %  {\footnotesize \begin{align}
    %     H_0:~& \mu_t = \mu_{t^\dagger} \text{ for all }t, t^\dagger \in \mathcal{T}_a \text{ and } \mu_t = \mu_{t^\dagger} \text{ for all } t, t^\dagger \in \mathcal{T}_{a'} \label{eq:h0-tvia} \\
    %     H_A:~& \mu_t \neq \mu_{t^\dagger} \text{ for some } t, t^\dagger \in \mathcal{T}_a \text{, and/or } \mu_t \neq \mu_{t^\dagger} \text{ for some } t, t^\dagger \in \mathcal{T}_{a'} \nonumber
    % \end{align}}
    \item and \textit{group treatment version irrelevance}:
     {\footnotesize \begin{align}
        H_0:~& \mu_t(\xg) = \mu_{t^\dagger}(\xg) \text{ for all }t, t^\dagger \in \mathcal{T}_a \text{ and } \mu_t(\xg) = \mu_{t^\dagger}(\xg) \text{ for all } t, t^\dagger \in \mathcal{T}_{a'}, \label{eq:h0-tvia} \\
        H_A:~& \mu_t(\xg) \neq \mu_{t^\dagger}(\xg) \text{ for some } t, t^\dagger \in \mathcal{T}_a \text{, and/or } \mu_t(\xg) \neq \mu_{t^\dagger}(\xg) \text{ for some } t, t^\dagger \in \mathcal{T}_{a'}. 
    \end{align}}
\end{itemize}
% \end{itemize}
One way to test these hypotheses jointly is to consider parameter combination $d_2 + d_3$
{\footnotesize \begin{align*}
d_{2}(a,g) + d_{3}(a,g) & = \sum_{t\in\ta} [e_{ta}(\xg) - e_{ta}] \mu_t + [e_{ta}(\xg) - e_{ta}] [\mu_t(\xg) - \mu_t] \\
 % & = \sum_{t\in\ta} [e_{ta}(\xg) - e_{ta}] \mu_t(\xg) \\
 & = \sum_{t\in\ta} [e_{ta}(\xg) - e_{ta}] [\mu_t(\xg) - \mu_a(\xg)],
\end{align*}}
where $\mu_a(\xg) = \sum_{t\in\ta} e_{ta} \mu_t(\xg)$. This means that rejecting $d_2 + d_3 = 0$ is sufficient to reject both \eqref{eq:h0-tah} and \eqref{eq:h0-tvia} and therefore treatment homogeneity. Similarly $\delta_2 + \delta_3$ and $\Delta_2 + \Delta_3$ can serve as tests of group treatment heterogeneity. Many other parameter (combinations) imply average or individualized treatment homogeneity tests. However, they are usually not of primary interest and therefore not explicitly discussed here.

\subsection{Alternative Centering} \label{sec_centering1}
% $e_a$ to $e_a(\xg \cup \mathcal{X}_{g'})$ and $\mu$ to $\mu(\xg \cup \mathcal{X}_{g'})$ or same with $\bigcup_{g\in\mathcal{G}}\xg$. Analogue to BO. Same implied hypothesis as above. However, scale different. power against other alternatives. [const effect for g1 g2, but not necces medium that is not analyzed e.g. if we have g1,g2,g3]
The decomposition of the synthetic estimand \eqref{eq:decomp-srct} and its resulting differences are centered around unconditional within aggregate propensities $e_{ta}$ and potential outcomes $\mu_t$. These allowed us to decompose and contrast heterogeneity estimands relative to the full population. For instance, $d_0$ would be a synthetic estimand corresponding to constant effects and propensities in the whole population. When assessing heterogeneity between groups $g$ and $g'$, we might instead want to build contrasts relative to a scenario where there is homogeneity across exactly the groups we compare. 
Consider two groups characterized by $\mathcal{X}_g$ and $\mathcal{X}_{g'}$. We could center the decomposition of the stratified RCT in \eqref{eq:decomp-srct} only around the respective averages over these groups, $e_{a}(\xg \cup \xgp)$ and $\mu(\xg \cup \xgp)$, i.e.

 {\footnotesize \begin{align}
     \sum_{t\in\ta} e_{ta}(\xg) \mu_t(\xg) 
     &=\sum_{t\in\ta} \underbrace{e_{ta}(\xg \cup \xgp) \mu_t(\xg \cup \xgp)}_{\text{baseline}} + \underbrace{e_{ta}(\xg \cup \xgp) [\mu_t(\xg) - \mu_t(\xg \cup \xgp)]}_{\text{outcome heterogeneity}} \notag \\  &\quad + \underbrace{[e_{ta}(\xg) - e_{ta}(\xg \cup \xgp)] \mu_t(\xg \cup \xgp)}_{\text{group targeting of average outcomes}} \notag \\ &\quad  + \underbrace{[e_{ta}(\xg) - e_{ta}(\xg \cup \xgp)] [\mu_t(\xg) - \mu_t(\xg \cup \xgp)]}_{\text{group targeting of group outcomes}},
 \end{align}}
 
where now heterogeneity and targeting are defined relative to the composite group $\xg \cup \mathcal{X}_{g'}$. When these groups form a complete partition, $\xg \cup \mathcal{X}_{g'} = \mathcal{X}$, this collapses back to the original decomposition \eqref{eq:decomp-srct}.
From a definition perspective centering the decompositions around other quantities is also possible. For estimation and inference, however, the centering quantities need to behave qualitatively similar to the conditional means and propensities, see Section \ref{sec_estimation1}. Analogous to the discussion in previous sections, these alternative decompositions will have a different economic interpretation and power against alternatives that match their respective weighting schemes.

\section{Proof of Theorem \ref{thm_asyN1}}
\label{sec_app_inference}
\subsection{Preliminaries and Definitions}
For generic $A$ and $B$, we often make use of the following decomposition {\footnotesize \begin{align*}
  \hat{A}\hat{B} -  AB = (\hat{A} - A)B + A(\hat{B} - B) + (\hat{A} - A)(\hat{B} - B).
\end{align*}}
For random variable $A$ and sample $i=1,\dots,n$, we denote  $E_n[A] = \frac{1}{n}\sum_i^nA_i$ and  $G_n[A] = \frac{1}{\sqrt{n}}\sum_{i}^n(A_i - E[A_i])$.
% {\footnotesize \begin{align*}
%     E_n[A] &= \frac{1}{n}\sum_i^nA_i, \\
%     G_n[A] &= \frac{1}{\sqrt{n}}\sum_{i}^n(A_i - E[A_i]).
% \end{align*}}
% Moreover, let $\1_t = \1(T=t)$ and {\footnotesize \begin{align*}
%     r_{e,n} = \sup_{\hat{e}\in \mathcal{E}_n,x\in \mathcal{X}}\frac{e_t(x)}{\hat{e}_t(x)}
% \end{align*}}
%All sums and suprema that run or are defined for entries in $\mathcal{T}_a$ only, e.g.~$\sup_t e_t(x) = \sup_{t\in\mathcal{T}_a} e_t(x)$ and $\sum_{t\neq s}\mu_t = \sum_{t\in\mathcal{T}_a:t\neq s}\mu_t$. 
We denote the efficient influence function for a average potential outcome $E[Y(t)] = \mu_t$ as {\footnotesize \begin{align*}
    \psi_t(\eta) &= \frac{\mathbbm{1}(T=t)(Y-\mu_t(X))}{e_t(X)} + \mu_t(X) - \mu_t 
    = \frac{\mathbbm{1}_t\varepsilon(t)}{e_t(X)} + \mu_t(X) - \mu_t, 
\end{align*}}
where $\1_t = \1(T=t)$ and $\varepsilon(t) = Y(t)-\mu_t(X)$. We omit the true nuisance function argument $\eta$ whenever irrelevant. Without loss of generality, we limit our attention to sequences where $\sup_{t\in\ta}Je_t \lesssim 1$. The extension to the complete case with lower bounded propensities as in Assumption A.4 is trivial as, by construction, there can only be a finite number of treatments appearing at this rate. We also define the sequence space $
     \ell^2(\mathcal{T}_a) = \{ g : \sum_{t\in\ta}|g(t)|^2 < \infty \} 
$
and inner product relative to counting measure $
    \ip{g_1(x),g_2(x)}_{\ell^2(\mathcal{T}_a)} = \sum_{t\in\ta} g_1(t,x)g_2(t,x)$,
where $g_1,g_2$ are generic functions (possibly indexed by some $x \subseteq \mathcal{X}$).

\subsection{IF-based Estimator and Expansion} \label{sec_proof_ifbasedestimator}
Let $\IF(\theta) = \IF(\theta,W,\eta)$ be the (vector of) centered orthogonalized moments or ``influence curves'' \cite{Kennedy2024SemiparametricReview} for any parameter $\theta$ using data $W = (Y,T,X)$ and nuisance functions $\eta \in \mathcal{H}$. Denote $\hat{\IF}(\theta) = \IF(\theta,W,\hat{\eta})$ where $\hat{\eta} \in \mathcal{H}_n$ are the cross-fitted nuisance parameters. 
For the proof, we first consider the parameter $d_0(a) = {d}_0(a)(e_a,\mu) = \ip{e_a,\mu}_{\ell^2(\ta)}$. The extension to all other decomposition parameters is then given in subsection \ref{app_proof_N_decompALL}.
% We first derive the asymptotic theory for the single parameter ${d}_0(a) = \ip{{e}_a,{\mu}}_{\ell^2(\ta)} = \sum_{t\in\mathcal{T}_a}e_{ta}\mu_t$. The extension to the other decomposition parameters and their joint convergence follows. 
The estimated decomposition parameter is given by $
    \hat{d}_0(a) = \ip{\hat{e}_a,\hat{\mu}}_{\ell^2(\ta)} $
The influence function estimators take the plug-in estimators using the empirical analogue for each component
$E_n[\hat{\IF}(\hat{\mu})] = 0$ and $E_n[\hat{\IF}(\hat{e}_a)] = 0$. 
We obtain the following expansion 

{\footnotesize \begin{align*}
    \hat{d}_0(a) - \ip{{e}_a,{\mu}}_{\ell^2(\ta)}
    &= \ip{\hat{e}_a - e_a,{\mu}}_{\ell^2(\ta)} + \ip{{e}_a,\hat{\mu} - \mu}_{\ell^2(\ta)} + \ip{(\hat{e}_a - e_a),(\hat{\mu} - \mu)}_{\ell^2(\ta)}, 
\end{align*}}
which yields {\footnotesize \begin{align*}
    \sqrt{n}(\hat{d}_0(a) - {d}_0(a))
    &= G_n\left[\hat{\IF}({d}_0(a))\right]\left( 1 + O_p\left(E_n[\mathbbm{1}(T\in\mathcal{T}_a)] - P(T\in\mathcal{T}_a)\right) \right)\\ &\quad + \sqrt{n}\ip{(\hat{e}_a - e_a),(\hat{\mu} - \mu)}_{\ell^2(\ta)}. 
\end{align*}}
The first term on the right hand side can be further expanded as {\footnotesize \begin{align*}
    G_n\left[\hat{\IF}({d}_0(a))\right]
    &= \underbrace{G_n\left[{\IF}({d}_0(a))\right]}_{\text{Leading Term}} + \underbrace{\left(G_n\left[\hat{\IF}({d}_0(a))\right] - G_n\left[{\IF}({d}_0(a))\right]\right)}_{\text{Empirical Process}} \\
    &\quad + \underbrace{\sqrt{n}\left(E\left[\hat{\IF}({d}_0(a))\right] - E\left[{\IF}({d}_0(a))\right] \right)}_{\text{Machine Learning Bias}}. 
\end{align*}}

\subsection{Bounding the Machine Learning Bias} \label{sec_proof_mlbias1}
%We now verify the conditions from \citeA{Chernozhukov2018}, Assumption 3.2(iii) 
We now bound the machine learning bias for the given influence function 
$ {\IF}({d}_0(a)) = \sum_{t\in\mathcal{T}_a}\psi_t(\eta)e_t$.
Whenever it does not cause confusion, we keep the conditioning on the cross-fitted folds implicit in what follows, i.e.~we treat $\hat{\eta}(X)$ as depending on $X$ only through the evaluation at $X$ itself and not through the nuisance model.
Let $r \in [0,1]$. First note that, for any $\hat{\eta} = \eta + r(\tilde{\eta} - \eta)$ we can decompose 

{\footnotesize \begin{align*}
    \psi_t(\hat{\eta}) - \psi_t(\eta)
    &= \frac{\mathbbm{1}_t(Y - \mu_t(X) - r(\tilde{\mu}_t(X) - \mu_t(X))}{e_t(X) + r(\tilde{e}_t(X) - e_t(X))} + \mu_t(X) + r(\tilde{\mu}_t(X) - \mu_t(X)) - \mu_t \\
    &\quad - \frac{\mathbbm{1}_t(Y - \mu_t(X))}{e_t(X)} - \mu_t(X) +  \mu_t \\
    % &= \frac{\mathbbm{1}_t(\varepsilon(t) - r(\tilde{\mu}_t(X) - \mu_t(X))e_t(X)}{(e_t(X) + r(\tilde{e}_t(X) - e_t(X)))e_t(X)} + \frac{r(\tilde{\mu}_t(X) - \mu_t(X))(e_t(X) + r(\tilde{e}_t(X) - e_t(X)))e_t(X)}{(e_t(X) + r(\tilde{e}_t(X) - e_t(X)))e_t(X)} \\
    % &\quad - \frac{\mathbbm{1}_t \varepsilon(t)(e_t(X) + r(\tilde{e}_t(X) - e_t(X)))}{(e_t(X) + r(\tilde{e}_t(X) - e_t(X)))e_t(X)} \\
    % &= \frac{\mathbbm{1}_t \varepsilon(t) e_t(X) - \mathbbm{1}_t \varepsilon(t)(e_t(X) + r(\tilde{e}_t(X) - e_t(X)))}{(e_t(X) + r(\tilde{e}_t(X) - e_t(X)))e_t(X)} - \frac{\mathbbm{1}_t r(\tilde{\mu}_t(X) - \mu_t(X)e_t(X)}{(e_t(X) + r(\tilde{e}_t(X) - e_t(X)))e_t(X)} \\
    % & \quad + \frac{r(\tilde{\mu}_t(X) - \mu_t(X))(e_t(X) + r(\tilde{e}_t(X) - e_t(X)))e_t(X)}{(e_t(X) + r(\tilde{e}_t(X) - e_t(X)))e_t(X)} \\
    % &= \mathbbm{1}_t \varepsilon(t) \frac{ e_t(X) - e_t(X) - r(\tilde{e}_t(X) - e_t(X))}{(e_t(X) + r(\tilde{e}_t(X) - e_t(X)))e_t(X)} - \frac{\mathbbm{1}_t r(\tilde{\mu}_t(X) - \mu_t(X)}{(e_t(X) + r(\tilde{e}_t(X) - e_t(X)))} \\
    % & \quad + \frac{r(\tilde{\mu}_t(X) - \mu_t(X)) e_t(X)}{(e_t(X) + r(\tilde{e}_t(X) - e_t(X)))} \\
    % & \quad + \frac{r(\tilde{\mu}_t(X) - \mu_t(X)) r(\tilde{e}_t(X) - e_t(X))e_t(X)}{(e_t(X) + r(\tilde{e}_t(X) - e_t(X)))e_t(X)} \\
    % &= \mathbbm{1}_t \varepsilon(t) \frac{- r(\tilde{e}_t(X) - e_t(X))}{(e_t(X) + r(\tilde{e}_t(X) - e_t(X)))e_t(X)} \\
    % & \quad + r \frac{(\tilde{\mu}_t(X) - \mu_t(X)) (e_t(X) - \mathbbm{1}_t)}{e_t(X) + r(\tilde{e}_t(X) - e_t(X))} \\
    % & \quad + r^2\frac{(\tilde{\mu}_t(X) - \mu_t(X)) (\tilde{e}_t(X) - e_t(X))}{(e_t(X) + r(\tilde{e}_t(X) - e_t(X)))} \\
    &= -\mathbbm{1}_t\varepsilon(t) r\frac{(\tilde{e}_t(X) - e(X))}{(e_t(X) + r(\tilde{e}_t(X) - e(X)))e_t(X)} - r \frac{(\tilde{\mu}_t(X) - \mu_t(X))(\mathbbm{1}_t - e_t(X))}{e_t(X) +  r(\tilde{e}_t(X) - e_t(X))} \\
    &\quad + r^2 \frac{(\tilde{\mu}_t(X) - \mu_t(X))(\tilde{e}_t(X) - e_t(X))}{e(X) +  r(\tilde{e}_t(X) - e_t(X))},
\end{align*}}
for all $t$. Plugging this back into the machine learning bias and taking expectations then yields 

{\footnotesize \begin{align*}
    E[{\IF}({d}_0(a)) - \hat{\IF}({d}_0(a))]
    &= r^2 \sum_{t\in \mathcal{T}_a}E\left[ \frac{(\tilde{\mu}_t(X) - \mu_t(X))(\tilde{e}_t(X) - e_t(X))}{e(X) +  r(\tilde{e}_t(X) - e(X))}e_t\right] \\
    &= r^2 \sum_{t\in \mathcal{T}_a}E\left[ \frac{(\tilde{\mu}_t(X) - \mu_t(X))(\tilde{e}_t(X) - e_t(X))}{1 +  r(\tilde{e}_t(X)/e_t(X) - 1)}\frac{e_t}{e_t(X)}\right],
\end{align*}}

as $E[\varepsilon(t)|T=t,X] = E[\varepsilon(t)|X] = 0$ by Assumption 1 as well as $E[\mathbbm{1}_t - e_t(X)|X] = 0$ by definition. Using Assumption A.4 we then obtain
{\footnotesize \begin{align*}
    \sqrt{n}\sup_{\hat{\eta} \in \mathcal{H}_n}&||E[{\IF}({d}_0(a)) - \hat{\IF}({d}_0(a))]|| \\
    &= \sqrt{n}\sup_{r\in [0,1]}r^2 {\bigg|\bigg|}\sum_{t\in \mathcal{T}_a}E\left[ \frac{(\tilde{\mu}_t(X) - \mu_t(X))(\tilde{e}_t(X) - e_t(X))}{1 +  r(\tilde{e}_t(X)/e_t(X) - 1)}\frac{e_t}{e_t(X)}\right] {\bigg|\bigg|} \\
    &\lesssim \sqrt{n} J \sup_{t,x} \frac{e_t}{e_t(x)} \sup_t E[(\tilde{\mu}_t(X) - \mu_t(X))^2]^{1/2}E[(\tilde{e}_t(X) - e_t(X))^2]^{1/2} \\
    &\lesssim \sqrt{n}Jr_{\mu,n,2}r_{e,n,2}, 
\end{align*}}
which is $o(1)$ by Assumption A.5.

\subsection{Bounding the Empirical Process}
Given a finite number of non-overlapping, independent folds for data $W_1,\dots,W_n$, denote fold data  $I_f$ of length $n_f$ and let $I_f^c$ denote the corresponding hold-out data. To bound the empirical process, we first consider convergence conditional on the hold-out sample. In particular conditional on $I_f^c$, we have that

{\footnotesize \begin{align*}
    \left(G_{n_f}\left[\hat{\IF}({d}_0(a))\right] - G_{n_f}\left[{\IF}({d}_0(a))\right]\right)^2
    &\lesssim_P E\left[\left(\sum_{t\in\mathcal{T}_a}(\psi_t(\hat{\eta}) - \psi_t(\eta))e_t\right)^2\bigg|I_f^c\right] \\
    &= \sum_{t\in\ta}\sum_{s\in\ta}e_te_s E[(\psi_t(\hat{\eta}) - \psi_t(\eta))(\psi_s(\hat{\eta}) - \psi_s(\eta))|I_f^c],
    \end{align*}}
by Chebyshev's inequality and independent data. 
Now let again $r \in [0,1]$ such that $\hat{\eta} = \eta + r(\tilde{\eta} - \eta)$. For the $t=s$ summand, we have that {\footnotesize \begin{align*}
    &E[(\psi_t(\hat{\eta}) - \psi_t(\eta))^2|I_f^c] \\
    &= r^2 E\left[\frac{\mathbbm{1}_t\varepsilon^2(t)(\tilde{e}_t(X) - e_t(X))^2}{(e_t(X) + r(\tilde{e}_t(X) - e_t(X)))^2e_t(X)^2}\bigg|I_f^c\right] + r^2 E\left[\frac{(\tilde{\mu}_t(X) - \mu_t(X))^2(\mathbbm{1}_t - e_t(X))^2}{(e_t(X) + r(\tilde{e}_t(X) - e_t(X)))^2}\bigg|I_f^c\right] \\
    &\quad + r^4 E\left[\frac{(\tilde{\mu}_t(X) - \mu_t(X))(\tilde{e}_t(X) - e_t(X))}{(e_t(X) + r(\tilde{e}_t(X) - e_t(X)))^2}\bigg|I_f^c\right] \\
    &= r^2 E\left[\frac{\sigma^2_t(X)(\tilde{e}_t(X) - e_t(X))^2}{(e_t(X) + r(\tilde{e}_t(X) - e_t(X)))^2e_t(X)}\bigg|I_f^c\right] + r^2 E\left[\frac{(\tilde{\mu}_t(X) - \mu_t(X))^2e_t(X)(1-e_t(X))}{(e_t(X) + r(\tilde{e}_t(X) - e_t(X)))^2}\bigg|I_f^c\right] \\
    &\quad + r^4 E\left[\frac{(\tilde{\mu}_t(X) - \mu_t(X))(\tilde{e}_t(X) - e_t(X))}{(e_t(X) + r(\tilde{e}_t(X) - e_t(X)))^2}\bigg|I_f^c\right] \\
    &\lesssim \sup_{t\in\ta,x}\frac{1}{e_te_t(x)^2}\frac{e_t(x)^2}{\hat{e}_t(x)^2} E[(\tilde{e}_t(X) - e_t(X))^2|I_f^c] + \frac{1}{e_t(x)}\frac{e_t(x)}{\hat{e}_t(x)}E[(\tilde{\mu}(X)- \mu(X))^2|I_f^c] \\
    &\quad + \frac{1}{e_t(x)^2}\frac{e_t(x)}{\hat{e}_t(x)}E[(\tilde{e}_t(X) - e_t(X))^2|I_f^c]^{1/2}E[(\tilde{\mu}_t(X) - \mu_t(X))^2|I_f^c]^{1/2},
\end{align*}}
where the second equality follows from Assumption 1 and the law of iterated expectations. For the cross terms $t \neq s$, note that $\mathbbm{1}_t\mathbbm{1}_s = 0$ and hence 

{\footnotesize \begin{align*}
    E&[(\psi_t(\hat{\eta}) - \psi_t(\eta))(\psi_s(\hat{\eta}) - \psi_s(\eta))|I_f^c] \\
    &= r^2E\left[\frac{(\mathbbm{1}_t - e_t)(\mathbbm{1}_s - e_s)(\tilde{\mu}_t(X) - \mu_t(X))(\tilde{\mu}_s(X) - \mu_s(X))}{(e_t(X) + r(\tilde{e}_t(X) - e_t(X)))(e_s(X) + r(\tilde{e}_s(X) - e_s(X)))}\bigg|I_f^c\right] \\
    &\quad + r^4E\left[\frac{(\tilde{\mu}_t(X) - \mu_t(X))(\tilde{\mu}_s(X) - \mu_s(X))(\tilde{e}_t(X) - e_t(X))(\tilde{e}_s(X) - e_s(X))}{(e_t(X) + r(\tilde{e}_t(X) - e_t(X)))(e_s(X) + r(\tilde{e}_s(X) - e_s(X)))}\bigg|I_f^c\right] \\
    &\lesssim \sup_{t,s\in \ta, t\neq s,x} \frac{e_t(x)e_s(x)}{\hat{e}_t(x)\hat{e}_s(x)}E[|(\tilde{\mu}_t(X) - \mu_t(X))(\tilde{\mu}_s(X) - \mu_s(X))|~|I_f^c] \\
    &\quad + \frac{1}{\hat{e}_t(x)}E[|(\tilde{\mu}_t(X) - \mu_t(X))(\tilde{\mu}_s(X) - \mu_s(X))(\tilde{e}_t(X) - e_t(X))(\tilde{e}_s(X) - e_s(X))|~|I_f^c].
\end{align*}}
Plugging this back into the original empirical process and looking at the worst nuisance case, we obtain conditional bound rates 

{\footnotesize \begin{align*}
  \sup_{\hat{\eta}\in \mathcal{H}_n} & \sum_{t\in\ta}\sum_{s\in\ta}e_te_s E[(\psi_t(\hat{\eta}) - \psi_t(\eta))(\psi_s(\hat{\eta}) - \psi_s(\eta))|I_f^c] \\
  &\lesssim \sup_{\hat{\eta}\in \mathcal{H}_n}\left(\sup_t Je_t^2 E[(\psi_t(\hat{\eta}) - \psi_t(\eta))^2|I_f^c] + \sup_{t,s\in\ta,s\neq t}J^2E[(\psi_t(\hat{\eta}) - \psi_t(\eta))(\psi_s(\hat{\eta}) - \psi_s(\eta))|I_f^c]\right) \\
  &\lesssim \frac{1}{J}\left( J^3 r_{e,n,2}^2 + J r_{\mu,n,2}^2 + J^2 r_{e,n,2}r_{\mu,n,2}\right) + (r_{\mu,n,2}^2 + J^2r_{\mu,n,2}r_{e,n,2}) \\
  &\lesssim (r_{\mu,n,2} + J r_{e,n,2})^2, 
\end{align*}}

where the second inequality follows from repeated application of the Cauchy-Schwarz inequality as well as Assumption A.4. Lastly, given a finite amount of folds, Lemma 6.1 of \citeA{Chernozhukov2018} yields that conditional convergence implies unconditional and thus, we obtain that 

{\footnotesize \begin{align*}
     G_{n}\left[\hat{\IF}({d}_0(a))\right] - G_{n}\left[{\IF}({d}_0(a))\right] \lesssim_P r_{\mu,n,2} + J r_{e,n,2},
\end{align*}}
which is $o(1)$  by Assumption A.5.

\subsection{Leading Term} \label{sec_proof_leadingterm1}
We now consider the limiting behavior of the leading term and show that it is asymptotically normal. We denote all inner product, sums and corresponding suprema without the treatment aggregation for readability, i.e.~$\ip{A,B} = \ip{A,B}_{\ell^2(\ta)}$, $\sum_tA_t = \sum_{t\in\ta}A_t$, and $\sup_tA_t = \sup_{t\in\ta}A_t$.  For simplicity, we consider the modified parameter $\ip{e,\mu}$ where $e = (e_1,\dots,e_J)'$. The proof for $\ip{e_a,\mu}$ is identical up to a $O(n^{-1/2})$ remainder as $e_{ta} = e_t/P(T \in \ta)$ and $P(T \in \ta)$ is bounded away from zero by Assumption A.3. We first find a rate for $\sigma_n^2 = E[\IF(\ip{e,\mu})^2]$. Simplifying the influence function and taking expectations yields

{\footnotesize {\footnotesize \begin{align*}
    \sigma_n^2 &= \sum_t e_t(1-e_t)\mu_t^2 + 2 \mu_t e_t E[\mathbbm{1}_t\psi_t] + e_t^2 E[\psi_t^2] \\
    &\quad+ \sum_{t,s\neq t}E[(\mathbbm{1}_t-e_t)(\mathbbm{1}_{s}- e_{s})]\mu_t\mu_s + e_tE[\psi_t(\mathbbm{1}_s - e_s)]\mu_s + e_sE[\psi_s(\mathbbm{1}_t - e_t)]\mu_t + e_te_s E[\psi_t \psi_s] \\
    &= \sum_t e_t(1-e_t)\mu_t^2 + 2 \mu_t e_t^2E[(\mu_t(X) - \mu_t)|T=t] + e_t^2 E\left[\frac{\sigma_t^2(X)}{e_t(X)} + (\mu_t(X) - \mu_t)^2\right] \\
    &\quad+ \sum_{t,s\neq t}(-e_te_s)\mu_t\mu_s + e_tE[(\mu_t(X) - \mu_t)|T=s]e_s\mu_s + e_sE[(\mu_s(X) - \mu_s)|T=t]e_t\mu_t\\ 
    &\quad+ e_te_s E[(\mu_t(X) - \mu_t)(\mu_s(X) - \mu_s)] \\
    &\lesssim J \left( \sup_{t} e_t + e_t^2 + e_t\left[\sup_{x\in\mathcal{X}}\frac{e_t}{e_t(x)} + 1\right] \right) + J^2 \sup_{t} e_t^2
\end{align*}}}

which is $O(1)$ by Assumptions A.1 and A.4.
Now we consider the self-normalized sum based on $\IF(\ip{e,\mu})$ and verify the Lindeberg condition for the triangular array with iid entries. For any $c > 0$, note that

{\footnotesize \begin{align*}
    \sum_{i=1}^n&\frac{E[\IF(\ip{e,\mu})^2 \mathbbm{1}(|\IF(\ip{e,\mu})| > \sigma_n \sqrt{n}c)]}{n\sigma_n^2} \\ 
    &\lesssim \sigma_n^{-1}\left(E[\ip{e,\IF(\mu)}^2 \mathbbm{1}(|\IF(\ip{e,\mu})| > \sigma_n \sqrt{n}c)] + E[\ip{\IF(e),\mu}^2 \mathbbm{1}(|\IF(\ip{e,\mu})| > \sigma_n \sqrt{n}c)] \right).
\end{align*}}
Note that $\ip{\IF(e),\mu}$ only contains uniformly bounded random variables and thus {\footnotesize \begin{align*}
     E[\ip{\IF(e),\mu}^2 \mathbbm{1}(|\IF(\ip{e,\mu})| > \sigma_n \sqrt{n}c)] 
     \lesssim J^2 \sup_t\mu_t^2 E\left[\mathbbm{1}(|\IF(\ip{e,\mu})| > \sigma_n \sqrt{n}c)\right] 
     =o(1).
\end{align*}} For the unbounded component we use the law of iterated expectations  
 {\footnotesize \begin{align*}
E&\left[\IF(\ip{e,\mu})^2 \mathbbm{1}(|\IF(\ip{e,\mu})| > \sigma_n \sqrt{n}c)\right] \\
&= E\left[\sum_t E[\IF(\ip{e,\mu})^2 \mathbbm{1}(|\IF(\ip{e,\mu})| > \sigma_n \sqrt{n}c)|T=t,X]e_t(X) \right] \\
&\leq J\sup_{t,x} E[\IF(\ip{e,\mu})^2 \mathbbm{1}(|\IF(\ip{e,\mu})| > \sigma_n \sqrt{n}c)|T=t,X=x]e_t(x). 
   \end{align*}}
We now proceed with decomposing the latter expectation and bound the relevant conditional moments. Denote $\mathbbm{1}_c = \mathbbm{1}(|\IF(\ip{e,\mu})| > \sigma_n \sqrt{n}c)$. For any $t \in \mathcal{T}_a$ and $x \in \mathcal{X}$, we have that

{\footnotesize \begin{align*}
    E&[\ip{e,\IF(\mu)}^2 \mathbbm{1}_c|T=t,X=x] 
    =E\left[\left(\sum_se_s\psi_s\right)^2\mathbbm{1}_c\bigg|T=t,X=x\right] \\
    &= e_t^2 E[\psi_t^2\mathbbm{1}_c|T=t,X=x] + 2\sum_{s\neq t}e_se_tE[\psi_t\psi_s\mathbbm{1}_c|T=t,X=x]  \\ &\quad + \sum_{r\neq t}\sum_{s\neq t}e_se_rE[\psi_r\psi_s\mathbbm{1}_c|T=t,X=x] \\
    &= e_t^2 E\left[\frac{\varepsilon(t)^2}{e_t(X)^2}\mathbbm{1}_c\bigg|T=t,X=x\right] + 2\sum_{s}e_te_sE\left[\frac{\varepsilon(t)}{e_t(X)}\mathbbm{1}_c\bigg|T=t,X=x\right](\mu_s(x) - \mu_s) \\
    &\quad + \sum_{r}\sum_{s}e_re_s(\mu_r(x) - \mu_r)(\mu_s(x) - \mu_s)E[\mathbbm{1}_c|T=t,X=x].
\end{align*}}
% {\footnotesize \begin{align*}
%     E&[\IF(\ip{e,\mu})^2 \mathbbm{1}(|\IF(\ip{e,\mu})| > \sigma_n \sqrt{n}c)|T=t,X=x] \\
%     &= E\left[\left(\frac{\varepsilon(t)}{e_t(X)} + \sum_{t\in\ta}\right)^2\mathbbm{1}(|\IF(\ip{e,\mu})| > \sigma_n \sqrt{n}c\bigg|T=t,X=x\right] \\
%     &= \sum_{t,s}(\mu_t(x) - \mu_t)(\mu_s(x) - \mu_s)E[\mathbbm{1}(|\IF(\ip{e,\mu})| > \sigma_n \sqrt{n}c|T=t,X=x] \\ &\quad + 2\sum_t (\mu_t(x) - \mu_t)E\left[\frac{\varepsilon(t)}{e_t(X)}\mathbbm{1}(|\IF(\ip{e,\mu})| > \sigma_n \sqrt{n}c\bigg|T=t,X=x\right]\\
%     &\quad + E\left[\frac{\varepsilon(t)^2}{e_t(X)^2}\mathbbm{1}(|\IF(\ip{e,\mu})| > \sigma_n \sqrt{n}c)\bigg|T=t,X=x\right]
% \end{align*}}
Now note that, conditional on $T=t$ and $X=x$, 
{\footnotesize \begin{align*}
    \mathbbm{1}_c &=  \mathbbm{1}\left(\left|\mu_t - \sum_se_s\mu_s + e_t\frac{\varepsilon(t)}{e_t(x)} + \sum_s (\mu_s(x) - \mu_s)\right| > \sigma_n\sqrt{n}c\right) \\
    &\leq \mathbbm{1}\left(|\varepsilon(t)| >  \bigg[\sigma_n\sqrt{n}c - \bigg|\sum_{s}e_s(\mu_t + \mu_s(x) - 2 \mu_s)\bigg|\bigg]\frac{e_t(x)}{e_t}\right) \\
    &=\mathbbm{1}(|\varepsilon(t)| > c_n(x)),
\end{align*}}

by the definition of the moment function and the triangle inequality.
% {\footnotesize \begin{align*}
%     \mathbbm{1}(|\IF(\ip{e,\mu})| &\leq \mathbbm{1}(|\varepsilon(t)/e_t(x)|  > c\sqrt{n}\sigma_n) \\
%     &= \mathbbm{1}(|\varepsilon(t)|  > [c\sqrt{n}\sigma_n - |\sum_t(\mu_t(x) - \mu)|]e_t(x)) \\
%     &= \mathbbm{1}(|\varepsilon(t)|  > c_n(x))
% \end{align*}}
Next note that $\varepsilon(t) \bigCI \mathbbm{1}_t | X=x$. We now consider sequences where $n^{-1/2}c_n(x) \lesssim 1$ which is implied by Assumptions A.1 and A.4.
Thus, for any $x \in \mathcal{X}$, we can bound the tail moments (including probability) using H\"older's and Markov's inequality. In particular, for any $m_1 \in [0,2+m]$, where $m$ is given by Assumption A.1, we obtain 

{\footnotesize \begin{align*}
    E&[|\varepsilon(t)|^{m_1}\mathbbm{1}(|\varepsilon(t)|  > c_n(x)|T=t,X=x] \\
    &\leq E[|\varepsilon(t)^{2+m}|X=x]^{{m_1}/(2+m)}P(|\varepsilon(t)| > c_n|X=x)^{(2+m-{m_1})/(2+m)} \\
    &\lesssim P(|\varepsilon(t)| > c_n|X=x)^{(2+m-{m_1})/(2+m)} \\
    &\leq \bigg(\frac{E[|\varepsilon(t)|^{2+m}|X=x]}{c_n^{2+m}}\bigg)^{(2+m-{m_1})/(2+m)} \\
    &\lesssim n^{-(2+m-{m_1})/2}.
\end{align*}}

% even probabiltiy {\footnotesize \begin{align*}
%     E[\mathbbm{1}(|\varepsilon(t)|  > c_n(X))|T=t,X=x]
%     &= P(|\varepsilon(t)|  > c_n(X)|X=x) \\
%     &\leq \frac{E[|\varepsilon(t)|^{2+m}|X=x]}{c_n(x)^{2+m}} \\
%     &\lesssim \bigg(\sqrt{\frac{J^2}{n}} \bigg)^{2+m}
% \end{align*}}
% due to Markov's inequality with moment assumption [REF] and $\sup_tJe_t(x) \lesssim 1$. For the first moment part, we similarly obtain {\footnotesize \begin{align*}
%     E&[|\varepsilon(t)|\mathbbm{1}(|\varepsilon(t)|  > c_n(X))|X=x]  \\
%     &\lesssim E[|\varepsilon(t)|^{2+m}]^{1/(2+m)}P(|\varepsilon(t)|  > c_n(X)|X=x)^{(1+m)/(2+m)} \\
%     &\lesssim \bigg[c_n(x)^{-2+m}\bigg]^{(1+m)/(2+m)}\\
%     &= c_n(x)^{-(1+m)} \\
%     &= \bigg(\sqrt{\frac{J^2}{n}}\bigg)^{(1+m)}
% \end{align*}}
% where the first step follows from H\"older's inequality. For the second moment, we analogously obtain {\footnotesize \begin{align*}
%     E[\varepsilon(t)^2\mathbbm{1}(|\varepsilon(t)|  > c_n(X))|X=x]
%     &\leq E[\varepsilon(t)^{(m+2)/2}]^{2/(2+m)}P(|\varepsilon(t)|  > c_n(X)|X=x)^{m/(2+m)} \\
%     &\lesssim \bigg[c_n(x)^{-(2+m)}\bigg]^{-m/(2+m)} \\
%     &= c_n(x)^{-m} \\
%     &= \bigg(\sqrt{\frac{J^2}{n}}\bigg)^{m}
% \end{align*}} 
Applying this to ${m_1}=0,1,2$ and putting all the conditional results back into the original equation then yields 

{\footnotesize \begin{align*}
E&\left[\IF(\ip{e,\mu})^2 \mathbbm{1}(|\IF(\ip{e,\mu})| > \sigma_n \sqrt{n}c)|T=t\right] \\
 &\lesssim  J\sup_{t,x} E[\IF(\ip{e,\mu})^2 \mathbbm{1}(|\IF(\ip{e,\mu})| > \sigma_n \sqrt{n}c)|T=t,X=x]e_t(x) \\
 &\lesssim \sup_{t,x}{Je_t(x)}(n^{-(2+m)/2} + n^{-(1+m)/2} +  n^{-m/2}) \\
 &\lesssim n^{-m/2}.
\end{align*}}
% {\footnotesize \begin{align*}
% E&[\IF(\ip{e,\mu})^2 \mathbbm{1}(|\IF(\ip{e,\mu})| > \sigma_n \sqrt{n}c)|T=t,X=x] \\
%  &\lesssim \sup_{t,x} J^2P(|\varepsilon(t)|  > c_n(X)|X=x) \\ 
%  &\quad + \frac{J}{e_t(x)}E\left[|\varepsilon(t)|\mathbbm{1}(|\varepsilon(t)|  > c_n(X)|X=x\right]\\
%      &\quad + J^2\sup_{t,x}\frac{1}{J^2e_t^2(x)}E\left[{\varepsilon(t)^2}\mathbbm{1}(|\varepsilon(t)| > c_n(X)|X=x\right] \\
%     &\lesssim J^2\frac{J^m}{n^{m/2}} \\
%     &\lesssim \frac{J^{2+m}}{n^{m/2}}
%    \end{align*}}
which is $o(1)$ as $m > 0$ by Assumption A.1. Thus, the Lindeberg condition applies. 

\subsection{Bounding the Remainder}
For the remainder note that, due to the ratio form of the conditional probability we have that {\footnotesize \begin{align*}
    \sqrt{n}\ip{\hat{e}_a - e_a,\hat{\mu} - \mu}_{\ell^2(\ta)}
    &=\sqrt{n}\ip{\hat{e} - e,\hat{\mu} - \mu}_{\ell^2(\ta)} \left(1 + O(E_n[\mathbbm{1}(T \in \mathcal{T}_a)] - P(T \in \mathcal{T}_a) \right) \\
    &= \sum_{t\in \mathcal{T}_a}E_n[\mathbbm{1}_t - e_t]G_n[\psi_t(\hat{\eta})]\left(1 + O_p(n^{-1/2})\right). 
\end{align*}}

By the previous two subsections, we have that $G_n[\psi_t(\hat{\eta})] \lesssim_P 1$. Moreover, by a standard CLT argument, we obtain that the empirical sample frequency estimators are superefficient, i.e.~%{\footnotesize \begin{align*}
    $\sqrt{\frac{n}{e_t(1-e_t)}}E_n[\mathbbm{1}_t - e_t] \overset{d}{\rightarrow} \mathcal{N}(0,1)$
%\end{align*}}
and thus, using Assumption A.4, we obtain 

{\footnotesize \begin{align*}
     \sqrt{n}\ip{\hat{e}_a - e_a,\hat{\mu} - \mu}_{\ell^2(\ta)} &\lesssim_P J\sup_t\sqrt{\frac{e_t}{n}} 
     \lesssim \sqrt{\frac{J}{n}},
\end{align*}}
which is again $o(1)$ by Assumption A.5.

\subsection{Generalization to All Decomposition Estimands} \label{app_proof_N_decompALL}
Derivations for decomposition estimands $d_{j}(a,g)$ with $j=1,2,3,4,4^{'},5$ are obtained analogously to $d_{0}(a)$ as they rely on influence functions that contain either scaled versions of the ones entering $d_{0}(a)$ with a.s.~bounded scaling factors such as indicators and probabilities or simple averages without or at most root-$n$ consistent nuisances (such as simple sample means) that can be controlled. We provide an explicit example of how to adapt the proof below. To bound the explicit varying denominator weights in the influence functions, however, we need to add Assumption A.3. Consider for instance  component $\ip{e_a(\xg),\mu(\xg)}_{\ell^2(\ta)}$. We have that 

{\footnotesize \begin{align*}
&\IF  \{\ip{e_a(\xg),\mu(\xg)}\}_{\ell^2(\ta)} %= \IF\{\sum_{t\in\ta} e_{ta}(\xg)\mu_t(\xg)\} 
= \sum_{t\in\ta} \IF\{e_{ta}(\xg)\mu_t(\xg)\} \\
&= \frac{\mathbbm{1}(X\in\xg)}{P(X\in\xg)} \bigg\{ \sum_{t\in\ta} \left[ \frac{\mathbbm{1}(T\in\mathcal{T}_a)\mathbbm{1}(T = t)}{P(T\in\mathcal{T}_a|X\in\xg)}  \mu_t(\xg) + e_{ta}(\xg) \left( \frac{\mathbbm{1}(T=t) (Y - \mu_t(X))}{P(T=t|X)} + \mu_t(X)\right) \right] \\
& \quad  - \left(1 + \frac{\mathbbm{1}(T\in\mathcal{T}_a)}{P(T\in\mathcal{T}_a|X\in\xg)} \right) { \sum_{t\in\ta} e_{ta}(\xg)\mu_t(\xg)} \bigg\}. 
\end{align*}}

We can see that this is the same influence function with (i) unconditional nuisances $e_{ta}$ and $\mu_t$ replaced by $e_{ta}(\xg)$ and $\mu_t(\xg)$ respectively as well as (ii) a different weight in front of the components. These weights depend only on indicators and probabilities. We use the efficient influence function estimators for the unknown probabilities, see Appendix \ref{sec_app_implement1}. To adapt the proof, we thus need to modify the remainder as well as the supremum weight outside the $\ta$-dependent sums. Solving for the sample analogue with estimated nuisances then yields the following decomposition 

 {\footnotesize   {\footnotesize \begin{align*}
    \sqrt{n}&(\hat{d}_j(a,g) - {d}_j(a,g)) \\
    &= G_n\left[\hat{\IF}({d}_j(a,g))\right]\left( 1 + O\big(|E_n[\IF(P(\ta))]| + |E_n[\IF(P(\xg))]|  + |E_n[\IF(P(\ta|\xg))]|\big) \right)\\
    &\quad  + \sqrt{n}\ip{(\hat{e}_a(\xg) - e_a(\xg)),(\hat{\mu}(\xg) - \mu(\xg))}_{\ell^2(\ta)} \\
    &= \underbrace{G_n\left[{\IF}({d}_j(a,g))\right]}_{\text{Leading Term}} + \underbrace{\left(G_n\left[\hat{\IF}({d}_j(a,g))\right] - G_n\left[{\IF}({d}_j(a,g))\right]\right)}_{\text{Empirical Process}} \\
    &\quad + \underbrace{\sqrt{n}\left(E\left[\hat{\IF}({d}_j(a,g))\right] - E\left[{\IF}({d}_j(a,g))\right] \right)}_{\text{Machine Learning Bias}} + o_p(1).
\end{align*}}}
 In particular, Assumption A.3 states that $\inf \{P(\xg),P(\ta),P(\ta|\xg)\} > \underline{c}$
 and thus all the influence function based frequency estimators above are root-$n$ consistent, i.e.~$E_n[\IF(\cdot)] = o_p(n^{-1/2})$. The proof for machine learning bias as well as empirical process is identical to the previous step with again $e_{ta}$ and $\mu_t$ replaced by their $\xg$ conditional counterparts. Thus, overall we obtain an asymptotic linearization with $o_p(1)$ remainder as for $d_{0}(a)$. The same applies to the remaining decomposition parameters. 

\subsection{Multivariate CLT} \label{app_proof_multiCLT}
We now stack all the influence functions in a single column vector $\IF(\theta)$ where $\theta \in \mathbb{R}^8$ is the the collection of all decomposition estimands from Appendix \ref{sec_app_implement1}. Any decomposition parameter or contrast can then be obtained via linear combinations of these components, i.e.~we consider the asymptotic distribution of estimates $c_n'\hat{\theta}$, i.e.~$
    [c_n'\Sigma c_n]^{-1/2}c_n'(\hat{\theta} - \theta)$, 
where $c_n \lesssim 1$ are bounded (sequences of) weight vectors in $\mathbbm{R}^{8}$ and $\hat{\theta}$ are obtained from solving all influence functions with respect to their target parameters with estimated, cross-fitted nuisances. The variance is given by $\Sigma = E[\IF(\theta)\IF(\theta)']$. Due to the leading term result regarding variance in Subsection \ref{sec_proof_leadingterm1} and the finite dimension of $\theta$ we have that $||c_n'\Sigma c_n||_2 \lesssim ||c_n||_2^2||\Sigma ||_2 \lesssim 1.$
Now we use an expansion as in Section \ref{sec_proof_ifbasedestimator} which can be bounded using the marginal results from the previous subsections. 

{\footnotesize \begin{align*}
    G_n\left[\hat{\IF}(\theta)\right]
    &= {G_n\left[{\IF}(\theta)\right]}{\left(G_n\left[\hat{\IF}(\theta)\right] - G_n\left[{\IF}(\theta)\right]\right)} 
     + {\sqrt{n}\left(E\left[\hat{\IF}(\theta)\right] - E\left[{\IF}(\theta)\right] \right)} \\
    &= {G_n\left[{\IF}(\theta)\right]} + O\left(\max_{i} \sqrt{8}\bigg|\bigg| \left(G_n\left[\hat{\IF}(\theta_i)\right] - G_n\left[{\IF}(\theta_i)\right]\right)\bigg|\bigg|\right) \\
    &\quad + O\left(\max_i \sqrt{8} \sqrt{n}\bigg|\bigg|\left(E\left[\hat{\IF}(\theta_i)\right] - E\left[{\IF}(\theta_i)\right] \right)\bigg|\bigg| \right) \\
    &= G_n\left[{\IF}(\theta)\right] + o_p(1),
\end{align*}}
by norm equivalence in the finite-dimensional vector space. We now consider the multivariate Lindeberg condition for the leading term. 
We make repeated use of the following bound for generic random vectors $A,B$ 

{\footnotesize \begin{align*}
    E[||A + B||_2^2\mathbbm{1}(||A + B|| > c)] 
    \leq 2 \left(E[||A||_2^2\mathbbm{1}(||A|| > c/2)] + E[||A||_2^2\mathbbm{1}(||B|| > c/2)]\right).
\end{align*}}
Consider now the leading term for the linear combination
{\footnotesize \begin{align*}
{\sqrt{n}}E_n\left[\left(c_n'\Sigma c_n\right)^{-1/2}c_n'\IF(\theta)\right] &= E_n[\omega_n \IF(\theta)],
\end{align*}}
where $\omega_n = n^{-1/2}\left(c_n'\Sigma c_n\right)^{-1/2}c_n'$. Thus, by definition of the variance for the iid sum, we have that $V[E_n[\omega_n \IF(\theta)]] = 1$. We can now evaluate the Lindeberg condition. In particular note that, for any $c > 0$, {\footnotesize \begin{align*}
    E[||\omega_n \IF(\theta)||_2^2\mathbbm{1}(||\omega_n \IF(\theta)|| > c)] 
    &\leq ||\omega_n||_2^2 E[||\IF(\theta)||_2^2\mathbbm{1}( ||\IF(\theta)|| > c/||\omega_n||)] \\
    & \leq ||\omega_n||_2^2 8 \max_j E[\IF(\theta_j)^2\mathbbm{1}(|\IF(\theta_j) > c/(8||\omega_n||)] \\
    & \lesssim \max_j E[\IF(\theta_j)^2\mathbbm{1}(|\IF(\theta_j) > c_{n,j}],
\end{align*}}
where $c_{n,j} \sim \sqrt{n}$ by definition of $\omega_n$ as well as $||c_n'\Sigma c_n || \lesssim ||c_n||_2^2 \max \Sigma_{jj} \lesssim 1$ by Subsection \ref{sec_proof_leadingterm1}. Now recall that $E[\IF(\theta_j)^2\mathbbm{1}(|\IF(\theta_j)| > c_{n,j})] = o(1)$ for all $j=1,\dots,8$. Thus the multivariate Lindeberg condition applies and hence %{\footnotesize \begin{align*}
    $[c_n'\Sigma c_n]^{-1/2}c_n'(\hat{\theta} - \theta) \overset{d}{\rightarrow} \mathcal{N}(0,1).$
%\end{align*}}

\subsection{Variance Estimation} \label{sec_app_variance1}
We now consider estimated variancecross-fitted nuisances. In particular define $ \hat{\Sigma} = E_n[\hat{\IF}(\hat{\theta})\hat{\IF}(\hat{\theta})']$.
% {\footnotesize \begin{align*}
%     \hat{\Sigma} = E_n[\hat{\IF}(\hat{\theta})\hat{\IF}(\hat{\theta})'].
% \end{align*}}
It is sufficient to show that $    ||\hat{\Sigma} - \Sigma||_2 = o_p(1)$ and then apply Slutzky's Theorem.
We adapt \citeA{Chernozhukov2018}, Theorem 3.2 to our setup with extending nuisance parameter space. Steps are identical except for some of the moment bounds. In particular, for their (A.25), we need a rate for $E[||\IF(\theta)||_2^q]$ under growing treatments. We again focus on a worst-case component $d_{0}(a)$ in $\theta$ that then provides a bound for the whole vector norm due to its finite dimension. In particular, let $q = 2+m$, then, for any $q > 2$, 

{\footnotesize \begin{align*}
    E[||\IF(\theta)||_2^q] 
    &\lesssim E[||\ip{\IF(\mu),e}_{\ell^2(\ta)}||_2^q] + E[||\ip{\mu,\IF(e)}_{\ell^2(\ta)}||_2^q].
\end{align*}}  
The two components can be bounded by rate and moment assumptions A.1 and A.4. We consider both the case where $q \in (2,4)$ as well as $q \geq 4$. For the first term, triangle inequality, law of total probability, and Jensen's inequality yields 

{\footnotesize \begin{align*}
E[||\ip{\IF(\mu),e}_{\ell^2(\ta)}||_2^q]
&\leq E[|\sum_{t\in\ta}\mathbbm{1}_t\varepsilon(t)|^q] + E[|\sum_{t\in\ta}e_t(\mu_t(X) - \mu_t)|^q] \\
&\leq \sum_{t\in\ta}E[|\varepsilon(t)|^q|T=t]e_t + J^{q-1}\sum_{t\in\ta}|e_t|^qE[|\mu_t(X) - \mu_t|^q] \\
&\lesssim J \sup_{t,x} E[|\varepsilon(t)|^q|X=x]e_t + J^{q-1}J \sup_{t,x}e_t^qE[|Y(t)|^q|X=x]
\end{align*}}
which is $O(1)$. For the second component note that again by Jensen's inequality
{\footnotesize \begin{align*}
    E[||\ip{\mu,\IF(e)}_{\ell^2(\ta)}||_2^q] &\leq J^{q-1} \sum_{t\in\ta}E[|\mu_t(\mathbbm{1}_t - e_t)|^q] \\
    &\lesssim J^q \sup_{t\in\ta}|\mu_t|^q(1-e_t)^qe_t + e_t^q(1-e_t) \\
    &\lesssim J^{q-1}.
\end{align*}}
We now combine this for $q \in (2,4)$ as well as $q \geq 4$, paralleling (A.25) in \citeA{Chernozhukov2018}. In particular, note that for $q > 4$, the optimal rate of the previous terms is obtained already at $q=4$. Thus, combining results yields modified (A.25) rate 

{\footnotesize \begin{align*}
    E_{n_f}[\IF(\theta)] &\lesssim_P \sqrt{\frac{J^3}{n}} + \left(\frac{J}{n}\right)^{1-2/q}J 
    \lesssim J \left(\frac{J}{n}\right)^{[(1/2)\wedge (1-2/q)]}. 
\end{align*}}
For the other moment bound in (A.27), we have to additionally control the second moment of the moment function nuisance relative to using true nuisances on the realization set. In particular, equivalently to the proof in Subsection \ref{sec_proof_mlbias1} we obtain that 

{\footnotesize \begin{align*}
    \sup_{\eta \in \mathcal{H}_n} E[||\hat{\IF}(\theta) - \IF(\theta)||_2^2|I_k^c]
    \lesssim (r_{\mu,n,2} + J r_{e,n,2})^2, 
\end{align*}}
where the rate again comes from the worst case component from Subsection \ref{sec_proof_mlbias1}. Combining rates together as in  \citeA{Chernozhukov2018}, Equation (A.23), and noting that $q = 2 + m$ with $m > 0$ from Assumption A.1, we obtain that

{\footnotesize \begin{align*}
   ||\hat{\Sigma} - \Sigma||_2 &= O_p\left(n^{-1/2} + J \left(\frac{J}{n}\right)^{\left[\frac{1}{2}\wedge \frac{m}{2+m}\right]} + r_{\mu,n,2} + Jr_{e,n,2} \right) \\
   &= O_p\left(J \left(\frac{J}{n}\right)^{\left[\frac{1}{2}\wedge \frac{m}{2+m}\right]} + r_{\mu,n,2} + Jr_{e,n,2} \right), 
\end{align*}}
which is $o_p(1)$ by Assumption A.5 and the statement in Theorem \ref{thm_asyN1}.

\section{Supplementary Material for Section \ref{sec:generalized1}} \label{sec_app_generalizedT}

\subsection{Decompositions}
\subsubsection{Auxiliary Results}\label{sec_app_aux_gen_covariance1}Note that, for any measurable $g_1,g_2:\mathcal{T}\times\mathcal{X}\rightarrow \mathbb{R}$, we have that, for any $t \in \mathcal{T}$, {\footnotesize \begin{align*}
    E&[g_1(t,X)g_2(t,X)|X\in \xg] \\
    &= Cov(g_1(t,X),g_2(t,X)|X\in \xg) + E[g_1(t,X)|X\in\xg]E[g_2(t,X)|X\in\xg] \\
    &= Cov(g_1(t,X),g_2(t,X)|X\in \xg) + g_{1,t}(\xg)g_{2,t}(\xg).
\end{align*}}
\subsubsection{Difference-in-Means}
{\footnotesize \begin{align*}
    E[Y|X\in\xg,T\in\ta] 
    &= E[E[Y|X,T]|X\in\xg,T\in\ta] \\
    &= \int_{\mathcal{X}}\left[\int_{\ta\backslash\{t_r\}}\mu_t(x)\frac{f(t|x)}{P(T\in\ta|X\in\xg)}d\lambda(t) \right]dP(x|X\in\xg) \\ &\quad+ \sum_{t\in\{t_r\}} e_{ta}(\xg) E[\mu_{t}(X)|T=t,X\in \xg].
\end{align*}}
For the first component, Fubini's Theorem together with auxiliary result \ref{sec_app_aux_gen_covariance1} yields

{\footnotesize \begin{align*}
&\int_{\ta\backslash\{t_r\}}\int_{\mathcal{X}}\mu_t(x)\frac{f(t|x)}{P(T\in\ta|X\in\xg)}dP(x|x\in\xg)d\lambda(t) \\
    &=\int_{\ta\backslash\{t_r\}}\left(E[\mu_t(X)|x\in\xg]\frac{E[f(t|X)|X\in\xg]}{P(T\in\ta|X\in\xg)} + \frac{Cov(f(t|X),\mu_t(X)|X\in\xg)}{P(T\in\ta|X\in\xg)}\right)d\lambda(t) \\
    &= \ip{\mu(\xg),f_{a}(\xg)}_{L^2(\ta\backslash \{t_r\},\lambda)} + \frac{\ip{Cov(f(X),\mu(X)|X\in\xg),1}_{L^2(\ta\backslash \{t_r\},\lambda)}}{P(T\in\ta|X\in\xg)},
\end{align*}}
where $f(x) = \{f(t|x):t\in\ta \backslash \{t_r\}\}$ and $f_{a}(\xg) = \{f_{ta}(\xg):t \in \ta\backslash \{t_r\} \} = \{f(t|T\in\ta,X\in\xg): t \in \ta\backslash \{t_r\}\}$. 
The discrete component can be decomposed analogously as  

{\footnotesize \begin{align*}
    \sum_{t\in\{t_r\}}& e_{ta}(\xg) E[\mu_{t}(X)|T=t,X\in \xg] \\
    &= \ip{e_{a}(\xg),\mu(\xg)}_{\ell^2(\{t_r\})} + \frac{\ip{Cov(e_{a}(X),\mu(X)|X\in\xg),1}_{\ell^2(\{t_r\})}}{P(T\in\ta|X\in\xg)}.
\end{align*}}
Combining results yields the first decomposition {\footnotesize \begin{align*}
    E&[Y|X\in\xg,T\in\ta] \\
    &=\ip{\mu(\xg),f_{a}(\xg)}_{L^2(\ta\backslash \{t_r\},\lambda)} + \frac{\ip{Cov(f(x),\mu(x)|x\in\xg),1}_{L^2(\ta\backslash \{t_r\},\lambda)}}{P(T\in\ta|x\in\xg)} \\
    &\quad + \ip{e_{a}(\xg),\mu(\xg)}_{\ell^2(\{t_r\})} + \frac{\ip{Cov(e_{a}(X),\mu(X)|X\in\xg),1}_{\ell^2(\{t_r\})}}{P(T\in\ta|X\in\xg)} \\
    &=\ip{e_{a}(\xg),\mu(\xg)}_{L^2(\ta\backslash \{t_r\},\lambda) \oplus \ell^2(\{t_r\})} + \frac{\ip{Cov(e_{a}(X),\mu(X)|X\in\xg),1}_{L^2(\ta\backslash \{t_r\},\lambda) \oplus \ell^2(\{t_r\})}}{P(T\in\ta|X\in\xg)}.
\end{align*}}  
\subsubsection{Adjusted Difference-in-Means}
Derivations for the adjusted DiM work similarly with modified starting point. For any $t \in \ta \backslash \{t_r\}$ and $x \subseteq X$, let $f_{ta}(x)$ be the conditional density of $T$ given $T \in \ta$ and $X\in x$ evaluated at $t$. Note that, for any $t \in \ta \backslash \{t_r\}$,

{\footnotesize \begin{align*}
    \int_{\mathcal{X}}&\mu_t(x)f_{ta}(x)dP(x|X\in\xg) 
    = E[\mu_t(X)f_{ta}(X)|X\in \xg] \\
    &= Cov(\mu_t(X),f_{ta}(X)|X\in\xg) + E[\mu_t(X)|X\in\xg]E[f_{ta}(X)|X\in\xg] \\
    &= Cov(\mu_t(X),f_{ta}(X)|X\in\xg) + \mu_t(\xg)f_{ta}(\xg) + \left(E[f_{ta}(X)|X\in\xg] - f_{ta}(\xg)\right)\mu_t(\xg).
\end{align*}}
We now can decompose the adjusted DiM. For the discrete part, we use the derivations as in Appendix \ref{app:deriv-both}. The definition of the adjusted DiM with the auxiliary result \ref{sec_app_aux_gen_covariance1} then yields 

{{\footnotesize \begin{align*}
    &E\left[E[Y|T\in\ta,X]|X\in\xg\right] \\
    &= \int_x\int_{\ta\backslash\{t_r\}}\mu_t(x)f_{ta}(x)d\lambda(t)dP(x|X\in\xg) + \sum_{t\in\{t_r\}} E[e_{ta}(X) \mu_{t}(X) |X\in \xg] \\
    &= \ip{f_{a}(\xg),\mu(\xg)}_{L^2(\ta\backslash \{t_r\},\lambda)} + \ip{E[f_{\ta}(X)|X\in\xg] - f_{a}(\xg),\mu(\xg)}_{L^2(\ta\backslash \{t_r\},\lambda)}  \\ &\quad+ \ip{Cov(f_{a}(X),\mu(X)|X\in\xg),1}_{L^2(\ta\backslash \{t_r\},\lambda)} + \ip{e_{a}(\xg),\mu(\xg)}_{\ell^2(\{t_r\})} \\ &\quad+ \ip{E[e_{a}(X)|X\in\xg] - e_{a}(\xg),\mu(\xg)}_{\ell^2(\{t_r\})} + \ip{Cov(e_{a}(X),\mu(X)|X\in\xg),1}_{\ell^2(\{t_r\})} \\
    &= \ip{e_{a}(\xg),\mu(\xg)}_{L^2(\ta\backslash \{t_r\},\lambda) \oplus \ell^2(\{t_r\})} + \ip{E[e_{a}(X)|X\in\xg] - e_{a}(\xg),\mu(\xg)}_{L^2(\ta\backslash \{t_r\},\lambda) \oplus \ell^2(\{t_r\})} \\ &\quad + \ip{Cov(e_{a}(X),\mu(X)|X\in\xg),1}_{L^2(\ta\backslash \{t_r\},\lambda) \oplus \ell^2(\{t_r\})}.
\end{align*}}}

\subsubsection{Synthetic Stratified Experiment}
We now decompose the synthetic stratified experiment estimand. We first define unconditional estimands over the composite space as ordered pairs %{\footnotesize \begin{align*}
 %   \mu &= (\{\mu(t) : t \in \ta \backslash \{t_r\}\}, \{\mu({t_r})\}), \\
 %   e_a &= (\{f_a(t) : t \in \ta \backslash \{t_r\}\}, \{e_{t_r}\}).
%\end{align*}} 
%For both inner products above, by linearity in first argument as well as symmetry, we obtain, for generic $A,B,a,b$ 
For all inner products, we have 

{\footnotesize \begin{align*}
    \ip{A,B} 
    &= \ip{a,b} + \ip{A-a,b} + \ip{a,B-b} + \ip{A-a,B-b}.
\end{align*}}
Applying this to the joint decomposition then yields {\footnotesize \begin{align*}
    &\ip{e_{a}(\xg),\mu(\xg)}_{L^2(\ta\backslash \{t_r\},\lambda) \oplus \ell^2(\{t_r\})} \\
    &= \ip{e_{a},\mu}_{L^2(\ta\backslash \{t_r\},\lambda) \oplus \ell^2(\{t_r\})}
    + \ip{e_{a}(\xg),\mu(\xg)-\mu}_{L^2(\ta\backslash \{t_r\},\lambda) \oplus \ell^2(\{t_r\})} \\
    &\quad + \ip{e_{a}(\xg) - e_a,\mu}_{L^2(\ta\backslash \{t_r\},\lambda) \oplus \ell^2(\{t_r\})} 
    + \ip{e_{a}(\xg) - e_a,\mu(\xg) - \mu}_{L^2(\ta\backslash \{t_r\},\lambda) \oplus \ell^2(\{t_r\})}.
\end{align*}}
The decompositions for differences between treatment aggregations and covariate aggregations follow similarly. 

\subsection{Estimation and Inference Large Sample Theory}
% We consider parameter $d_{0}(a)$ in what follows for simplicity, but the theory readily extends to all decomposition parameters as well as their linear combinations as in Theorem [REF]. 
% We define the target parameter in the decomposition as before but now living in the general treatment and aggregation spaces $\mathcal{T}$ and $\ta$ respectively. However, we still use the discretized estimator as in Section [REF]. 
% In particular, assume we now use  $\tilde{J} = J^* + J$ treatments for analysis where $J = J_n$ is used for $\{t_r\} = \{t_r\}_{r=1}^J$ discrete treatments with non-zero probability mass as in Section [REF] and $J^* = J^*_n$ uses the same estimator  but for $J^*$ user-defined partitions of the treatment aggregation space minus the $J$ components $\{t_r\}$. In particular, we define the partitioning $\{v_j\}_{j=1}^{J^*}$ such that $v_j \cap v_k = \emptyset$ if $j\neq k$ and $\bigcap_{j=1}^{J^*} v_j = \ta \backslash \{t_r\}$. 
 Pseudo-true parameter and target estimand yield 
 % are given by {\footnotesize \begin{align*}
 %     d^*_{0}(a) &= \sum_{j=1}^{J^*}E[E[Y|T\in v_j,X]]P(T\in v_j|T \in \ta) + \sum_{t\in\{t_r\}}E[E[Y|T=t,X]]e_{ta}, \\
 %     d_{0}(a) &= \int_{\ta \backslash \{t_r\}} E[E[Y|T=t,X]]f(t|T\in \ta)d\lambda(t) + \sum_{t\in\{t_r\}}E[E[Y|T=t,X]]e_{ta}.
 % \end{align*}}
%Thus

{\footnotesize \begin{align*}
    d^*_{0}(a) - d_{0}(a) 
    &= \frac{1}{P(T\in \ta)}\left(\int_{\ta \backslash \{t_r\}} P_{f,J^*}\mu(t)P_{J^*}f(t)d\lambda(t) - \int_{\ta \backslash \{t_r\}} \mu(t) f(t) d\lambda(t) \right),
\end{align*}}
where $P_{J^*,f}\mu$ and $P_{J^*}f$ are the orthogonal projections. % of $\mu_t = E[E[Y|T=t,X]]$ and $f(t)$  on the subspace of piecewise constant functions in a ($f$-weighted) $L^2(\ta \backslash \{t_r\},\lambda)$ sense.
% : {\footnotesize \begin{align*}
% P_{f,J^*}\mu &= \arg \min_{\tilde{\mu} \in V_{J^*}}\int_{\ta \backslash \{t_r\}} |\mu(t) - \tilde{\mu}(t)|^2f(t)d\lambda(t), \\
% P_{J^*}f &= \arg \min_{\tilde{f} \in V_{J^*}}\int_{\ta \backslash \{t_r\}} |f(t) - \tilde{f}(t)|^2d\lambda(t), \\
% \mathcal{V}_{J^*} &= {\bigg\{ } \sum_{j=1}^{J^*}c_j\mathbbm{1}(t \in v_j) \bigg| c_j \in \mathbb{R} {\bigg \} }.
% \end{align*}}
Note that both projection errors must also be in the respective $L^2$ spaces. Thus we obtain 

{\footnotesize \begin{align*}
    d^*(0) - d_{0}(a) 
    &\lesssim \frac{1}{P(T\in \ta)}\left(\ip{P_{f,J^*}\mu - \mu ,f}_{L^2(\ta \backslash \{t_r\},\lambda)} + \ip{P_{f,J^*}\mu,P_{J^*}f - f}_{L^2(\ta \backslash \{t_r\},\lambda)} \right) \\
    &\lesssim \ip{P_{f,J^*}\mu - \mu ,1}_{L^2(\ta \backslash \{t_r\},f,\lambda)} + ||\mu||_{L^2(\ta \backslash \{t_r\},\lambda)}||f - P_{J^*}f||_{L^2(\ta \backslash \{t_r\},\lambda)} \\
    &\lesssim ||  \mu - P_{f,J^*}\mu||_{L^2(\ta \backslash \{t_r\},f,\lambda)} + ||f - P_{J^*}f||_{L^2(\ta \backslash \{t_r\},\lambda)}, 
\end{align*}}
where $\ip{g_1,g_2}_{L^2(T,f,\lambda)} = \int_Tg_1(t)g_2(t)f(t)d\lambda(t)$ is the $f$-weighted inner product. Theorem \ref{thm_general_asyN1} follows then directly by adding and subtracting $d_0(a)$ using Theorem \ref{thm_asyN1} together with the bounded variance. Corollary \ref{corr_asyN_smooth1} follows from standard approximation rates for local partitions of smooth functions.

\section{Supplementary Material for Section \ref{sec_indirect1}} \label{sec_app_powerJ1}

\subsection{Theory}

We consider the asymptotic approximation for large $J$ to various testing procedures. To do so, we first make some simplifying assumptions to provide analytically tractable formulations. This can be extended at the expense of carrying around constant weight vectors and matrices throughout. First note that, when $\mu_0(x) = 0$, the strong group effect homogeneity null hypothesis \eqref{eq_strongeffecthomogeneity} can be written as 

{\footnotesize \begin{align*}
   H_0: \mu_t(\xg) = \mu_t(\xgp) \text{ for all } t \in \ta 
\end{align*}}
We additionally make the assumption that, for all $x \in \mathcal{X}$, $e_t(x) = e_t$ and $V[Y(t)|X=x] = \sigma_t^2$. 
Assuming known nuisances, the influence function estimators of the group mean are given by

{\footnotesize \begin{align*}
    \hat{\mu}_t(\xg) -\mu_t(\xg) 
    &= E_n\left[\psi_t\frac{\mathbbm{1}(X \in \xg)}{P(X \in \xg)}\right]= E_n\left[\left(\frac{\mathbbm{1}_t\varepsilon(t)}{e_t(X)} + \mu_t(X) - \mu_t\right)\frac{\mathbbm{1}(X \in \xg)}{P(X \in \xg)}\right].
\end{align*}}
%In practice $P(X\in\xg)$ is estimated. However, they can be replaced by their root-n consistent counterparts without affecting the approximation. 
The centered mean difference parameter can then be written as
{\footnotesize \begin{align*}
   [\hat{\mu}_t(\xg) - \hat{\mu}_t(\xgp)]  - [\mu_t(\xg)  - \mu_t(\xgp)]
    &= E_n\left[\left(\frac{\mathbbm{1}_t\varepsilon(t)}{e_t(X)} + \mu_t(X) - \mu_t \right)\mathbbm{a}_{g,g'}\right]. 
\end{align*}}
where  {\footnotesize \begin{align*}
    \mathbbm{a}_{g,g'} &= \frac{\mathbbm{1}(X \in \xg)}{P(X \in \xg)} - \frac{\mathbbm{1}(X \in \xgp)}{P(X \in \xgp)}.
\end{align*}}
Now note that we have {\footnotesize \begin{align*}
    \sqrt{e_t n}\left([\hat{\mu}_t(\xg) - \hat{\mu}_t(\xgp)]  - [\mu_t(\xg)  - \mu_t(\xgp)]\right)
    &= G_n\left[\left(\frac{\mathbbm{1}_t\varepsilon(t)}{\sqrt{e_t}} + \sqrt{e_t}(\mu_t(X) - \mu_t)\right)\mathbbm{a}_{g,g'} \right] \\
    &=G_n\left[\mathbbm{1}_t\varepsilon(t)\mathbbm{a}_{g,g'}\right] + O_p(J^{-1/2}),
\end{align*}}
due to Assumptions A.1 and A.3.
Consider now the variance of the leading term {\footnotesize \begin{align*}
    E\left[\frac{\mathbbm{1}_t\varepsilon(t)^2}{\sqrt{e_t}^2}\mathbbm{a}^2_{g,g'}\right] 
    &= \sigma_t^2 E[\mathbbm{a}_{g,g'}^2].
\end{align*}}
For the covariances $s\neq t$, we have that
{\footnotesize \begin{align*}
    n&\sqrt{e_te_s} Cov([\hat{\mu}_t(\xg) - \hat{\mu}_t(\xgp)]  - [\mu_t(\xg)  - \mu_t(\xgp)],[\hat{\mu}_s(\xg) - \hat{\mu}_s(\xgp)]  - [\mu_s(\xg)  - \mu_s(\xgp)]) \\
    &= \frac{n\sqrt{e_te_s}}{n^2}\sum_{i=1}^n Cov(\mathbbm{a}_{g,g'}\psi_t,\mathbbm{a}_{g,g'}\psi_s) \\
    &= \sqrt{e_te_s} E[({\mu}_t(X) -\mu_t)({\mu}_s(X) -\mu_s)\mathbbm{a}_{g,g'}^2],
\end{align*}}
due to independence. We now make the following unit-variance-type assumption for simplification: $
    \sigma_t^2 E[\mathbbm{a}_{g,g'}^2] = 1$ and $
    \sqrt{e_te_s} E[({\mu}_t(X) -\mu_t)({\mu}_s(X) -\mu_s)\mathbbm{a}_{g,g'}^2] = 1/J$.
This is without loss of generality as the $1/J$ covariance rate is in line with Assumption A.4. Hence the joint distribution of the vector in large samples is characterized by 

{\footnotesize \begin{align*}
    \sqrt{\frac{n}{J}}\left([\hat{\mu}(\xg) - \hat{\mu}(\xgp)]  - [\mu(\xg)  - \mu(\xgp)]\right) \overset{d}{\rightarrow} \mathcal{N}(0,S),
\end{align*}}
where $S = I_J(1-1/J) + \iota\iota'/J$ is the variance covariance matrix.
For simplicity, we also make the assumption of an exact normal distribution in what follows. This makes all results only first-order approximations if the vectors are not exactly normal. For each $t$-specific entry we can then decompose the process into a system component whose influence vanishes at rate $1/\sqrt{J}$ and a $t$-specific idiosyncratic component.

{\footnotesize \begin{align*}
    \sqrt{\frac{n}{J}}\left([\hat{\mu}_t(\xg) - \hat{\mu}_t(\xgp)]  - [\mu_t(\xg)  - \mu_t(\xgp)]\right) 
    &= \sqrt{\frac{n}{J}}(\hat{m}_t - m_t) \overset{d}{=} \frac{Z_0}{\sqrt{J}} + Z_t\sqrt{1-1/J},
\end{align*}}
where $Z_0,Z_1,Z_2,\dots,Z_J$ are independent standard normal random variables. We denote the stacked vector shorthand as {\footnotesize \begin{align*}
     \sqrt{\frac{n}{J}}\left([\hat{\mu}(\xg) - \hat{\mu}(\xgp)]  - [\mu(\xg)  - \mu(\xgp)]\right) 
     &=  \sqrt{\frac{n}{J}}\left( \hat{m} - m \right) \overset{d}{=} \frac{Z_0}{\sqrt{J}} + {Z}\sqrt{1-1/J},
\end{align*}}
where $Z = (Z_1,\dots,Z_J)'$. We now consider the following test statistics for (implications of) the strong null 
{\footnotesize \begin{align*}
    \hat{t}_{\ell_2}^2 &= \frac{n}{J}\hat{m}'S^{-1}\hat{m}, \quad
     \hat{t}_{\ell_{\infty}} = \sup_t\left|\sqrt{\frac{n}{J}}\frac{\hat{m}_t}{S_{tt}^{-1/2}}\right|, \quad
     \hat{t}_{\Delta} = \frac{\sqrt{n}\hat{\Delta}_1(g,g')}{\Sigma_{{\Delta}_1}^{1/2}},
\end{align*}}
where the formula for the variance of the decomposition $\Sigma_{{\Delta}_1}$ is the corresponding component of the expected squared influence function of $\Delta_1$ as in Appendix \ref{sec_app_implement1}.  For all test-statistics, we compare them against the critical value of the true distribution under the strong null, i.e.~we perform a test with exact size control under normality.

For the the local power analysis, we consider sequences where the group differences are local to zero, i.e.~they are assumed to be small relative to the total sample size, but potentially large relative to effective treatment specific sample size. In particular we assume that, for all $t \in \ta$,$
    \sqrt{n}m_t \rightarrow \xi_t \in \mathbb{R}$,
or $\sqrt{n}m \rightarrow \xi$ in vector notation. We now derive the local power curves. 
\paragraph{\textbf{Auxiliary Results:}} We first provide some auxiliary results. 
Note that, by the properties of standard normal variables, we have that, jointly 

{\footnotesize \begin{align*}
   \begin{pmatrix}
  \frac{n}{J}(\hat{m} - m)S^{-1}(\hat{m} - m) \\
  \sqrt{\frac{n}{J}}S^{-1/2}(\hat{m} - m)
\end{pmatrix} \overset{d}{=} \begin{pmatrix}
    \sum_t(Z_t^*)^2 \\
    Z^*_1 \\
    \vdots \\
    Z^*_J
\end{pmatrix} = \begin{pmatrix}
    \sum_t(Z_t^*)^2 \\
    Z^*
\end{pmatrix},
\end{align*}}
where $Z^* = (Z_1^*,\dots,Z_J^*)'$ are independent standard normal. Thus, for $J\rightarrow \infty$,

{\footnotesize \begin{align*}
    \begin{pmatrix}
    \frac{(\sum_t(Z_t^*)^2 - J)}{\sqrt{2J}}  \\
    Z^* 
\end{pmatrix} \overset{d}{\rightarrow} \mathcal{N}(0,I_{J+1}),
\end{align*}}
as $E[(Z_t^*)^2Z_s] = 0$ for all $t \neq s$ by the properties of the standard normal distribution. Next, due to the Morrison-Sherman-Woodbury formula and the definition of $S$

{\footnotesize \begin{align*}
    (JS)^{-1} &= \left(I_J(J-1) + \iota\iota'\right)^{-1} = \frac{1}{J-1}\left[I_J - \frac{1}{(2J-1)}\iota\iota'\right].
\end{align*}}
\paragraph{\textbf{Wald Test:}}
We can decompose {\footnotesize \begin{align*}
     \hat{t}_{\ell_2}^2
     &= \frac{n}{J}(\hat{m} - m + m)'S^{-1}(\hat{m} - m + m) \\
     &= \frac{n}{J}(\hat{m} - m)'S^{-1}(\hat{m} - m) + 2\frac{n}{J}m'S^{-1}(\hat{m} - m) + \frac{n}{J}m'S^{-1} m \\
     &= {Z^*}'Z^*  + 2\sqrt{{n}}\mu'(JS)^{-1/2}S^{-1/2}\sqrt{\frac{n}{J}}(\hat{m} - m) + \frac{n}{J}m'm - \frac{n}{J-1}\frac{m'\iota\iota'm}{(2J - 1)}. 
\end{align*}}
Hence, we obtain an upper bound of the power under local alternatives as {\footnotesize \begin{align*}
    P_{\xi}\left( \hat{t}_{\ell_2}^2 > CV_{\chi^2(J),1-\alpha}\right) 
    &\leq P\left(  {Z^*}'Z^*  + 2 \xi' S^{-1/2} Z +  > CV_{\chi^2(J),1-\alpha} - \frac{||\xi||_2^2}{J} \right) \\
    &\leq P\left( \frac{{Z^*}'Z^*  - J}{\sqrt{2J}}  + 2 \xi' \frac{1}{\sqrt{2J}}(JS)^{-1/2} Z^*   > \frac{CV_{\chi^2(J),1-\alpha} - J}{\sqrt{2J}} - \frac{1}{2J}\frac{||\xi||_2^2}{J} \right).
\end{align*}}
Now note that {\footnotesize \begin{align*}
    \xi'(JS)^{-1}\xi \leq ||\xi||_2^2 ||(JS)^{-1}||_2 \leq \frac{||\xi||_2^2}{J}
\end{align*}} This together with joint convergence above, we have that, for large $J$ {\footnotesize \begin{align*}
     P_{\xi}\left( \hat{t}_{\ell_2}^2 > CV_{\chi^2(J),1-\alpha}\right) 
     &\leq P\left( Z^{**}  > \frac{z_{1-\alpha}}{\sqrt{1 + {2||\xi||^2_2}/{J}}} - \frac{1}{\sqrt{2J}}\frac{{||\xi||_2^2}/{J}}{\sqrt{1 + {2||\xi||^2_2}/{J}}} \right) + o(1),
\end{align*}}
where $Z^{**} \sim N(0,1)$. 
\paragraph{\textbf{Supremum Test:}}
Denote as $CV_{Gumbel(a,b),\alpha}$ the $\alpha$-quantile of the $Gumbel(a,b)$ distribution. We first show that, for large $J$, we have that 

{\footnotesize \begin{align}
    \frac{\sup_{t\neq 0}|Z_t + Z_0/\sqrt{J}| - a_J}{b_J} \overset{d}{=} Gumbel(0,1) + o(1), \label{eq_app_gumbel1}
\end{align}}
where %{\footnotesize \begin{align*}
    $a_J = \sqrt{2 \log J} - \frac{\log \log J + \log 4\pi}{2 \sqrt{2 \log J}}$ and
$b_J = \frac{1}{\sqrt{2 \log J}}$.
%\end{align*}}
% Moreover, note that, for large $J$, we also have that {\footnotesize \begin{align}
%     \frac{\sup_t|Z_t + Z_0/\sqrt{J}| - a_J}{b_J} \overset{d}{=} Gumbel(0,1)
% \end{align}}
To do so, we verify Lemma 3.2 of \citeA{Berman1964LimitSequences}. Lemma 3.1 then implies the distributional result. Note that $\{Z_0/\sqrt{J} + Z_j\sqrt{1-1/J}\}_{j>0}$ is a mean zero, unit variance Gaussian sequence with covariance $r_j = E[Z_jZ_0] = 1/J$. Lemma 3.2 of \citeA{Berman1964LimitSequences} then corresponds to showing that

{\footnotesize \begin{align*}
    \sum_{j=1}^J|r_j|(J-j)(1-r_j^2)^{-1/2}J^{-2/(1+|r_j|)}(\log J)^{1/(1+|r_j|)} = o(1).
\end{align*}}
Plugging in $r_j$ and simplifying yields {\footnotesize \begin{align*}
     \sum_{j=1}^J&J^{-1}(J-j)(1-J^{-2})^{-1/2}J^{-2/(1+J^{-1})}(\log J)^{1/(1+J^{-1})} \\
     &= J^{-1}(1-J^{-2})^{-1/2}J^{-2/(1+J^{-1})}(\log J)^{1/(1+J^{-1})}(J^2 - J^2/2) \\
     &\lesssim (\log J)^{1/(1+J^{-1})} J^{-1 - 2/(1+J^{-1}) + 2} 
\end{align*}}
which is $o(1)$ if $J > 1$, which verifies \eqref{eq_app_gumbel1}. 
Next, we derive the local power of the supremum test.
Let $CV_{J,1-\alpha}$ denote the $1-\alpha$ critical value of $\sup_{t \neq 0}|Z_t\sqrt{1-1/J} + Z_0/\sqrt{J}|$. We can bound the local power as follows 

{\footnotesize \begin{align*}
    P_{\xi}\left( \hat{t}_{\ell_\infty} > CV_{J,1-\alpha}\right) 
    &= P_{\xi}\left( \sup_{t \neq 0}\left|\sqrt{\frac{n}{J}}\hat{m}_t\right| > CV_{J,1-\alpha}\right) \\
    &\leq P_{\xi}\left( \sup_{t \neq 0}\left|\sqrt{\frac{n}{J}}(\hat{m}_t - m_t)\right| > CV_{J,1-\alpha} - \frac{\sqrt{n}\sup_t|m_t|}{\sqrt{J}}\right) \\
    &= P_{\xi}\left( \sup_{t \neq 0}\left|Z_t\sqrt{1-1/J} + Z_0/\sqrt{J}\right| > CV_{J,1-\alpha} - \frac{||\sqrt{n}m||_{\infty}}{\sqrt{J}}\right) \\
    &=  P\left( \frac{\sup_{t \neq 0}\left|Z_t\sqrt{1-1/J} + Z_0/\sqrt{J}\right| - a_J}{b_J} >  \frac{CV_{J,1-\alpha} - a_J}{b_J} - \frac{||\xi||_{\infty}}{\sqrt{J}b_J}\right) + o(1) \\
    &= P\left( Gumbel(0,1) >  CV_{Gumbel(0,1),1-\alpha} - \sqrt{\frac{2\log J }{J}}||\xi||_{\infty}\right) + o(1).
\end{align*}}

\paragraph{\textbf{Decomposition Test:}}
Now consider the simple two-sided test for decomposition parameter based on Theorem \ref{thm_asyN1}. Its power given local alternatives is given by 

{\footnotesize \begin{align*}
P_{\xi}&\left(\left|\frac{\sqrt{n}\hat{\Delta}_1}{\Sigma_{{\Delta}_1}^{1/2}}\right| > z_{1-\alpha/2} \right) \\
&= P_{\xi}\left(\sqrt{n}\frac{\hat{\Delta}_1}{\Sigma_{{\Delta}_1}^{1/2}} > z_{1-\alpha/2} \right) + P\left(\sqrt{n}\frac{\hat{\Delta}_1}{\Sigma_{{\Delta}_1}^{1/2}} < z_{\alpha/2} \right) \\
    &= P_{\xi}\left(\sqrt{n}\frac{(\hat{\Delta}_1-\Delta_1)}{\Sigma_{{\Delta}_1}^{1/2}} + \sqrt{n}\frac{\Delta_1}{\Sigma_{{\Delta}_1}^{1/2}} > z_{1-\alpha/2} \right) \\
    &\quad + P_{\xi}\left(\sqrt{n}\frac{(\hat{\Delta}_1-\Delta_1)}{\Sigma_{{\Delta}_1}^{1/2}} + \sqrt{n}\frac{\Delta_1}{\Sigma_{{\Delta}_1}^{1/2}}< z_{\alpha/2} \right) \\
    &= P\left(Z^* + \frac{1}{\Sigma_{{\Delta}_1}^{1/2}}\sum_te_{ta}\xi_t > z_{1-\alpha/2}\right) + P\left(Z^* + \frac{1}{\Sigma_{{\Delta}_1}^{1/2}}\sum_te_{ta}\xi_t  < z_{\alpha/2}\right) + o(1)\\
    &=1- \Phi\left(z_{1-\alpha/2} -\ip{e_a,\xi}_{\ell^2(\ta)} \right) + \Phi\left(z_{\alpha/2} - \ip{e_a,\xi}_{\ell^2(\ta)} \right) + o(1),
\end{align*}}
where the last step follows from the simplifying assumption that all potential outcome mean differences have variance one, see also the variance of the leading term derivation in Section \ref{sec_proof_leadingterm1}.

\subsection{Simulation} \label{app:power-simul}

The Figure \ref{fig_powerplots1} is obtained by applying the different tests to 10,000 draws from the following stylized data generating process for different values of $J$: $T \in \{0,1,...,J\}$, $P(T = 0) = 1/2, P(T = j) = (2J)^{-1}$ for $j \in \{1,...,J\}$, $a(T) = \mathbbm{1}(T > 0)$, $X \sim Bernoulli(1/2)$, $g(X) = \mathbbm{1}(X=1)$, $\mu_0(X) = 0$, $\mu_j(1) = \frac{j-1}{J-1}$ for $j \in \{1,...,J\}$, $\xi_j = 
c  \text{for } j = 1, \dots, round(J^a)$ and $\xi_j= 0  \text{for } j = round(J^a) + 1, \dots, J$, $\mu_j(0) = \mu_j(1) - \xi_j$ for $j \in \{1,...,J\}$, $Y = \sum_{g=0}^1 \sum_{j=0}^J \mathbbm{1}(X=g) \mathbbm{1}(T=j) \mu_j(g) + \varepsilon $, where $\varepsilon \sim \mathcal{N}(0,1)$.
% {\footnotesize\begin{itemize}
%     \item $T \in \{0,1,...,J\}$
%     \item $P(T = 0) = 1/2, P(T = j) = (2J)^{-1}$ for $j \in \{1,...,J\}$
%     \item $a(T) = \mathbbm{1}(T > 0)$
%     \item $X \sim Bernoulli(1/2)$
%     \item $g(X) = \mathbbm{1}(X=1)$
%     \item $\mu_0(X) = 0$
%     \item $\mu_j(1) = \frac{j-1}{J-1}$ for $j \in \{1,...,J\}$
%     \item $\xi_j = \begin{cases}
% c & \text{for } j = 1, \dots, round(J^a) \\
% 0 & \text{for } j = round(J^a) + 1, \dots, J
% \end{cases}$
%     \item $\mu_j(0) = \mu_j(1) - \xi_j$ for $j \in \{1,...,J\}$
%     \item $Y = \sum_{g=0}^1 \sum_{j=0}^J \mathbbm{1}(X=g) \mathbbm{1}(T=j) \mu_j(g) + \varepsilon $ where $\varepsilon \sim \mathcal{N}(0,1)$
% \end{itemize}}
The design parameters are:
%{\footnotesize\begin{itemize}
  $a = 1, c = 2/5$: dense case displayed in Figure \ref{fig:plot1} and
    $a = 1/2, c = 1/2$: sparse case displayed in Figure \ref{fig:plot2}
%\end{itemize}}
The simulations apply the estimator from Section \ref{sec_estimation1} with true nuisance parameters for simplicity. The full code is provided as \texttt{Power\_simulation.R} in the replication package.

%     \begin{table}[!h]
%     \centering 
%     \caption{Local Power Comparison}
%     \begin{tabular}{l|c}
%       Metric for Test   &  Approximate Local Power when $\sqrt{n}m \rightarrow \xi \in \mathbb{R}^{J}$ \\ \hline \\[-0.5ex]
%       Weighted $\ell^2$ (Wald) &    $ 1 - \Phi\left( \frac{z_{1-\alpha}}{\sqrt{1 + {2||\xi||^2_2}/{J}}} - \frac{1}{\sqrt{2J}}\frac{{||\xi||_2^2}/{J}}{\sqrt{1 + {2||\xi||^2_2}/{J}}} \right)$ \\
%       $\ell^\infty$ (Supremum) &  $1 - F_G\left(F^{-1}_{G,{1-\alpha}} - \sqrt{\frac{2\log J }{J}}||\xi||_{\infty}\right)$  \\
%       $\Delta_1(g,g')$ (Decomposition)  & $1 - \Phi\bigg(z_{1-\alpha/2} -\ip{e_t,\xi}_{\ell^2(\ta)} \bigg) + \Phi\bigg(z_{\alpha/2} - \ip{e_t,\xi}_{\ell^2(\ta)} \bigg)$  \\[1ex] \hline
%     \end{tabular} 
%    \begin{justify} \footnotesize
%     $\Phi(\cdot)$ and $z_{\alpha}$ are the cumulative distribution function and $\alpha$-quantile of the standard normal distribution. $F_G(\cdot)$ and $F^{-1}_{G,\alpha}$ are the the cdf and $\alpha$-quantile function of the $Gumbel(0,1)$ distribution. Here $<e_t,\xi>_{\ell^2(\ta)} = \sum_{t\in \ta}e_t\xi_t$. $J = |\ta|$.
% \end{justify}
% \end{table}

{
}

\section{Decomposition Parameter Influence Functions} \label{sec_app_implement1}

%\subsection{Decomposition Parameter Influence Functions}

Table \ref{tab_IF_text1} defines the influence functions of eight primitive parameters underlying all decomposition parameters. Here we define them for later reference: 
     $\theta_{a,g,t,1} := e_{ta}(\xg) \mu_t(\xg)$
     , $\theta_{a,g,t,2} := E[e_{ta}(X)|X\in\xg]\mu_t(\xg)$
    , $\theta_{a,g,t,3} := E[e_{ta}(X)\mu_t(X)|X\in \xg]$
     , $\theta_{a,g,t,4} := e_{ta}\mu_t(\xg)$
    , $\theta_{a,g,t,5} := e_{ta}(\xg)\mu_t$ 
    % \item $\theta_{tag6} := Cov(e_{ta}(X),\mu_t(X)|X\in \xg) = E[e_{ta}(X)\mu_t(X)|X\in \xg] - e_{ta}(\xg) \mu_t(\xg) = \theta_{a,g,t,3} - \theta_{a,g,t,1}$ 
    , $\theta_{a,g,t,6} := e_{t}(\xg) \mu_t(\xg)$
    , $\theta_{a,g,t,7} := E[e_t(X)\mu_t(X)|X\in \xg]$
    % \item $\theta_{tg3} := Cov(e_t(X),\mu_t(X)|X\in \xg) = E[e_t(X)\mu_t(X)|X\in \xg] - e_{t}(\xg) \mu_t(\xg) = \theta_{a,g,t,7} - \theta_{a,g,t,6}$
    , $\theta_{a,g,t,8} := e_{ta}\mu_t$.  
% {\footnotesize\begin{enumerate}
%     \item $\theta_{a,g,t,1} := e_{ta}(\xg) \mu_t(\xg)$
%     \item $\theta_{a,g,t,2} := E[e_{ta}(X)|X\in\xg]\mu_t(\xg)$
%     \item $\theta_{a,g,t,3} := E[e_{ta}(X)\mu_t(X)|X\in \xg]$
%     \item $\theta_{a,g,t,4} := e_{ta}\mu_t(\xg)$
%     \item $\theta_{a,g,t,5} := e_{ta}(\xg)\mu_t$ 
%     % \item $\theta_{tag6} := Cov(e_{ta}(X),\mu_t(X)|X\in \xg) = E[e_{ta}(X)\mu_t(X)|X\in \xg] - e_{ta}(\xg) \mu_t(\xg) = \theta_{a,g,t,3} - \theta_{a,g,t,1}$ 
%     \item $\theta_{a,g,t,6} := e_{t}(\xg) \mu_t(\xg)$
%     \item $\theta_{a,g,t,7} := E[e_t(X)\mu_t(X)|X\in \xg]$
%     % \item $\theta_{tg3} := Cov(e_t(X),\mu_t(X)|X\in \xg) = E[e_t(X)\mu_t(X)|X\in \xg] - e_{t}(\xg) \mu_t(\xg) = \theta_{a,g,t,7} - \theta_{a,g,t,6}$
%     \item $\theta_{a,g,t,8} := e_{ta}\mu_t$    
% \end{enumerate}}
Their respective influence functions $\Psi_{\theta_{a,g,t,p}}$ for $p\in \{1,...8\}$ are derived using the discretization approach discussed in \citeA{Kennedy2024SemiparametricReview}. %in %Supplementary Appendix \ref{app:if-detail}. 
Now we can write all $d$ parameters as combination of the eight primitives:
     $d_0(a) = \sum_{t\in\ta} \theta_{a,g,t,8}$
    , $d_1(a,g) = \sum_{t\in\ta} \theta_{a,g,t,4} - \theta_{a,g,t,8}$
    , $d_2(a,g) = \sum_{t\in\ta} \theta_{a,g,t,5} - \theta_{a,g,t,8}$
    , $d_3(a,g) = \sum_{t\in\ta} \theta_{a,g,t,1} - \theta_{a,g,t,5} - \theta_{a,g,t,4} + \theta_{a,g,t,8}$
    , $d_4(a,g) = \sum_{t\in\ta} (\theta_{a,g,t,7} - \theta_{a,g,t,6}) / P(T\in\mathcal{T}_a,X\in\xg)$
    , $d_{4'}(a,g) = \sum_{t\in\ta} \theta_{a,g,t,3} - \theta_{a,g,t,3}$
    , $d_{5}(a,g) = \sum_{t\in\ta} \theta_{a,g,t,2} - \theta_{a,g,t,1}$. 
% {\footnotesize\begin{enumerate}
%     \item $d_0(a) = \sum_{t\in\ta} \theta_{a,g,t,8}$
%     \item $d_1(a,g) = \sum_{t\in\ta} \theta_{a,g,t,4} - \theta_{a,g,t,8}$
%     \item $d_2(a,g) = \sum_{t\in\ta} \theta_{a,g,t,5} - \theta_{a,g,t,8}$
%     \item $d_3(a,g) = \sum_{t\in\ta} \theta_{a,g,t,1} - \theta_{a,g,t,5} - \theta_{a,g,t,4} + \theta_{a,g,t,8}$
%     \item $d_4(a,g) = \sum_{t\in\ta} (\theta_{a,g,t,7} - \theta_{a,g,t,6}) / P(T\in\mathcal{T}_a,X\in\xg)$
%     \item $d_{4'}(a,g) = \sum_{t\in\ta} \theta_{a,g,t,3} - \theta_{a,g,t,3}$
%     \item $d_{5}(a,g) = \sum_{t\in\ta} \theta_{a,g,t,2} - \theta_{a,g,t,1}$
% \end{enumerate}}
Their respective influence functions follow by applying the chain rule, e.g. $\Psi_{d_1(a,g)} := \sum_{t\in\ta} \Psi_{\theta_{a,g,t,4}} - \Psi_{\theta_{a,g,t,8}}$. The only parameter that is more involved in this regard is $d_4(a,g)$. Its IF is obtained via chain rule $\Psi_{d_4(a,g)} = \sum_{t\in\ta}  \Psi_{\theta_{a,g,t,7}} / e_{ag} - \Psi_{\theta_{a,g,t,6}}  / e_{ag} - \Psi_{e_{ag}}  (\theta_{a,g,t,7} - \theta_{a,g,t,6}) /  e_{ag}^2$,
where $\Psi_{e_{ag}} = \mathbbm{1}(T\in\mathcal{T}_a,X\in\xg) - e_{ag}$.
Similarly, the influence functions of $\delta$ parameters with $j \in \{0,1,2,3,4,4',5\}$ follow as $\Psi_{\delta_j(a,a',g)} = \Psi_{d_j(a,g)} - \Psi_{d_j(a',g)}$, and for $\Delta$ parameters as $\Psi_{\Delta_j(a,a',g,g')} = \Psi_{\delta_j(a,a',g)} - \Psi_{\delta_j(a,a',g')}$.

\section{Application} \label{app:app}

\begin{figure}[!htbp]
    \centering
    \caption{Decomposition of Adjusted Gender Differences in Job Corps} \label{fig:decomp-adim}
        \centering
        \includegraphics[width=\textwidth]{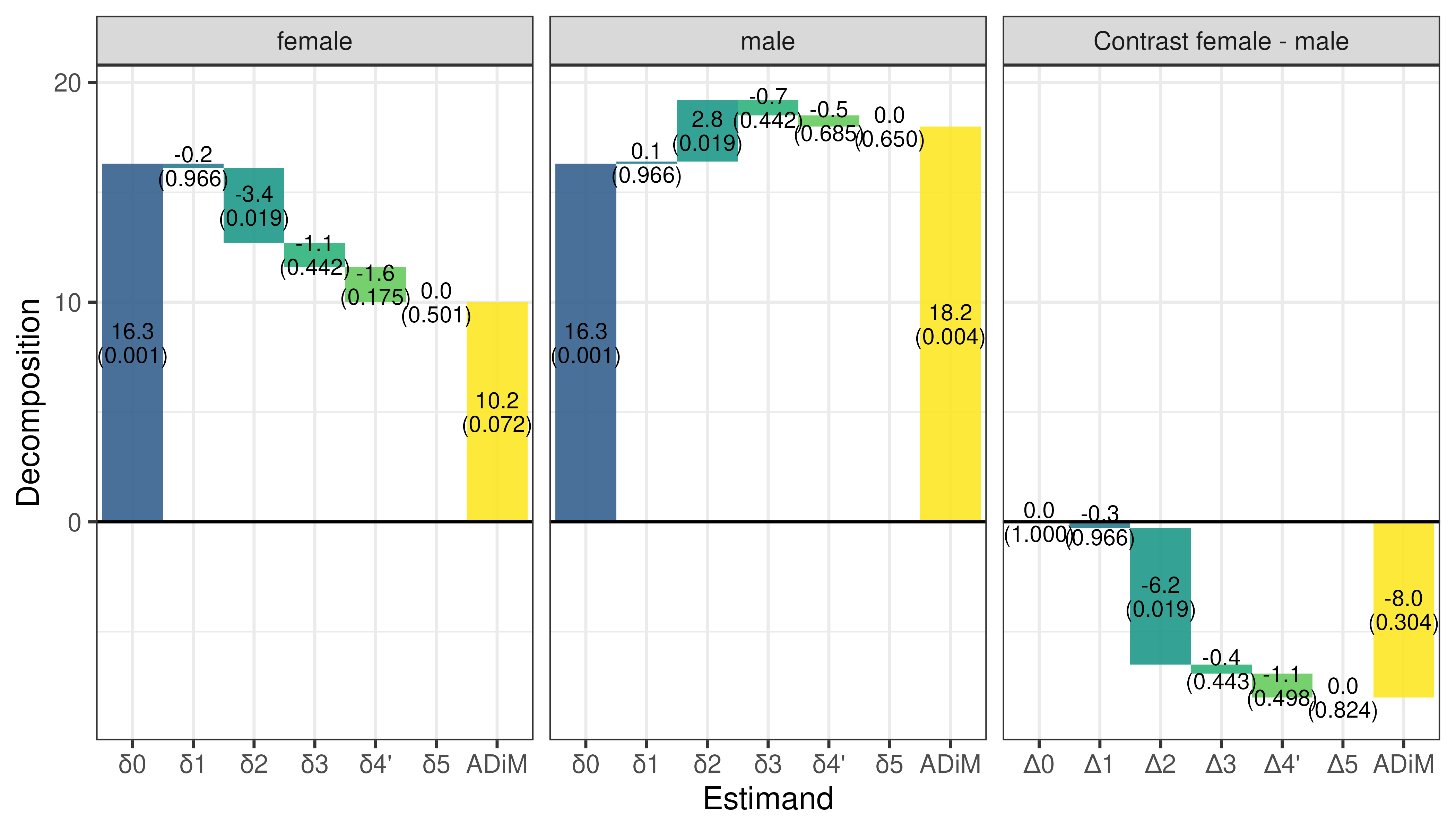}
    \subcaption*{\textit{Notes:} The numbers in the bars show the point estimate and the $p$-value in parentheses. They are obtained using the estimation and inference procedures discussed in Section \ref{sec_estimation1}.}
\end{figure}

Figure \ref{fig:decomp-adim} shows the decomposition of the adjusted differences-in-means provided in Section \ref{sec:adim-decomp}. As expected in a randomized controlled trial, $\Delta_{4'}$ is very close to $\Delta_{4}$ and $\Delta_{5}$ is basically zero.

\section{Extension to Aggregated Instrumental Variables} \label{app:IV1}
\subsection*{Setup and Identifying Assumptions}

We now extend the framework to the conditional LATE/IV setup when the conditionally independent and excluded instrument is potentially multi-valued, but aggregates are used. We express adjusted group Wald estimands relative to a synthetic stratified experiment that preserves the identifying content of the aggregated instrument while eliminating within-group version-level selection. The resulting representation clarifies which forms of heterogeneity are relevant for group differences in LATE under aggregated instruments and how assignment mechanisms and potential-outcome schedules jointly determine Wald estimands.

Let $T\in\mathcal T$ denote a multi-valued effective instrument and let $Z=a(T)\in\{0,1\}$ denote the aggregated instrument. Let $\mathcal T_a$ and $\mathcal T_{a'}$ denote the induced partition. Let $X$ denote covariates and $g(X)$ a coarse grouping with support $X_g$. For each instrument value $t\in\mathcal T$, define potential treatment $D(t)$. Potential outcomes for treatment level $d$ and instrument level $t$ are  defined as $Y(d,t)$. Throughout, assume:

\noindent {\textbf{Assumption IV (LATE Adapted to Aggregate Instrument)}}
%\begin{assump}[Regularity, Treatment, and Learning Rates]
\textit{
\begin{enumerate}[itemsep=0pt] \singlespacing
\item[IV.1)] Exclusion:
$Y(t,d)=Y(d)$ for all $t,d$.
\item[IV.2)] Conditional instrument independence:
$\{Y(t),D(t):t\in\mathcal T\} \perp T\mid X$.
\item[IV.3)] Monotonicity: For all $t\in\mathcal T_a$ and $t'\in\mathcal T_{a'}$,
$D(t) \ge D(t')$
\item[IV.4)] Relevance: On a set of positive probability we have
$E[D\mid Z=1,X] \neq E[D\mid Z=0,X].$
\end{enumerate}
}

% \textbf{Assumption IV.1 (Conditional instrument independence).}
% \begin{align*}
% \{Y(t),D(t):t\in\mathcal T\} &\perp T\mid X.
% \end{align*}

% \textbf{Assumption IV.2 (Exclusion).}
% $Y(t,d)=Y(d)$ for all $t,d$.

% \textbf{Assumption IV.3 (Monotonicity with respect to aggregation).}
% For all $t\in\mathcal T_a$ and $t'\in\mathcal T_{a'}$,
% \begin{align*}
% D(t) &\ge D(t') \quad \text{a.s.}
% \end{align*}

% \textbf{Assumption IV.4 (Relevance).} On a set of positive probability we have
% \begin{align*}
% E[D\mid Z=1,X] &\neq E[D\mid Z=0,X].
% \end{align*}

These assumptions adapt the standard Angrist--Imbens LATE framework to an underlying multi-valued effective instrument and an observed aggregation. Conditional instrument independence and exclusion are standard in the IV literature. 
Exclusion implies that observed outcomes satisfy $Y(D(t),t) = Y(D(t)) \equiv Y(t)$. Monotonicity is imposed at the level of the aggregated instrument, which is natural when aggregation reflects increasing encouragement intensity. Relevance is the usual first-stage condition and is required only for the aggregated instrument, making the assumptions no stronger than those typically invoked in applied LATE analyses with coarsened or aggregated instruments. These assumptions ensure that Wald estimands constructed from the aggregated instrument admit a causal LATE-type interpretation corresponding to the complier population induced by shifts in the distribution of effective instrument versions between aggregates.  
 
 %ensure that Wald estimands constructed from the aggregated instrument admit a LATE interpretation while allowing for heterogeneity across effective instrument versions. However, it is obtained via an instrument that is an aggregate and thus the complying subpopulation is defined with respect to this aggregate that is deterministic in its higher dimensional effective instrument. 

\subsection*{Primitives and Notation}

% For each $t\in\mathcal T$ and outcome $R\in\{Y,D\}$, define
% \begin{align*}
% \mu_t^R(x) &= E[R(t)\mid X=x], &
% e_{ta}(x) &= P(T=t\mid X=x,T\in\mathcal T_a).
% \end{align*}
% Let $e_a(x)=(e_{ta}(x))_{t\in\mathcal T_a}$ and $\mu^R(x)=(\mu_t^R(x))_{t\in\mathcal T_a}$. For vectors $u,v$ of equal dimension, define the inner product $\langle u,v\rangle=\sum_{t\in\mathcal T_a}u_t v_t$.

For each $t\in\mathcal T$ and outcome $R\in\{Y,D\}$, define

{\footnotesize\begin{align*}
\mu^R(x) &= (\mu_t^R(x))_{t\in\mathcal T}, \quad \mu_t^R(x)=E[R(t) | X=x], \quad 
\mu^R = E[\mu^R(X)],\\
e(x) &= (e_t(x))_{t\in\mathcal T}, \quad e_t(x)=P(T=t | X=x) , \quad  e_t = E[e_t(X)].
\end{align*}}
Let $\langle u,v\rangle=\sum_{t\in\mathcal T}u_tv_t$ denote the inner product on $\mathbb R^{|\mathcal T|}$. For a partition cell $S\in\{\mathcal T_a,\mathcal T_{a'}\}$, define the (normalized) within-cell version propensity vector
$e_S(x)\in\mathbb R^{|\mathcal T|}$ by

{\footnotesize\begin{align*}
(e_S(x))_t &=
\begin{cases}
\dfrac{e_t(x)}{P(T\in S | X=x)}, & t\in S,\\[6pt]
0, & t\notin S,
\end{cases}
\end{align*}}
whenever $P(T\in S | X=x)>0$. Thus $e_S(x)$ equals the conditional version distribution
$P(T=t | X=x, T\in S)$, embedded in $\mathbb R^{|\mathcal T|}$ by zeros outside $S$. 
All inner products and covariance objects below are taken in $\mathbb R^{|\mathcal T|}$ using these representations. For notational consistency with the main paper, we write $e_a()$ as its zero-extended within-aggregation propensity instead of $e_{\mathcal{T}_a}()$.

\subsubsection*{Synthetic Stratified Experiment E1}

Fix the joint distribution of $(X,\{Y(t),D(t)\}_{t\in\mathcal T})$. Define a counterfactual assignment mechanism for the instrument, denoted \textbf{E1}, that differs from the observed data-generating process only in the conditional distribution of $T$ given covariates $X$. Under E1, instrument versions are assigned according to

{\footnotesize\begin{align*}
P^{E1}(T=t\mid X=x)
&=
\begin{cases}
e_{ta}(\mathcal X_g) P(T\in\mathcal T_a\mid X=x), & t\in\mathcal T_a,\\[4pt]
e_{ta'}(\mathcal X_g) P(T\in\mathcal T_{a'}\mid X=x), & t\in\mathcal T_{a'}.
\end{cases}
\end{align*}}
%where still $g = g(X)$. %and 
%{\footnotesize\begin{align*}
%e_{ta}(\mathcal X_g) &= P(T=t\mid T\in\mathcal T_a, X\in X_g)
%\end{align*}
%denotes the group-level within-aggregate version propensity.
Thus, E1 preserves the conditional distribution of the aggregated instrument $Z$ given $X$, while randomizing the effective instrument $T$ independently of all remaining covariates conditional on $(Z,g)$, i.e.~$T \bigCI X \mid (Z,g)$ \text{under E1}.
By construction, E1 preserves the marginal distribution of $X$, the conditional distribution of $Z$ given $X$, and the full schedule of potential outcomes $\{Y(t),D(t)\}_{t\in\mathcal T}$. It modifies only the within-aggregate assignment of instrument versions by eliminating any dependence on covariates beyond the group indicator. For any outcome $R\in\{Y,D\}$,

{\footnotesize\begin{align*}
E^{E1}[R\mid T\in\mathcal T_a, X\in X_g]
&=
\langle e_a(\mathcal X_g), \mu^R(\mathcal X_g)\rangle,
\end{align*}}
where %$\mu^R(x) = \{\mu_t^R(x), t \in \mathcal{T}\}$ with 
$\mu_t^R(x)=E[R(t)\mid X=x]$ are conditional reduced form/first stage potential outcomes. This coincides exactly with the leading term in the (adjusted) group mean decomposition \eqref{eq:decomp-cm1}. All remaining terms in that decomposition therefore quantify deviations of the observed assignment mechanism from E1. This matches the idea behind the hypothetical estimand obtained from a synthetic experiment in Section \ref{sec:decomp-single-mean}.

\subsubsection*{Statistical Estimands}

For outcome $R$, define the adjusted aggregate contrast

{\footnotesize\begin{align*}
\delta_R^{adj}(g)
&=
E[E[R\mid T\in\mathcal T_a,X]\mid X\in X_g]
-
E[E[R\mid T\in\mathcal T_{a'},X]\mid X\in X_g].
\end{align*}}
Decompose $\delta_R^{adj}(g)$ as

{\footnotesize\begin{align*}
\delta_R^{adj}(g)
&=
\delta_R^{E1}(g)+\delta_R^{IT}(g)+\delta_R^{CA}(g),
\end{align*}}
where

{\footnotesize\begin{align*}
\delta_R^{E1}(g)
&=
\langle e_a(\mathcal X_g),\mu^R(\mathcal X_g)\rangle
-
\langle e_{a'}(\mathcal X_g),\mu^R(\mathcal X_g)\rangle,\\
\delta_R^{IT}(g)
&=
\langle \operatorname{Cov}(e_a(X),\mu^R(X)\mid X\in X_g),\mathbf 1\rangle
-
\langle \operatorname{Cov}(e_{a'}(X),\mu^R(X)\mid X\in X_g),\mathbf 1\rangle,\\
\delta_R^{CA}(g)
&=
\langle E[e_a(X)\mid X\in X_g]-e_a(\mathcal X_g),\mu^R(\mathcal X_g)\rangle
-
\langle E[e_{a'}(X)\mid X\in X_g]-e_{a'}(\mathcal X_g),\mu^R(\mathcal X_g)\rangle.
\end{align*}}
The adjusted group Wald is

{\footnotesize\begin{align*}
\mathrm{Wald}^{adj}(g)
&=
\frac{\delta_Y^{adj}(g)}{\delta_D^{adj}(g)},
\qquad
\delta_D^{adj}(g)\neq 0.
\end{align*}}
Define the synthetic-experiment Wald

{\footnotesize\begin{align*}
\mathrm{Wald}^{E1}(g)
&=
\frac{\delta_Y^{E1}(g)}{\delta_D^{E1}(g)},
\qquad
\delta_D^{E1}(g)\neq 0.
\end{align*}}
An exact ratio expansion yields

{\footnotesize\begin{align*}
\mathrm{Wald}^{adj}(g)
&=
\mathrm{Wald}^{E1}(g)
+
\frac{
\delta_Y^{IT}(g)-\mathrm{Wald}^{E1}(g)\delta_D^{IT}(g)
+
\delta_Y^{CA}(g)-\mathrm{Wald}^{E1}(g)\delta_D^{CA}(g)
}{
\delta_D^{adj}(g)
}.
\end{align*}}

\subsection*{Decomposition of the Synthetic Wald Estimand}
% Define population reference objects
% {\footnotesize\begin{align*}
% e_a &= E[e_a(X)], &
% \mu^R &= E[\mu^R(X)].
% \end{align*}
For $k=0,1,2,3$, define

{\footnotesize\begin{align*}
\delta_{k,R}(g)
&=
d_{k,R}(a,g)-d_{k,R}(a',g),
\end{align*}}
with $d_{k,R}(a,g)$ as in \eqref{eq:decomp-srct} with outcome $R \in \{Y,D\}$. Then

{\footnotesize\begin{align*}
\delta_R^{E1}(g)
&=
\sum_{k=0}^3\delta_{k,R}(g).
\end{align*}}
Define the baseline synthetic-experiment Wald

{\footnotesize\begin{align*}
\mathrm{Wald}_0^{E1}
&=
\frac{\delta_{0,Y}}{\delta_{0,D}},
\qquad
\delta_{0,D}\neq 0.
\end{align*}}
A second exact ratio expansion yields

{\footnotesize\begin{align*}
\mathrm{Wald}^{E1}(g)
&=
\mathrm{Wald}_0^{E1}
+
\sum_{k=1}^3
\frac{
\delta_{k,Y}(g)-\mathrm{Wald}_0^{E1}\delta_{k,D}(g)
}{
\delta_D^{E1}(g)
}.
\end{align*}}
For each $k\ge 1$, the contribution of channel $k$ to $\mathrm{Wald}^{E1}(g)$ is zero if and only if

{\footnotesize\begin{align*}
\delta_{k,Y}(g) &= \mathrm{Wald}_0^{E1}\,\delta_{k,D}(g).
\end{align*}}
Thus, vanishing of a Wald component does not require absence of heterogeneity; proportionality between reduced-form and first-stage components is sufficient.

The $k=1$ component captures heterogeneity in version-specific outcome and compliance schedules evaluated at a common version mix. The $k=2$ component captures heterogeneity in version assignment holding schedules fixed. The $k=3$ component captures alignment between version-mix heterogeneity and schedule heterogeneity. Individualized targeting and composition adjustment capture deviations of the observed assignment mechanism from the synthetic stratified experiment. In all cases, components affect the Wald only through non-proportional shifts in reduced form and first stage.

All estimands are continuous functions of objects that are identical to our original decomposition primitives with the instrument replacing the treatment and outcomes $R \in \{D,Y\}$ instead of $Y$ only. Thus, the debiased ML estimators and their asymptotic distributions can be obtained analogously via the delta method under analogous assumptions as for the decomposition parameters obtained under Assumption A.1 -- A.6.

\end{appendices}

\end{document}